# Deep learning—a first meta-survey of selected reviews across scientific disciplines, their commonalities, challenges and research impact

Jan Egger[1,2,3,4], Antonio Pepe[1,2], Christina Gsaxner[1,2,3], Yuan Jin[1,2,5], Jianning Li[1,2,4,6] and Roman Kern[7,8]

[1] Institute of Computer Graphics and Vision, Faculty of Computer Science and Biomedical Engineering, Graz University of Technology, Graz, Austria
[2] Computer Algorithms for Medicine Laboratory, Graz, Austria
[3] Department of Oral and Maxillofacial Surgery, Medical University of Graz, Graz, Austria
[4] Institute for AI in Medicine (IKIM), University Medicine Essen, Essen, Germany
[5] Research Center for Connected Healthcare Big Data, Zhejiang Lab, Hangzhou, Zhejiang, China
[6] Research Unit Experimental Neurotraumatology, Department of Neurosurgery, Medical University of Graz, Graz, Austria
[7] Knowledge Discovery, Know-Center, Graz, Austria
[8] Institute of Interactive Systems and Data Science, Graz University of Technology, Graz, Austria

## ABSTRACT

Deep learning belongs to the field of artificial intelligence, where machines perform tasks that typically require some kind of human intelligence. Deep learning tries to achieve this by drawing inspiration from the learning of a human brain. Similar to the basic structure of a brain, which consists of (billions of) neurons and connections between them, a deep learning algorithm consists of an artificial neural network, which resembles the biological brain structure. Mimicking the learning process of humans with their senses, deep learning networks are fed with (sensory) data, like texts, images, videos or sounds. These networks outperform the state-of-the-art methods in different tasks and, because of this, the whole field saw an exponential growth during the last years. This growth resulted in way over 10,000 publications per year in the last years. For example, the search engine PubMed alone, which covers only a sub-set of all publications in the medical field, provides already over 11,000 results in Q3 2020 for the search term 'deep learning', and around 90% of these results are from the last three years. Consequently, a complete overview over the field of deep learning is already impossible to obtain and, in the near future, it will potentially become difficult to obtain an overview over a subfield. However, there are several review articles about deep learning, which are focused on specific scientific fields or applications, for example deep learning advances in computer vision or in specific tasks like object detection. With these surveys as a foundation, the aim of this contribution is to provide a first high-level, categorized meta-survey of selected reviews on deep learning across different scientific disciplines and outline the research impact that they already have during a short period of time. The categories (computer vision, language processing, medical informatics and additional works) have been chosen according to the underlying data sources (image, language, medical, mixed). In addition, we review the common

Corresponding author
Jan Egger, egger@tugraz.at

architectures, methods, pros, cons, evaluations, challenges and future directions for every sub-category.

## INTRODUCTION

Deep learning belongs to the field of artificial intelligence, where machines execute tasks that usually require human intelligence. Deep learning is trying to achieve this by mimicking the learning of a human brain. Imitating the physiological structure of a brain, which consists of billions of neurons and connections between them, a deep learning algorithm consists of an artificial neural network of interconnected neurons (*McCulloch & Pitts, 1990*; *LeCun, Bengio & Hinton, 2015*). Also, similarly to the learning process of humans with their senses, deep neural networks are fed with sensory or sensor data like texts, images, videos or sounds (*Ravì et al., 2016*). These networks outperform the state-of-the-art methods in different tasks and, thanks to this, the whole field saw an exponential growth (*Wang et al., 2018a*; *Gibson et al., 2018*; *Pepe et al., 2020*). This resulted in way over 10,000 publications per year, in the last years. For example, alone the search engine PubMed (https://pubmed.ncbi.nlm.nih.gov/), which covers only a sub-set of all publications in the medical field, returns over 11,000 results for the search term deep learning in Q3 2020, and around 90% of these publications are from the last 3 years only. Consequently, a complete overview over the field of deep learning is already impossible to obtain and, in the near future, it will probably become difficult even for single sub-fields. However, there are several review or survey articles about deep learning, which focus on specific scientific fields or applications, for example, covering only deep learning approaches from computer vision (*Voulodimos et al., 2018*; *Guo et al., 2016*), or specific tasks like object detection (*Liu et al., 2020*; *Zhao et al., 2019*; *Jiao et al., 2019*) or object segmentation (*Garcia-Garcia et al., 2018*; *Minaee et al., 2020*). With these surveys as foundation, the aim of this contribution is to provide a first categorized and high-level meta-survey of selected works of deep learning reviews or surveys. On the top level, four main categories have been chosen for this contribution, namely: computer vision, (natural) language processing, medical informatics and additional works. The reasons behind this course of action was the underlying characteristics of data sources and to have about the same number of reviews for every main category with a well-balanced distribution. Although the last category could be further divided, this would lead to main categories with a small number of reviews; even only one review for some niche fields. Table 1 gives an overview of the four main categories and the number of screened reviews for each of them. Further, it presents the sum of the overall references and citations per category, to provide an impression of how comprehensive and influential the fields are. The subsequent tables from the single categories present more details for each of the

**Table 1 Overview of published reviews in deep learning divided into the categories: Computer vision, language processing, medical informatics and additional deep learning surveys.** The table presents also the sum of the overall references and citations for the single categories.

| Categories | Number of publications | Years | Number of references | Citations (until August 2020) | Preprints |
|---|---|---|---|---|---|
| Computer vision | 18 | 2016–2020* | 3,624 | 3,923 | Yes |
| Language processing | 14 | 2016–2020 | 2,109 | 2,490 | Yes |
| Medical informatics | 12 | 2016–2020 | 2,210 | 6,722 | No |
| Additional works | 17 | 2016–2020 | 3,481 | 4,171 | Yes |
| Sum | 61 | – | 11,424 | 17,306 | – |

**Note:**
[*] Note, there is one survey that was published in a journal in 2021, but existed already as preprint in 2020, hence we included it within this meta-survey.

main categories. The tables present the sub-categories and the corresponding publications, and again, also the number of references and citations for each of these sub-categories. Hence, these selected works of deep learning reviews or surveys across scientific disciplines depict the research impact they already had within a relatively short time. Note that the deep learning reviews selected for this contribution present themselves mostly an overview of (selected) deep learning works in a specific field and categorize them in sub-sections or areas. Therefore, this course of action is also applied to this meta-survey. The reason for this is that deep learning algorithms have often been applied to completely different datasets and modalities, which makes it difficult to combine them in a systematic survey as it can be seen in the referenced reviews.

## Search strategy

For this meta-survey a search in IEEE Xplore Digital Library, Scopus, DBLP, PubMed, Web of Science and Google Scholar for the keyword 'Deep Learning' together with any keyword between 'Review', 'Survey' was performed. Based on titles and abstracts, all records, which were not actual review or survey contributions, were excluded. This ultimately resulted in a total number of 61 review or survey publications about deep learning, which will be covered within this meta-survey. Summarized, this high-level meta-survey gives a snapshot overview of published deep learning reviews (status as of August 2020) and a compact overview of the search results can be found in the subsequent Tables of the corresponding sections of this contribution. Note that this meta-survey includes a few selected preprints. However, some of these have already up to one hundred or even several hundreds of citations, and hence have proven to be of high interest for the community and it can be expected that they will be published in a peer-reviewed venue sooner or later. These reviews were included as they cover specific and interesting research areas that have not been covered elsewhere yet.

## Manuscript outline

The core of this meta-survey explores exclusively reviews and surveys on deep learning. Because some of the included reviews cover up to several hundred publications themselves, only high-level summaries and excerpts are given to keep the manuscript concise for the reader. Hence, every review publication is summarized in around 100 to 200 words and, thus, every sub-category has around 100 to up to a few hundred words, depending on the amount of review contributions in this area. The classification and arrangement of the presented deep learning reviews should enable the interested reader to dive deeper into specific categories and sub-categories by pointing to the associated publications. We also mention already reviewed deep learning architectures, like CNNs. However, all architectures, methods, *etc*., will be outlined in detail within the next sections, called: *Going deeper: common architectures, methods, evaluations, pros, cons, challenges and future directions* of the categories. Hence, the second parts of this first section also outline the fundamental deep learning concepts. Summarized, the following main sections of this meta-survey are organized as follows: Section two introduces the deep learning reviews or surveys on a high (meta) level divided into four main categories: *computer vision*, *language processing*, *medical informatics* and *additional works*, and presents the common architectures, methods, pros, cons, evaluations, challenges and future directions for every sub-category. Section "Conclusion and Discussion" concludes and discusses the contributions and outlines areas of future directions, also on a high (meta) level.

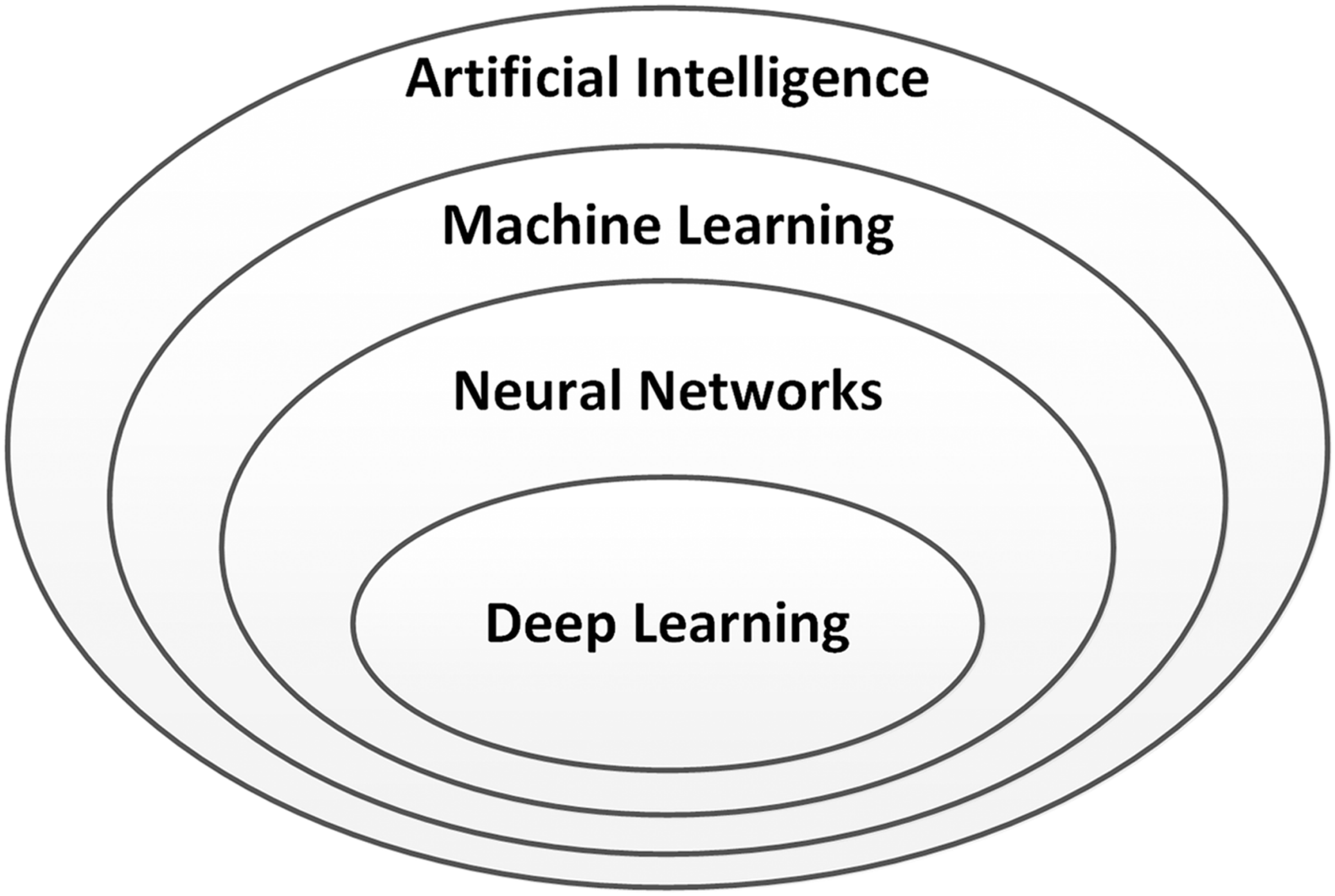

**Figure 1** **Relationship between deep learning and artificial intelligence.**
Full-size DOI: 10.7717/peerj-cs.773/fig-1

Furthermore, for readers with a particular interest towards the medical field, a systematic meta-survey about medical deep learning surveys, which are only partially covered within this contribution, is also available (*Egger et al., 2020*).

## Fundamental deep learning concepts

Deep learning (*Sze et al., 2017*; *Goodfellow et al., 2016*) is an important part of the discipline of artificial intelligence (AI), which was coined by John McCarthy. Figure 1 shows the relationship of deep learning to the whole field of artificial intelligence.

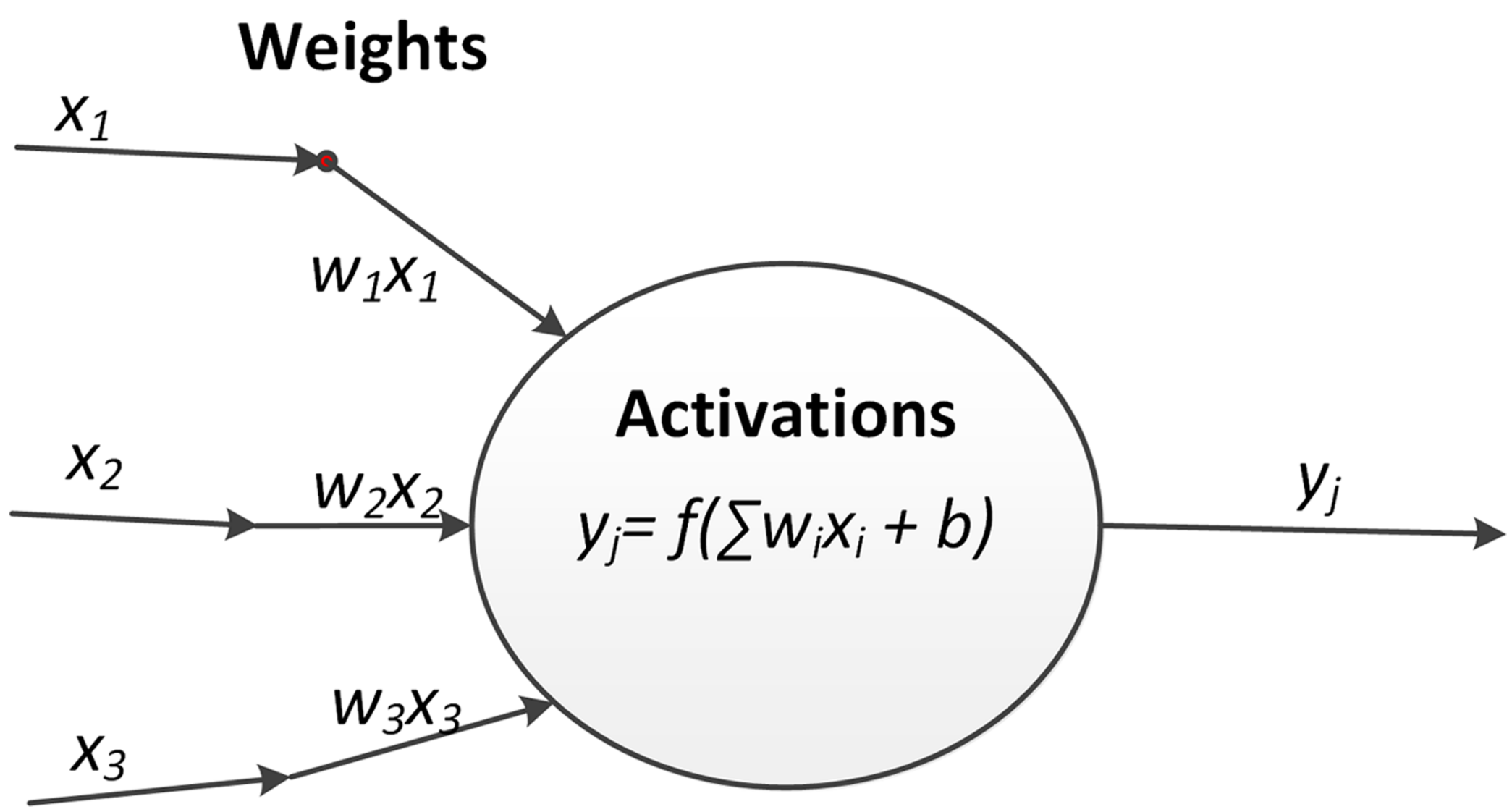

**Figure 2** **The outputs of neuron related to the inputs.** Full-size DOI: 10.7717/peerj-cs.773/fig-2

As one of the main subfields of AI, the study of machine learning, which was defined by Arthur Samuel in 1959, makes it possible for machines to learn certain patterns in data, without being further taught or programmed by humans. This process is also referred to as training. As a result, the machines can complete many tasks without hand-crafted approaches created by humans. Within the field of machine learning, neural networks, as the name suggests, aim to emulate how a human brain works, even though only in a highly abstracted way. Like the real brain, the neural networks comprise mainly neurons and synapses, which are usually called artificial neurons and connections.

## Neuron and neural network

As shown in Fig. 2, a neuron in machine learning is simply a mathematical function. The inputs are multiplied by weights $w$ and summed together. Additionally, a bias $b$ may be added. This weighted sum is then passed to a function $f$, which is usually non linear and the output of the neuron. During the forward propagation of data through a neural network, this procedure is applied to each neuron:

$$y_j = f(net_j) = f\left(\sum_{i=1}^{n} w_{ij}x_i + b_j\right) \tag{1}$$

where $x_i$ and $y_j$ are the inputs and output of the neuron and $w_{ij}$ and $b_j$ are the weights and bias terms. At the beginning of the training process these terms are typically randomly initialized. The input $net_j$ represents the weighted sum of outputs from previous neurons. A common activation functions is the logistic Sigmoid function:

$$y_j = \frac{1}{1 + e^{-net_j}} \tag{2}$$

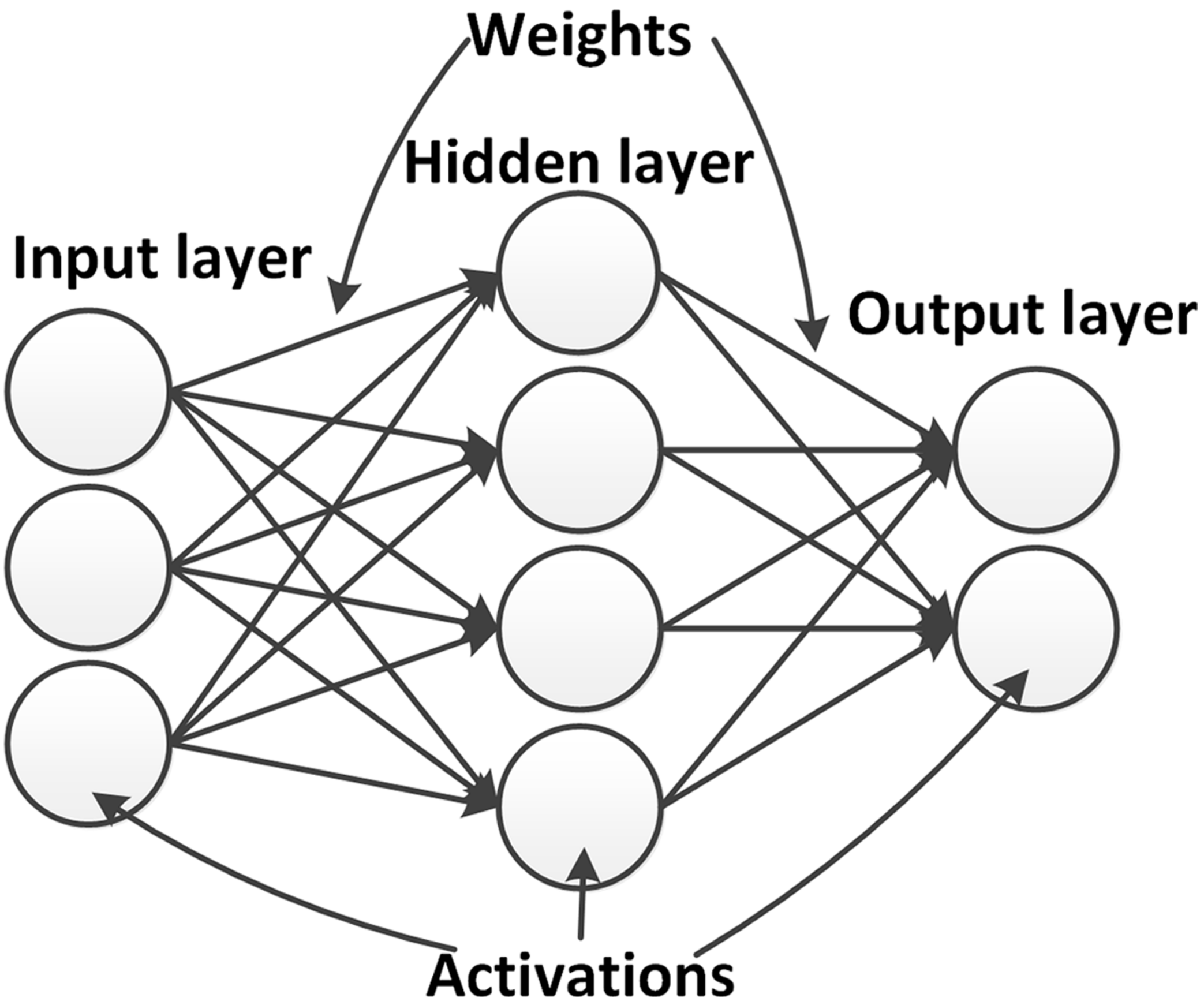

**Figure 3 Example of a simple neural networks.** The network contains one input layer, one hidden layer and one output layer. Activations (neurons) in different layers are connected by weights.
Full-size DOI: 10.7717/peerj-cs.773/fig-3

It can also be proven that the derivative of Eq. (2) function can be simply computed as:

$$\frac{\partial y_j}{\partial net_j} = y_j \cdot (1 - y_j) \tag{3}$$

Figure 3 shows an example of a neural network architecture in its simplest form. Within the neural network, neurons are arranged in layers. A network always has an input and output layer, whose configurations are determined by the dimensions of input and output data. Between them, a number of hidden layers is introduced, which, during the training process, are used to model the relationship between the input and output. In so-called deep neural networks (DNNs), a high number of hidden layers are introduced, which is the reason why DNNs are usually capable of learning features of high complexity. The propagation of the inputs to the output layer is called the *forward propagation*.

During the training process, the architecture of neural networks remains unchanged, while the weights of connections and the biases of neurons are adapted depending on the difference between network outputs and real data. In the field of computer vision, for example, the inputs can be pixels of images, while the outputs can be labels for the entire images (image classification) or labels for the individual pixels (image segmentation).

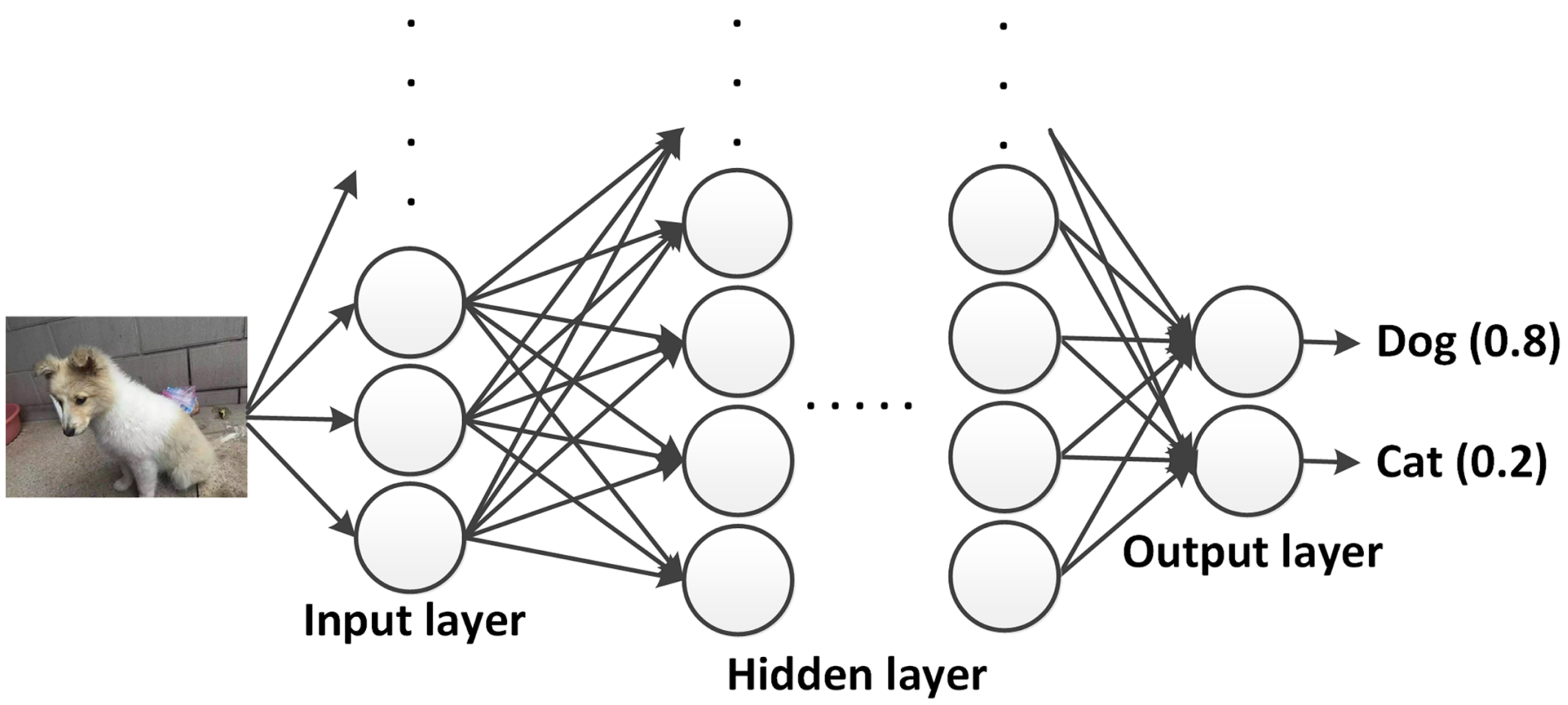

**Figure 4 Example of a DNN.** The image data is provided to the input layer, weighted and propagated to further hidden layers. The output layer provides the prediction about what kind of animal (dog or cat) is shown in the input image (photographer source credit: Yuan Jin).
Full-size DOI: 10.7717/peerj-cs.773/fig-4

An example for the forward propagation using DNNs is shown in Fig. 4. An image is used as an input to the network, where each pixel is weighted and propagated through the hidden layers. Each output activation gives a value, which is the probability of the shown animal to belong to a specific class or label. The highest value determines the most likely answer. The difference between the network output and the desired output is called the loss. The goal of the training process is to minimize the loss by means of changing the weights and biases, for which two main schemes are introduced in the next subsection.

## Backpropagation and gradient descent

In this section, two basic concepts for the training of deep learning models are introduced, namely backpropagation and gradient descent. For the reason of simplification, a simple network, as shown in Fig. 5, is introduced. It consists of only one input and one output layer, without any hidden layer. After the forward propagation, the outputs are calculated and compared with the correct values. The motivation during the training process is to update the model parameters such that the output moves closer to the *truth*. To this end, the current error or loss of the network is passed reversely through the model, which is why this step is called backward propagation or backpropagation.

The first step in the backpropagation is to calculate the output loss. There is a large range of loss functions that are commonly used in deep networks, depending on the application. As an example, we show the usage of a squared error loss function:

$$E = \frac{1}{2}\sum_{i=1}^{n}(t_i - y_i)^2 \quad (4)$$

Here, $t_i$ are the expected outputs, $y_i$ the actual outputs of output neurons, and $E$ is the resulting overall error or loss. The coefficient of 1/2 is added to offset the derivative exponent.

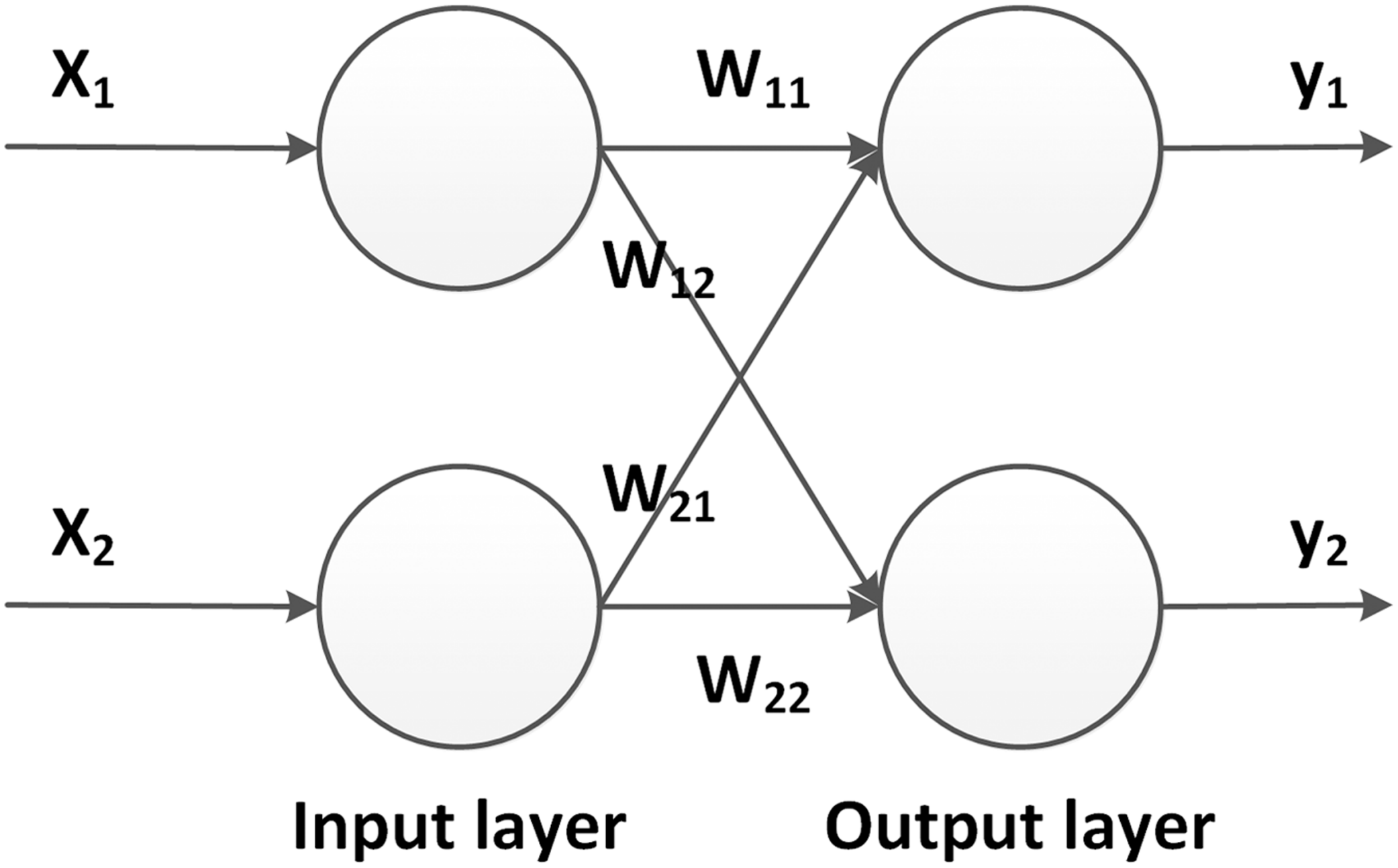

**Figure 5** Simplified neural network. Full-size DOI: 10.7717/peerj-cs.773/fig-5

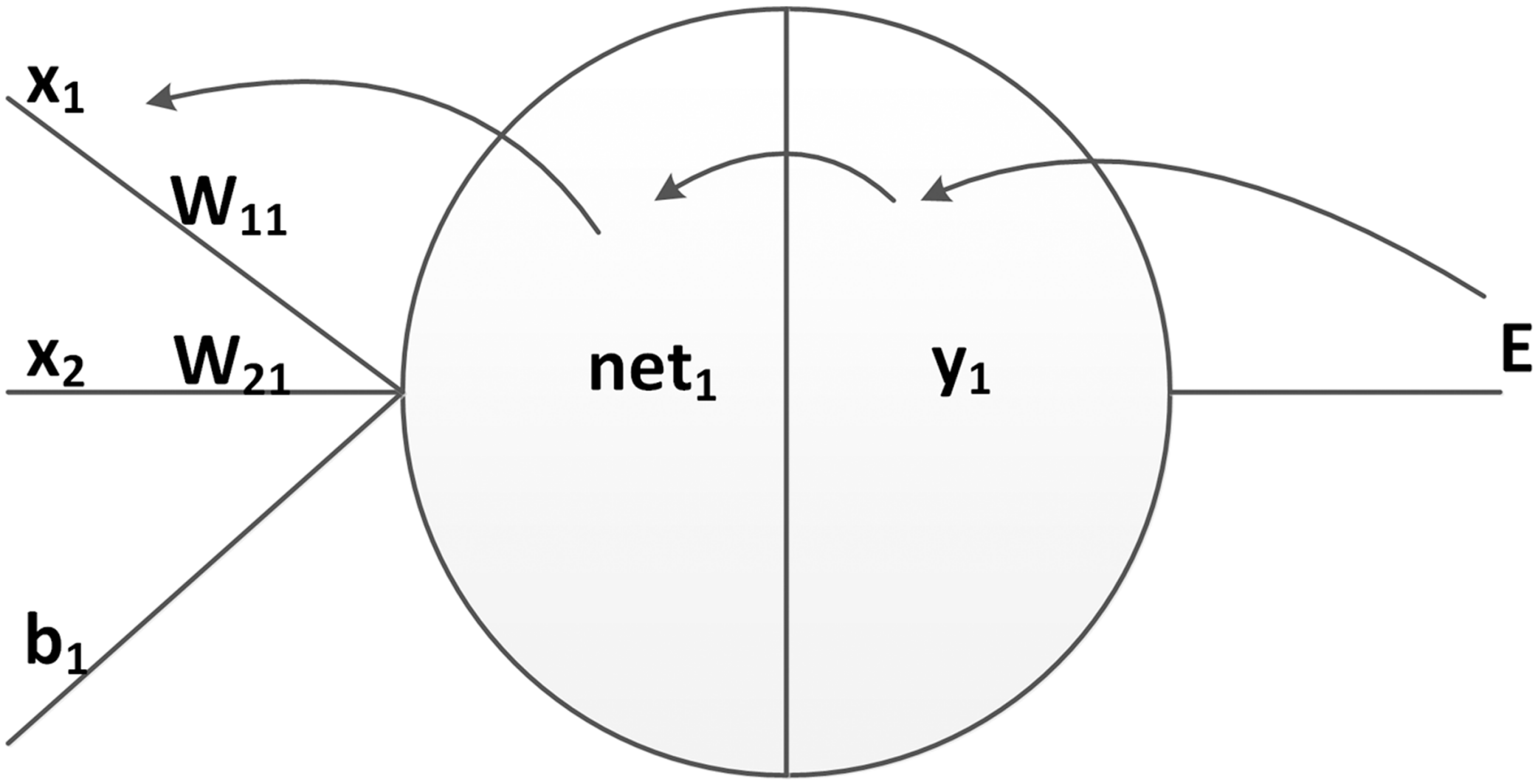

**Figure 6** The decomposition of a partial derivation. Full-size DOI: 10.7717/peerj-cs.773/fig-6

During error backpropagation, the gradient of the loss function with respect to the network parameters is calculated. As shown in Fig. 6, the partial derivative of the loss relative to each weight can be computed by using the chain rule twice:

$$\frac{\partial E}{\partial w_{ij}} = \frac{\partial E}{\partial y_j}\frac{\partial y_j}{\partial net_j}\frac{\partial net_j}{\partial w_{ij}} \tag{5}$$

It is assumed that this neuron is the output neuron, which means the output of neuron $o_j$ is the same as the output of neural network $y_j$. Under the assumption of a logistic activation function and a squared error loss, the three partial derivatives on the right-hand side in Eq. (5) are computable using:

$$\frac{\partial E}{\partial y_1} = \frac{\partial}{\partial y_1}\frac{1}{2}(t_1 - y_1)^2 = y_1 - t_1 \tag{6}$$

$$\frac{\partial y_j}{\partial net_j} = y_j(1 - y_j) \tag{7}$$

$$\frac{\partial net_j}{\partial w_{ij}} = \frac{\partial}{\partial w_{ij}}\left(\sum_{k=1}^{n} w_{kj} o_k\right) = \frac{\partial o_i}{\partial w_{ij}} = o_i \tag{8}$$

Note that in Eq. (5), only one term in the sum $net_j$ depends on $w_{ij}$, which leads to Eq. (8).

If this neuron is in an inner layer of the network, the calculation of the derivative of $E$ with respect to output $y$ is calculated. Considering $E$ as a function of all neurons $L = u,v,\ldots,w$ receiving their input from neuron $j$, Eq. (6) can be changed to:

$$\frac{\partial E(o_j)}{\partial o_j} = \frac{\partial E(net_u, net_v, \ldots, net_w)}{\partial o_j} \tag{9}$$

A recursive expression for the derivative is obtained:

$$\frac{\partial E}{\partial o_j} = \sum_{l \in L}\left(\frac{\partial E}{\partial net_l}\frac{\partial net_l}{\partial o_j}\right) = \sum_{l \in L}\left(\frac{\partial E}{\partial o_l}\frac{o_l}{\partial net_l}\frac{\partial net_l}{\partial o_j}\right) = \sum_{l \in L}\left(\frac{\partial E}{\partial o_l}\frac{o_l}{\partial net_l} w_{jl}\right) \tag{10}$$

Then, the derivative with respect to $o_j$ can be calculated if all the derivatives with respect to the outputs $ol$ of the next layer in the network are computed. A general solution to the derivative in Eq. (5) can be generated with Eqs. (6)–(8) and (10):

$$\frac{\partial E}{\partial w_{ij}} = \frac{\partial E}{\partial o_j}\frac{\partial o_j}{\partial net_j}\frac{\partial net_j}{\partial w_{ij}} = \frac{\partial E}{\partial o_j}\frac{\partial o_j}{\partial net_j} o_i = o_i \delta_j \tag{11}$$

with

$$\delta_j = \frac{\partial E}{\partial o_j}\frac{\partial o_j}{\partial net_j}\begin{cases} (o_j - t_j) f(net_j)(1 - f(net_j)), & \text{if } j \text{ is an output neuron,} \\ \left(\sum_{l \in L} \delta_l w_{lj}) f(net_j\right)(1 - f(net_j)), & \text{if } j \text{ is an inner neuron.} \end{cases} \tag{12}$$

After all derivatives of the loss with respect to the network parameters, and consequently, the gradient of the error function, have been calculated, the network parameters are updated using the gradient descent method. The basic idea of gradient descent is to move to the opposite direction of the gradient, to find its (local) minimum.

Figure 7 shows a simplified example of using a derivative to find the direction of the gradient descent. The weights are changed according to the following equation:

$$w_{ij}^{n+1} = w_{ij}^{n} - \beta \frac{\partial E}{\partial w_{ij}} \tag{13}$$

in which $\beta$ is called the learning rate.

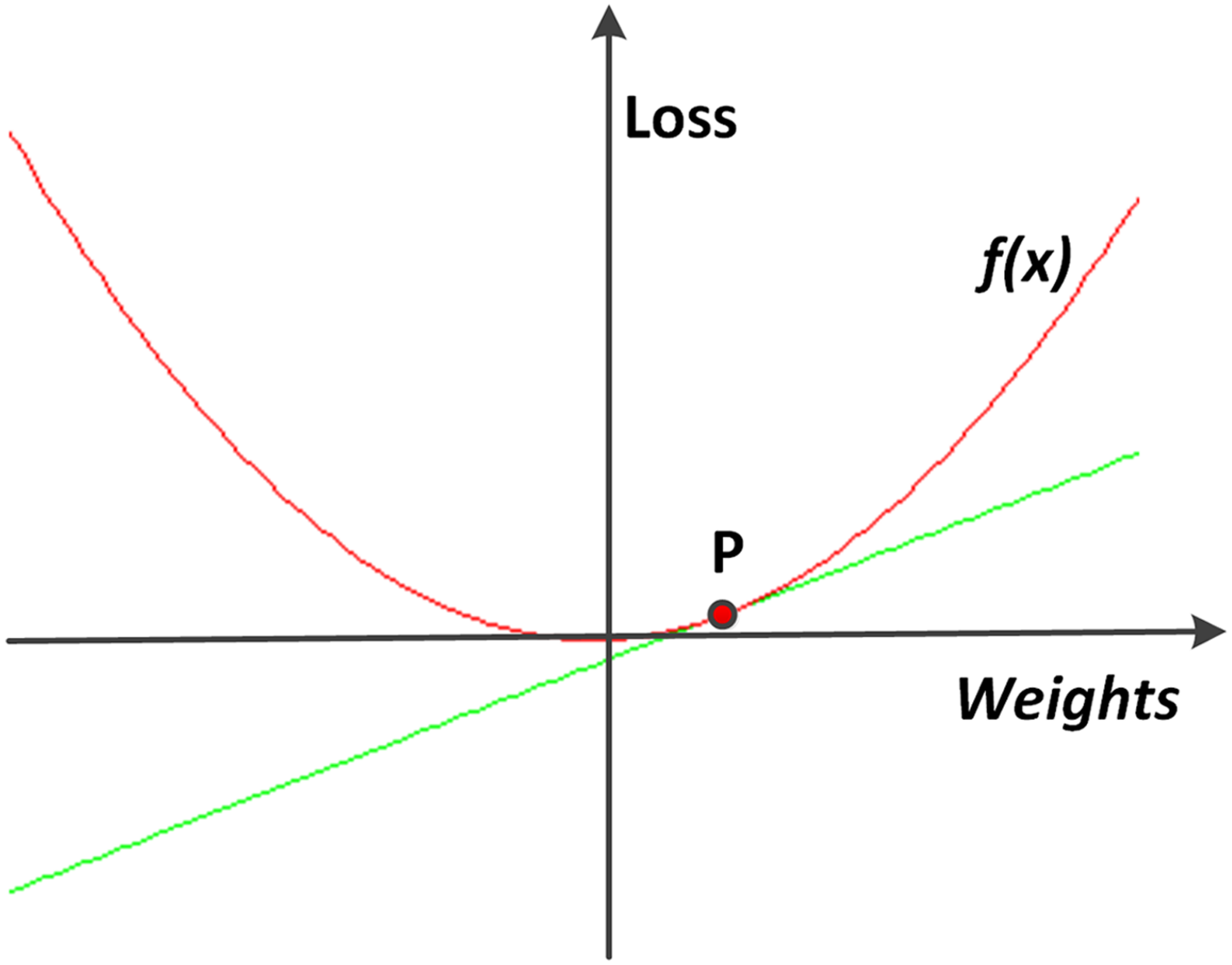

**Figure 7 An example of gradient descent.** To find the direction on the point *P*, along which the loss function *f* can decline, a derivative line of *f* along the point *P* is plotted. In this simple example, it gets obvious that the value of *f* decreases as the value of *x* decreases.

Full-size DOI: 10.7717/peerj-cs.773/fig-7

To sum up, in the forward propagation, the input data are propagated from the input layer to the output layer and the network provides the computing results (Fig. 8). In the backward propagation, the loss function is propagated from the output layer to the input layer, while the weights of each neuron is updated. One iteration of the training process ends when all weights of the network are updated. One complete training process usually involves numerous iterations during which the weights are updated systematically to move the network outputs closer to the expected ground truth.

## Convolutional neural networks

Convolutional neural networks (CNNs) (*Krizhevsky, Sutskever & Hinton, 2012*) are a special type of artificial neural networks (ANNs), which usually include more than one convolutional layer. CNNs are widely used in image processing (but also for example in language processing, however, they are far less popular there), since they have superior ability of information handling for a large amount of data. As the name describes, CNNs are based on convolution (shown in Figs. 9 and 10). The mathematical expression of a 2D convolution of an image is:

$$g(i,j) = l \times k(i,j) = \sum_{j=0}^{m-1} \sum_{i=0}^{n-1} l(m,n)k(i-m, j-n), \tag{14}$$

in which *k(i, j)* is the input image, *l* is the convolution or filter kernel and *g(i, j)* is the convolved/filtered image. Equation (14) shows the basic form of a 2D convolution, which can also describe the similar calculation of a 3D convolution as shown in Fig. 11.

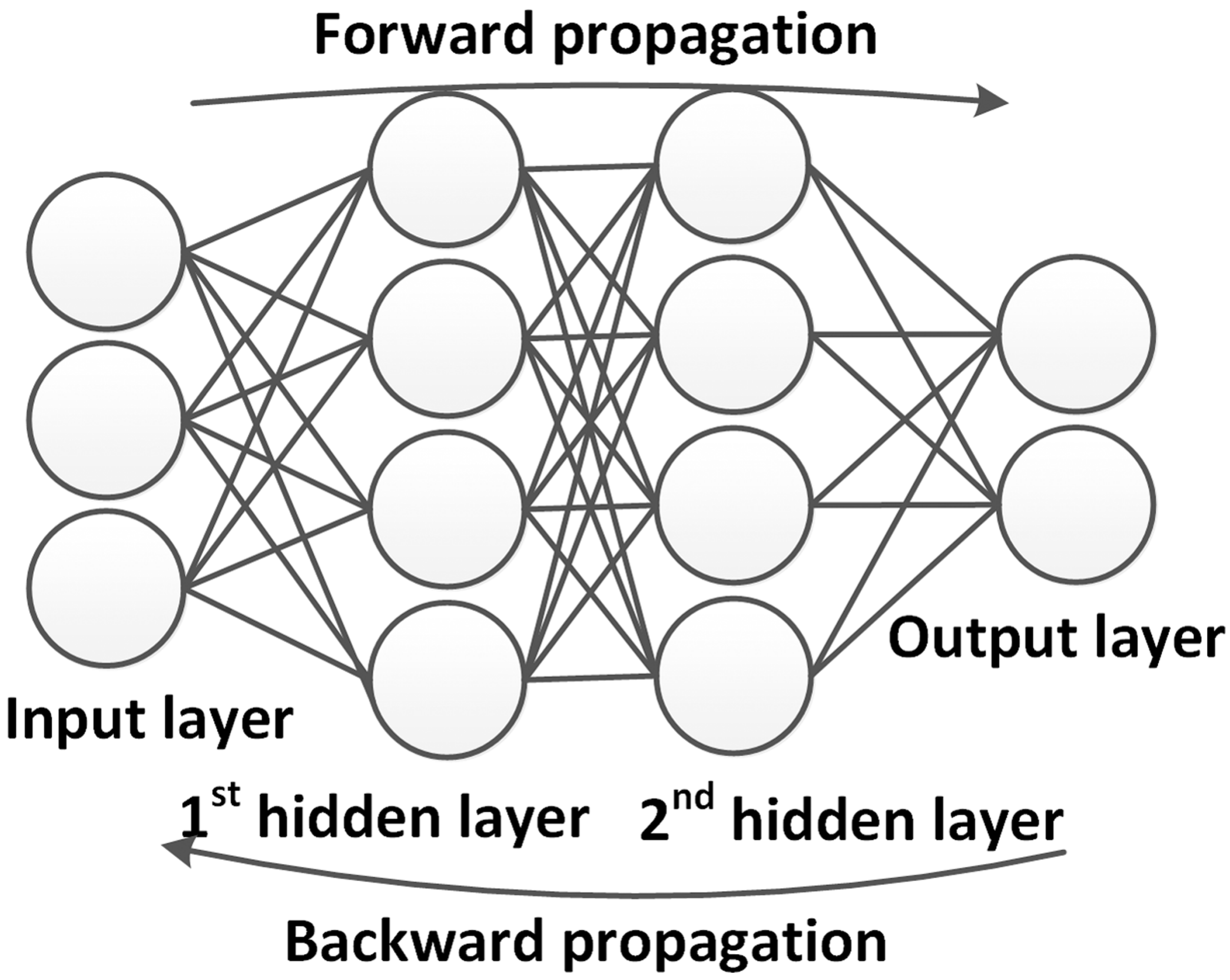

**Figure 8** **Example of the training process for DNNs.** Full-size DOI: 10.7717/peerj-cs.773/fig-8

Like traditional neural networks, a convolutional neural network consists of an input and an output layer, as well as several hidden layers which usually consist of convolutional layers in combination with other layer types, such as pooling layers and fully connected layers. The input images of CNNs are usually in the size of *image height* × *image width* × *image depth*. After the convolution, the images are abstracted to so-called feature maps and passed to the next layer. A convolutional layer has the following parameters:

- The number of input and output channels;
- Convolutional kernels defined by their shape.

During the training process of CNNs, the features maps can become very large, which results in a big amount of computational resources required. One of the methods to solve this problem is the pooling layer. Pooling layers reduce the dimensions of feature maps from convolutional layers and pass the reduced data to the next layer. Different functions for pooling are available, such as max pooling, average pooling and sum pooling, where max pooling is most commonly used.

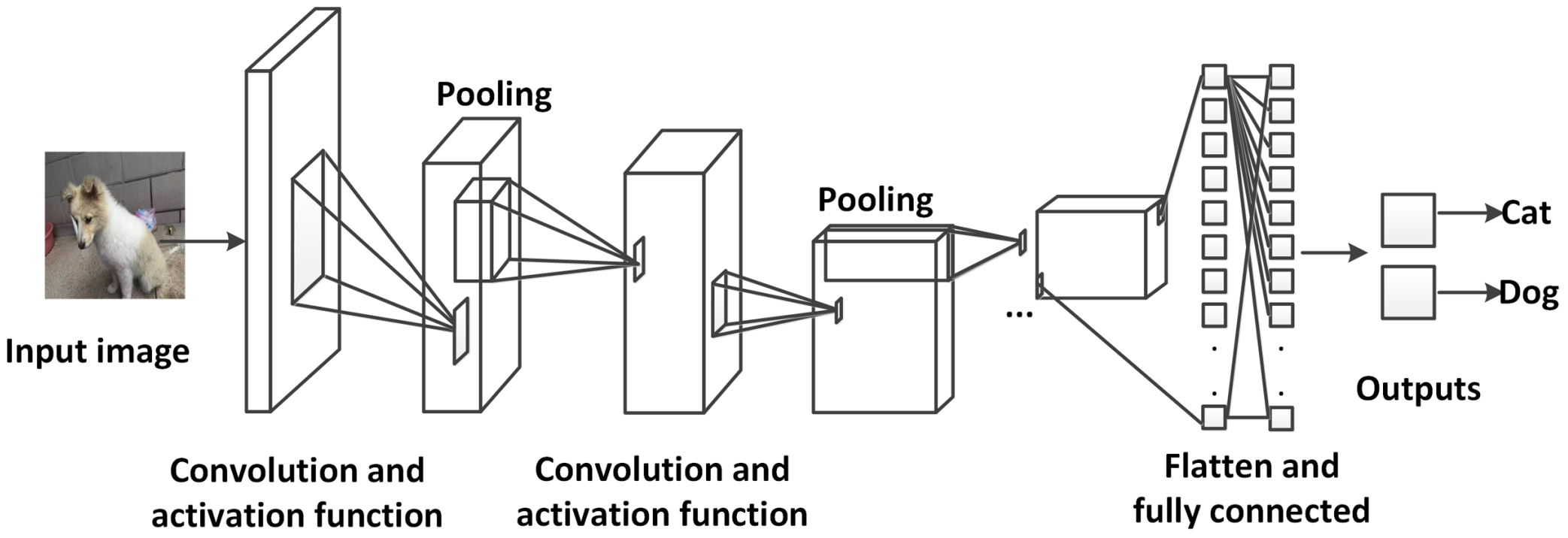

**Figure 9 An example of a convolutional neural network applied for image classification.** The images are inputted to the network, propagated through several convolutional layers, pooling layers and one fully connected layer. Finally, the possibility of the input image representing a cat or a dog is outputted (photographer source credit: Yuan Jin). Full-size DOI: 10.7717/peerj-cs.773/fig-9

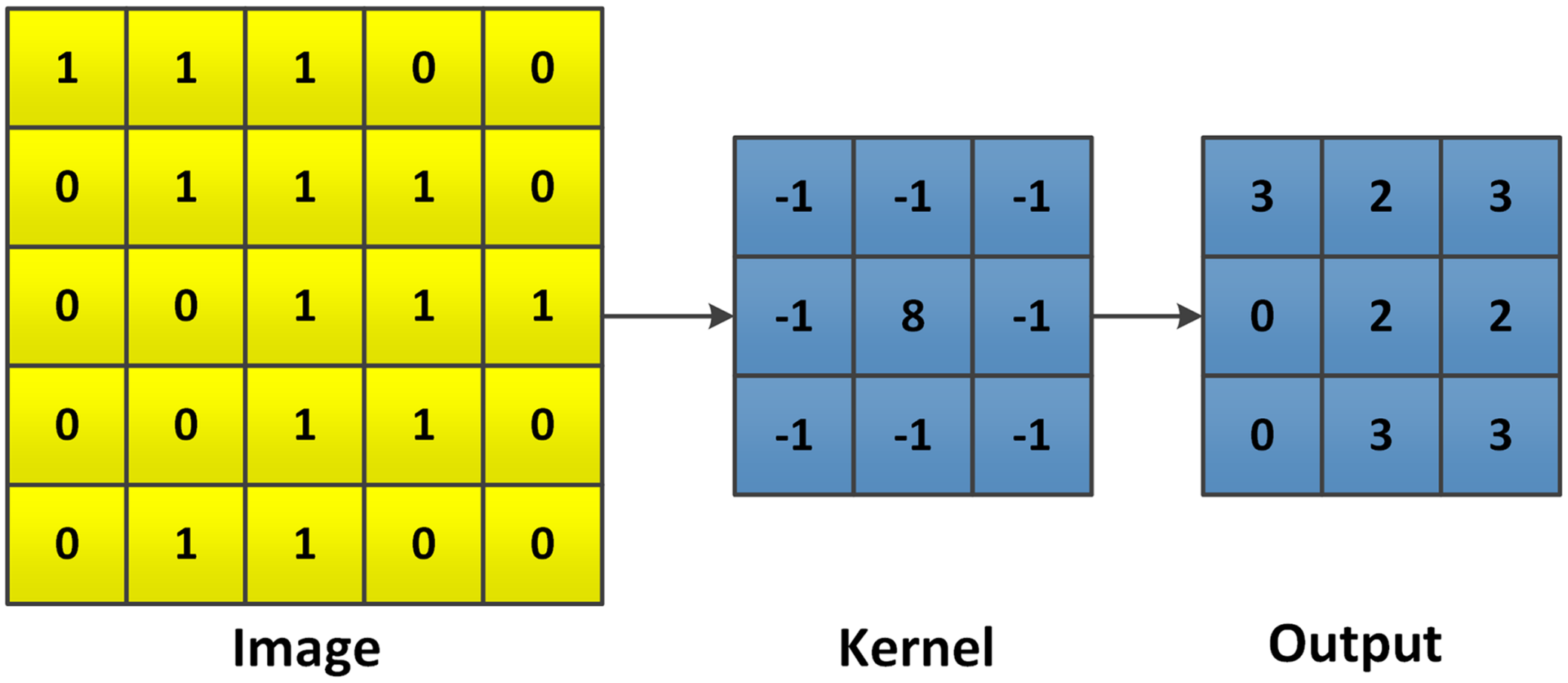

**Figure 10 Example of a general 2D convolution.** At first, the image is flipped for both, the rows and columns. Then the kernel slides over the flipped image, each element is multiplied by its corresponding pixel in the flipped image and summed up. The size of the output depends on both, the input image and the kernel. Full-size DOI: 10.7717/peerj-cs.773/fig-10

As shown in Fig. 12, with max pooling or average pooling operations, the $4 \times 4$ feature map is transformed to a $2 \times 2$ feature. The stride controls how the pooling filter is moved over the input volume.

At the end of a CNN, there are usually one or several fully connected layers, which connect every neuron in the last layer to every neuron in itself. The fully connected layers are used to concatenate the feature maps into the desired output values (*e.g.*, using the same example as above, the probability of an image showing a cat or a dog).

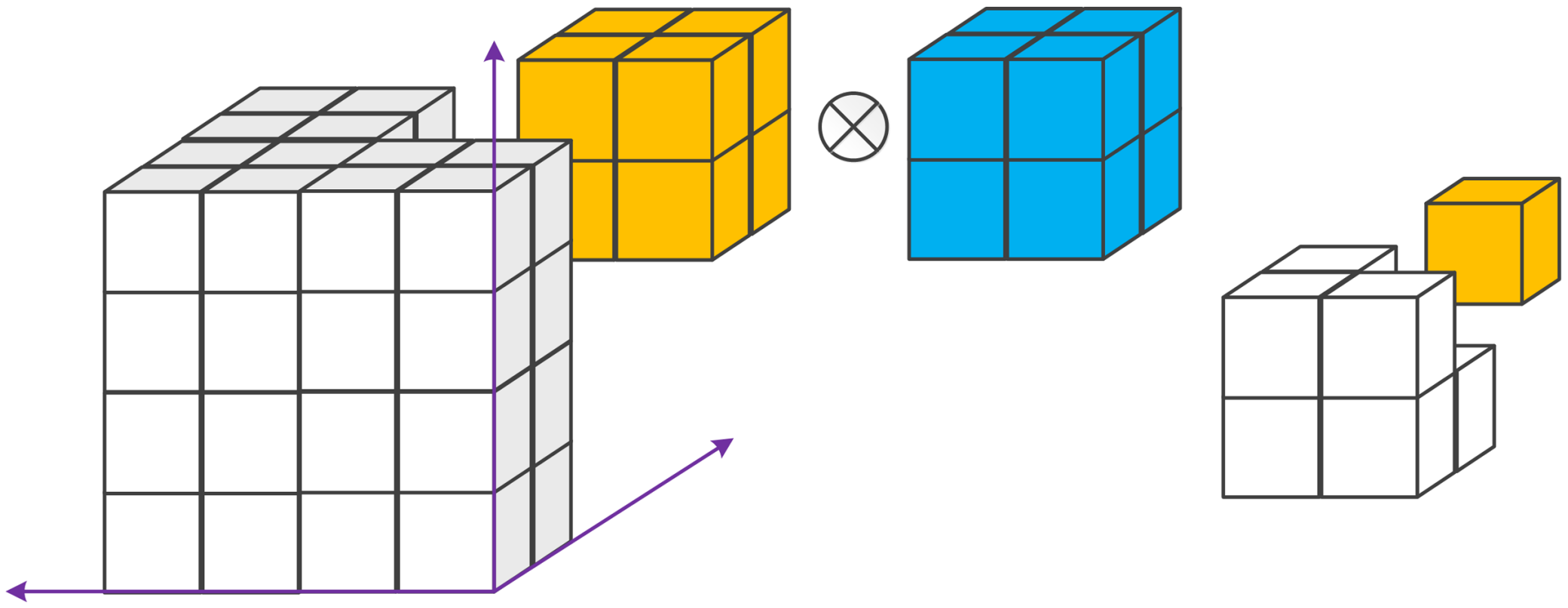

**Figure 11** **Example of a 3D convolution.** It applies a 3D filter to the dataset and the filter moves in 3-direction (x, y, z) to calculate the outputs. Both input and output data are a 3D volume, which is represented by cubes. Full-size DOI: 10.7717/peerj-cs.773/fig-11

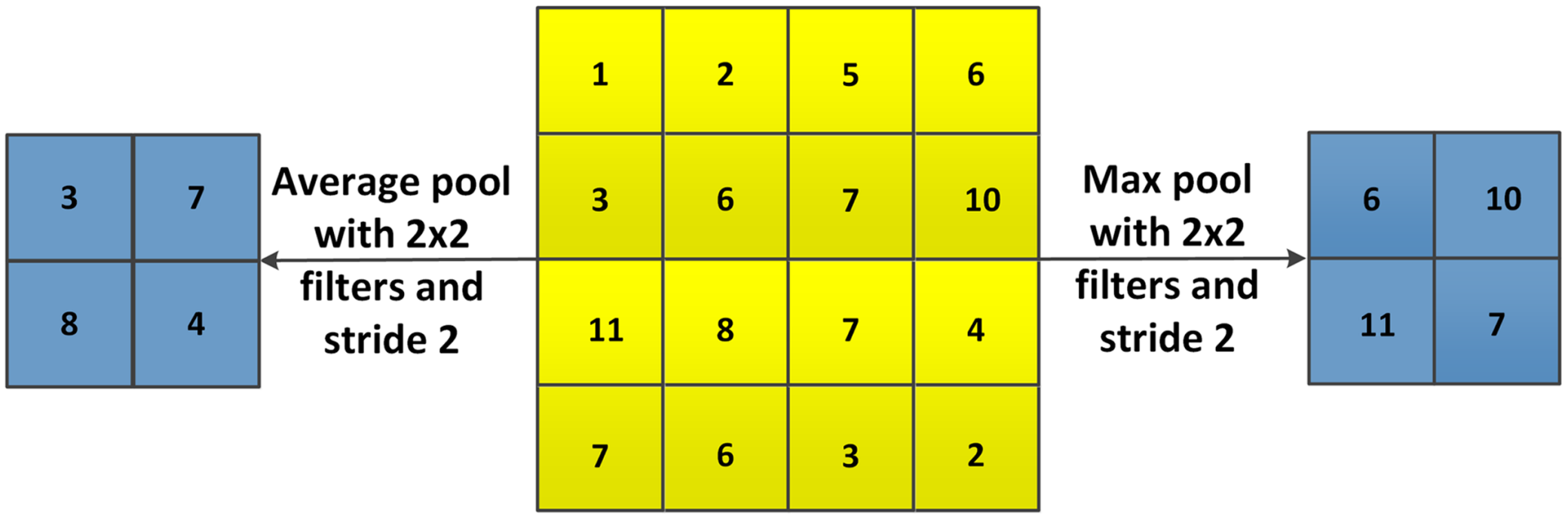

**Figure 12** **Example of max pooling and average pooling.** Full-size DOI: 10.7717/peerj-cs.773/fig-12

Currently, CNNs have been widely used in the field of imaging processing. In the following section, a specific CNN model mainly for medical image segmentation will be introduced.

## Fully convolutional neural networks

As one of the most common tasks in medical imaging, automatic segmentation is challenging because of the huge difference between different patients (anatomy and pathology). In this field, however, neural networks have shown great advantages to learn image features automatically from the medical images and corresponding ground truths (*Hesamian et al., 2019*). In addition, the development of fully convolutional neural networks (FCNs) [19] further improved the advantages of deep learning in the area of image segmentation and particularly semantic segmentation.

Semantic segmentation is to understand what is in the image on a pixel level, which can be also defined as to label each pixel of an image with a corresponding class. Figure 13 shows an example for a semantic segmentation. Fully convolutional neural networks were developed by *Long, Shelhamer & Darrell, 2015* based on normal convolutional neural networks. In FCNs, the final fully connected layer of CNNs are replaced by convolutional layers, so that the images are not downsized and the output will not be a single label as in CNNs. Instead, a pixel-wise output can be calculated to represent each pixel in the input image. Figure 14 shows a typical setup of an FCN for a semantic segmentation. Based on the improvement of FCNs, there are currently many deep learning methods in computer visions that are applied for segmentation tasks in medical imaging. Most of these methods are based on FCNs that learn the features of spatial dimensions from the original images.

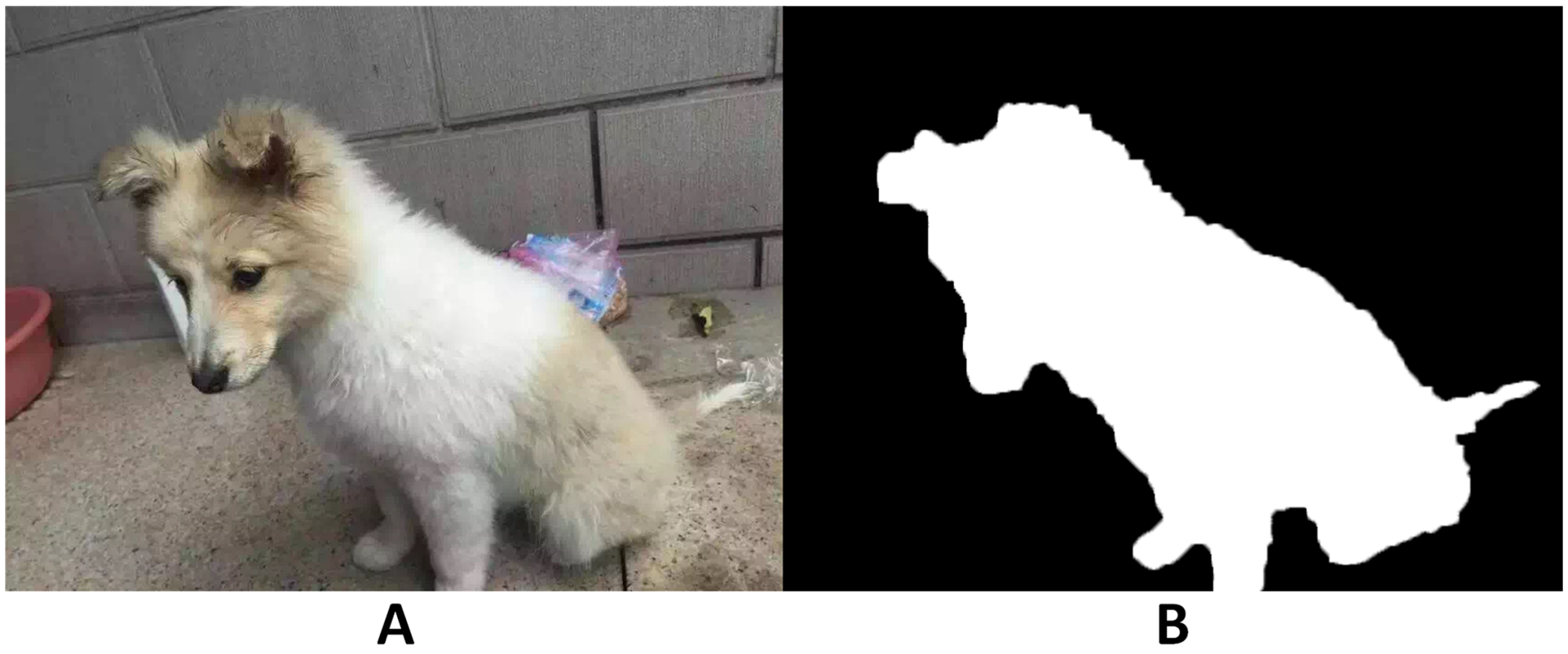

**Figure 13** **Example of a semantic segmentation.** The left image (A) is the original image and the right image (B) is the pixel-wise segmented image revealing a dog and its contour (white, photographer source credit: Yuan Jin). Full-size DOI: 10.7717/peerj-cs.773/fig-13

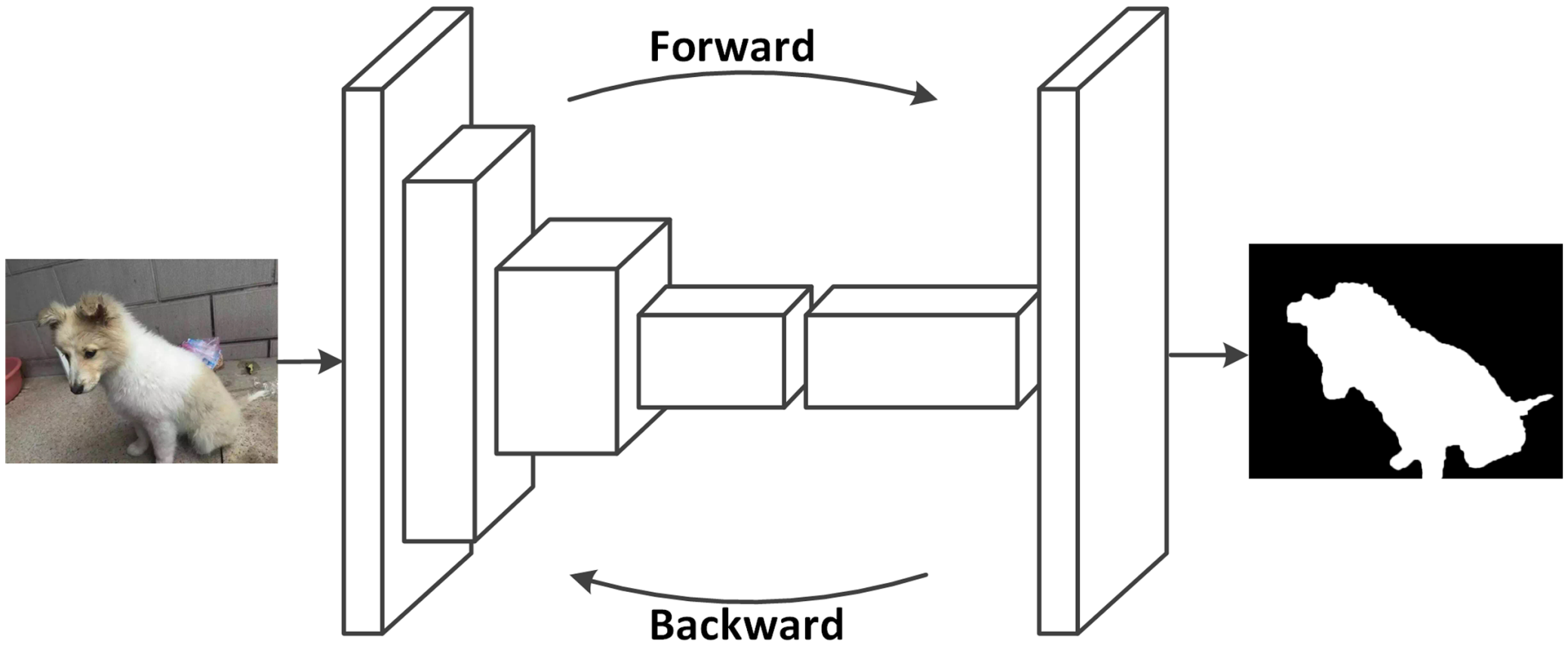

**Figure 14** **FCNs can learn to make dense predictions for pixel-wise predictions (photographer source credit: Yuan Jin).** Full-size DOI: 10.7717/peerj-cs.773/fig-14

## U-Net

One of the most well known FCN models for medical image segmentation is the U-Net, proposed by *Ronneberger, Fischer & Brox (2015)*, using the concept of deconvolution introduced in *Zeiler & Fergus (2014)*. As shown in Fig. 15, this model has two main parts for downsampling and upsampling. The downsampling part is similar to the structure of a CNN, while the upsampling part, usually known as the expansion phase, is built of an upsampling layer followed by a deconvolution layer. The former enlarge the dimensions of the feature maps that are reduced in the downsampling layer, while the latter corresponds to the convolutional layers. One important idea of U-Net is the connection between layers in the downsampling and upsampling parts at the same "levels" These connections provide high-resolution features of spatial dimensions from the downsampling part to the upsampling part. It means that the model can be trained with multi-scale features from different layers. This structure of deep learning model is suitable for the automatic segmentation of large size images.

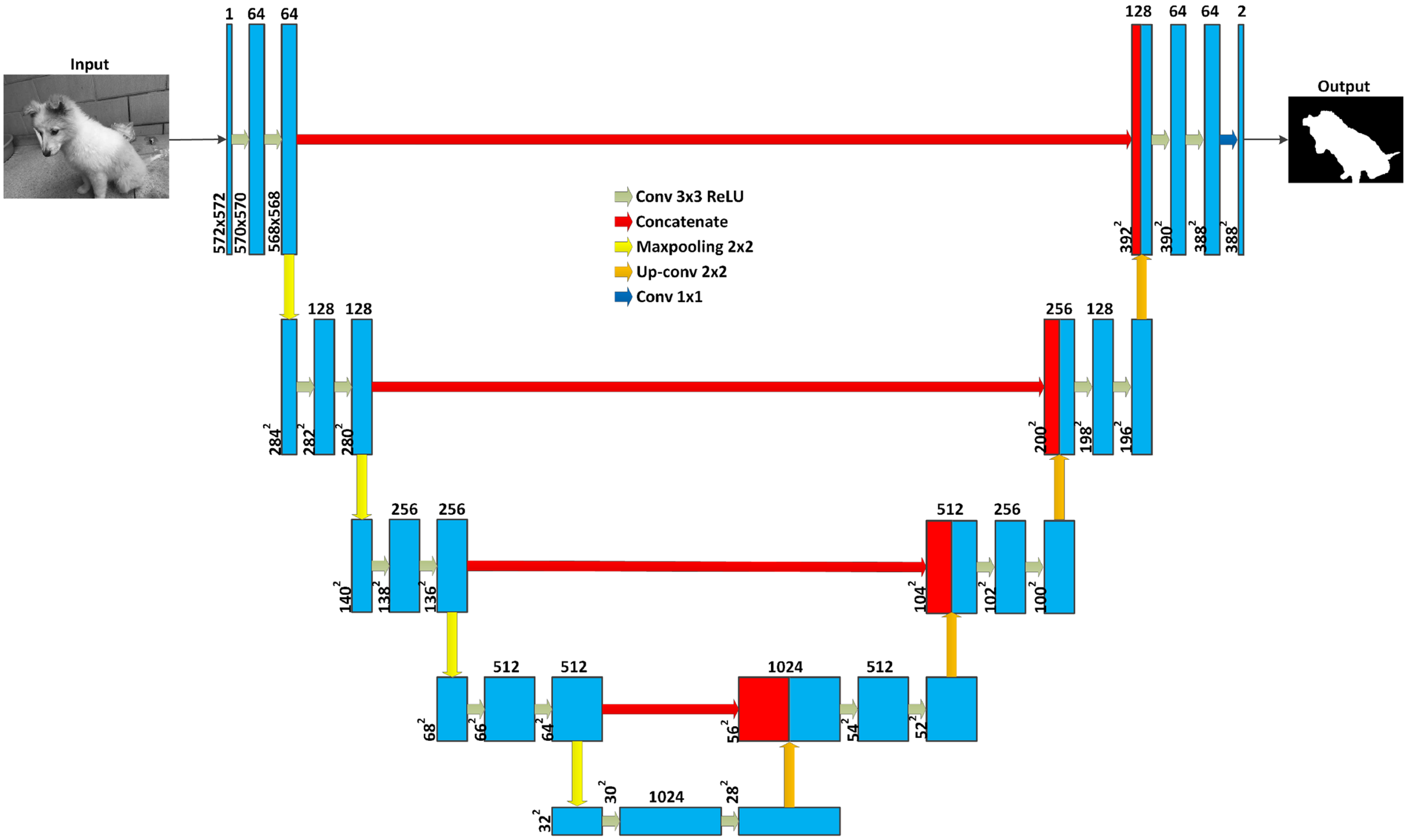

**Figure 15 U-Net architecture.** Each blue box corresponds to a multi-channel feature map. The number of channels is denoted on top of the box. White boxes represent copied feature maps. The arrows represent different operations (photographer source credit: Yuan Jin).
Full-size DOI: 10.7717/peerj-cs.773/fig-15

## 3D U-Net

Based on the main idea of the U-Net, *Çiçek et al. (2016)* developed the 3D version of the U-Net model [23], which is able to learn from sparsely annotated volumetric images. The 3D U-Net is developed by replacing all 2D operations with their 3D counterparts. Figure 16 shows the architecture of the 3D U-Net, which is similar to the architecture of the original 2D U-Net (Fig. 15).

There are two kind of applications for this model: (1) When semi-automatically applied, a part of the original image is manually segmented. The network learns from these segmented parts and provides a dense 3D segmentation. (2) For a fully automatic application, a trained model is assumed to exist. The model is trained with existing data and segments new (unseen) volumetric images.

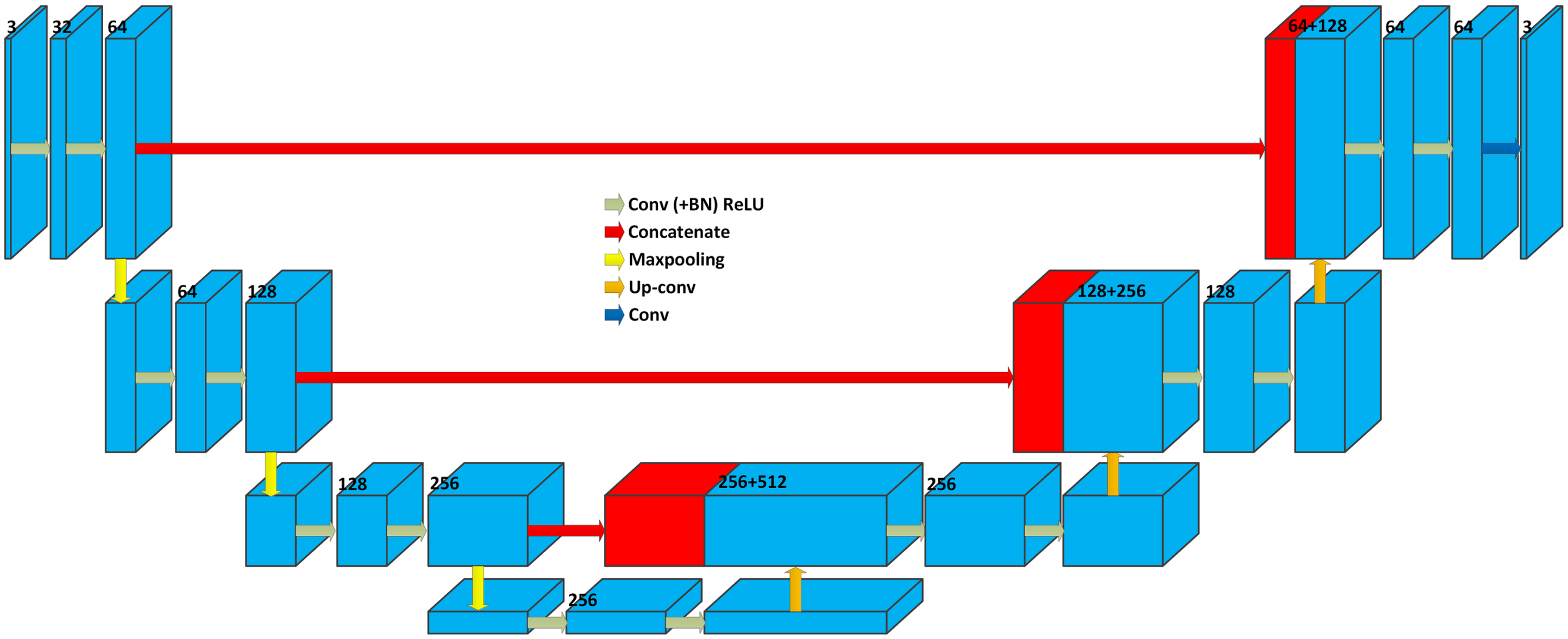

**Figure 16** **3D U-Net architecture.** Blue boxes represent feature maps. The number of channels is denoted above each feature map.
Full-size DOI: 10.7717/peerj-cs.773/fig-16

3D U-Nets are widely used in the field of medical imaging, as there are many types of medical imaging applications based on 3D data acquisitions. In general, 2D methods are usually not optimal for producing 3D image labels (segmentations), because of the missing information (and interpolations) between the single 2D slices. Manual expert segmentations (still considered the gold standard), however, are in general done on a slice-by-slice basis in 2D, because a 3D segmentation, handling several 2D slices simultaneously, is mentally often too demanding for the annotator (especially in larger structures, which need some considerable amount of time to be segmented). This often results in a step-like effect of the manual segmentations, because an overall smoothing in 3D is missing and hard to incorporate manually (Fig. 17).

We focused in this section on the main computer vision architectures as an example, because computer vision is currently the most crowded and popular field (and three of the top five computer science conferences are related to computer vision, according to: https://www.guide2research.com/topconf/), especially when taking also the medical imaging field into account. Note that the surveys we present in this meta-survey cover over 11,000 references, which makes it impractical to introduce all deep learning-based concepts and architectures. However, for most surveys, we outline what kind of deep learning schemes or architectures have been reviewed and pretty much all surveys have a background section about these in their contributions. This enables the interested reader to dive deeper into the specific deep learning background by directly accessing the corresponding survey.

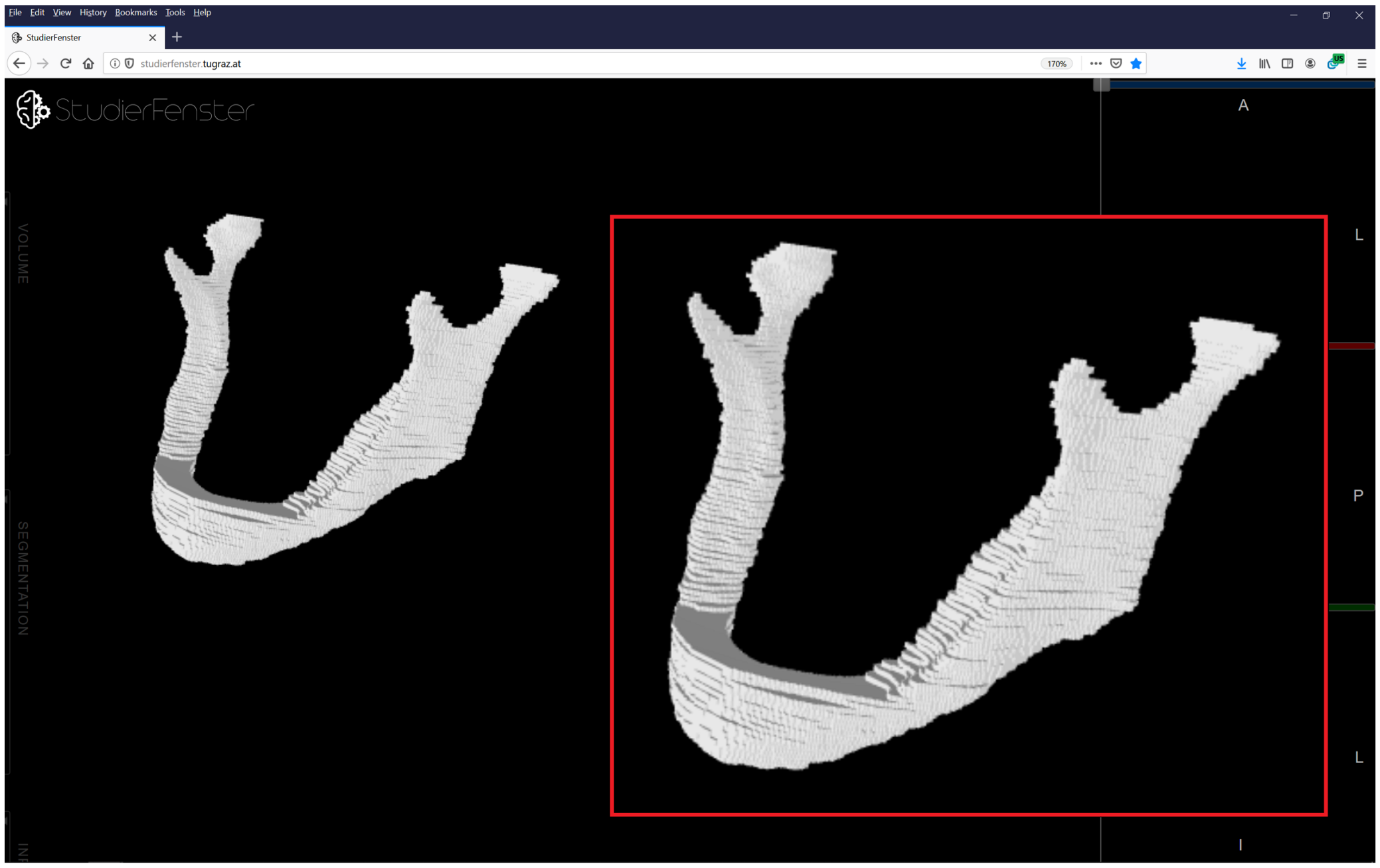

**Figure 17** **Visualization of a manual segmentation of the lower jawbone.** Because the manual segmentation has been done slice-by-slice in 2D, the overall 3D segmentation shows a step-like effect between the single 2D slices. Data and manual segmentation taken from *Wallner, Mischak & Egger (2019)*, visualization done with Studierfenster (www.studierfenster.at) (*Wild, Weber & Egger, 2019*).
Full-size DOI: 10.7717/peerj-cs.773/fig-17

## DEEP LEARNING: A SURVEY SUMMARY OF SELECTED REVIEWS ACROSS SCIENTIFIC DISCIPLINES

This section presents selected review and survey publications in deep learning. For a better overview, the publications are arranged in four categories:

- computer vision,
- language processing,
- medical informatics,
- and additional works.

The first category introduces deep learning reviews and surveys in the field of computer vision. Among others, this includes publications about object detection, image segmentation, object recognition and a survey about inpainting with generative adversarial networks. The second category presents deep learning review publications in (natural) language processing. This covers areas like language understanding but also language generation and answer selection. The third category presents deep learning reviews in the medical field, thus covering reviews about different aspects of medical image processing, medical imaging and computer-aided diagnosis. Ultimately, the last category of this section closes with additional deep learning reviews in areas like big data, networking, multimedia, agriculture and reviews that cover multiple scientific areas or applications. According to these sections and sub-sections, the tables of this manuscript are divided into the same categories and sub-categories, and present also the current citations for every publication according to Google Scholar (status as of mid-August 2020). Note that the review about data augmentation is listed in the first category about computer vision topics, because it mainly discusses approaches for images. Some review publications would fit in more than one main category. For example, reviews about medical image analysis, could also fit into the category computer vision. However, the final assignment and arrangement was made also under the consideration of balancing the number of publications per category.

The rationale behind choosing the four main categories are the underlying data sources and their different data generation process. Computer vision deals with image and video data, natural language processing deals with textual data, medical informatics with medical data and the last section covers surveys that deal with image and text (multimedia) data or span across different disciplines. Unfortunately, the boundaries between these categories are not always clear, *e.g.*, medical informatics connects to both computer vision and language processing. Hence, there are various other possibilities to structure such a meta-survey (or surveys in general), for example, by data modalities (*e.g.* image *vs*. language), by domains, by used architectures, by time of publication, by impact, just to name a few. But all these arrangements have their advantages and disadvantages, and are not unique, except maybe an arrangement by time of publication or impact (*e.g.* citations, which also very likely changes over time). However, the underlying medical data is distinct to conventional image data in computer vision with its specific characteristics. In addition, the medical field requires further knowledge on top of the technical expertise (*e.g.*, a medical image might require different pre-processing/deep learning than generic images). Moreover, deep learning had a massive impact in the medical field, as can be seen in the presented works and citations, which justifies for its own category (also, health is a fundamental right and probably the most important aspect in life). For illustration, a similar situation is also found for research on causality, where similar techniques are used across multiple fields, but the medical field established its own terminology (*e.g.*, standardization *vs*. adjustment formula). Finally, the order of the main categories also arises from the data sources, starting with computer vision and language processing working with fundamental data like images, videos and text. Ultimately, the survey starts with computer vision, because it is currently a more active research field than language processing.

Note that we do not claim to provide a complete meta-survey of all existing surveys on deep learning, which would go far beyond the scope of this contribution. We see our meta-survey more as a sort of *baseline* or *building block* contribution for deep learning in the computer science community, where meta-surveys are, in general, still somewhat a rarity, in comparison to other fields, like the medical domain, where meta-surveys are quite common. In addition, we did not find a deep learning survey for every task at the time of publication. Hence, there are some important tasks missing in our meta-survey, *e.g.* machine translation in NLP. which should to be tackled by the research community in the near future. However, our survey tables can provide here a compact overview, which tasks have been covered by deep learning surveys so far and which need more attention by the community. Our intention is to present selected works across scientific disciplines, divided into categories according to the underlying data sources, to give an impression about the impact deep learning has on a very broad level.

Table 2 List of published reviews in deep learning in the category *computer vision*.

| Computer vision | Publications | Number of references | Citations (until August 2020) | Preprints |
|---|---|---|---|---|
| General computer vision | *Voulodimos et al. (2018)* | 114 | 550 | No |
| | *Guo et al. (2016)* | 216 | 950 | No |
| Object detection | *Liu et al. (2020)* | 332 | 269 | No |
| | *Zhao et al. (2019)* | 230 | 491 | No |
| | *Jiao et al. (2019)* | 317 | 45 | No |
| Image segmentation | *Garcia-Garcia et al. (2018)* | 126 | 127 | No |
| | *Minaee et al. (2020)* | 172 | 24 | Yes |
| Face recognition | *Masi et al. (2018)* | 81 | 220 | No |
| | *Li & Deng (2020)* | 253 | 189 | No |
| | *Wang & Deng (2018)* | 305 | 11 | Yes |
| Action/motion recognition | *Herath, Harandi & Porikli (2017)* | 161 | 339 | No |
| | *Wang et al. (2018b)* | 182 | 122 | No |
| Biometric recognition | *Sundararajan & Woodard (2018)* | 176 | 66 | No |
| | *Minaee et al. (2019)* | 282 | 8 | Yes |
| Image super-resolution | *Wang, Chen & Hoi (2020)* | 214 | 74 | No |
| Image captioning | *Hossain et al. (2019)* | 161 | 118 | No |
| Data augmentation | *Shorten & Khoshgoftaar (2019)* | 140 | 274 | No |
| Generative adversarial networks | *Wang, She & Ward (2019)* | 162 | 46 | Yes |
| Sum | – | 3,624 | 3,923 | – |

## Deep learning reviews in computer vision

This sub-section deals with the deep learning reviews in the area of computer vision. It is divided into ten sub-categories and the number of references, and citations (according to Google Scholar and status as of mid-August 2020) for each of these categories is given in Table 2:

- general computer vision,
- object detection,
- image segmentation,
- face recognition,
- action/motion recognition,
- biometric recognition,
- image super-resolution,
- image captioning,
- data augmentation,
- and generative adversarial networks.

The commonality in the field of computer vision is to process and get some kind of understanding from digital images or videos. This can be, for example, the detection or segmentation (outlining) of people, animals or objects in images or videos. Computer vision algorithms lead to numerous real world applications, like automatic face or licence plate detection and recognition, or even self-driving cars, just to name a few. However, making the algorithms reliable enough for their specific task remains challenging. A missed person on a group photo for automatic tagging on a social network website may not be very dramatic, but a missed pedestrian by a self-driving car can have fatal consequences. Especially the fact that every image or video is slightly different makes it very hard for algorithms to generalize and there is no guarantee that new (exceptional) cases will not be missed or wrongly analysed/classified by an algorithm. To measure the performance of computer vision algorithms, common metrics like the Dice Similarity Coefficient (*Sampat et al., 2006*) or the Hausdorff distance (*Huttenlocher, Klanderman & Rucklidge, 1993*), *e.g.* for segmentation tasks, are used in the community to present their results. However, if research groups work on own data collections, comparison to other works are difficult and the presented metrics can be seen more like a proof of concept or common trend towards a solution for a task. A step towards a more objective evaluation is competing on common databases, like *ImageNet*, but also this course of action cannot replace real-life scenarios. The key challenge for all computer vision algorithms is the transition to real world applications, that work reliable in practice also for new, unseen data without major failures.

Note that data augmentation and generative adversarial networks are actually universal techniques that can also be used in other areas than computer vision, like language processing. However, the two surveys we chose for our contribution review only works for images and computer vision, hence, we present them in this first section.

### General computer vision

A review about selected deep learning methods that have been used in the general area of computer vision is presented by *Voulodimos et al. (2018)*. They introduce convolutional neural networks, deep Boltzmann machines, deep belief networks and stacked denoising autoencoders, alongside with their history and application tasks they have been used for. They conclude their review with an outlook on how future deep learning-based methods can be designed for tasks in computer vision and the challenges that arise in doing so.

Another general review on deep learning in computer vision comes from *Guo et al. (2016)*. They also start their review with a comprehensive overview of different deep learning architectures, like CNNs, covering neural network layer types, like convolutional layers and pooling layers, but also training strategies, like dropout, DropConnect and pre-training/fine-tuning. For the deep learning-based architectures, they discuss characteristics, advantages and disadvantages. They conclude their review with current trends and challenges, and provide future directions for a theoretical understanding, human-level vision, training with limited data, time complexity and more powerful models in deep learning.

### *Object detection*

Object detection is one of the most basic, but as well challenging problems in computer vision. Object detection deals with the localization of objects from predefined categories, like cats, dogs, *etc*., in natural images. *Liu et al. (2020)* outline more than 300 research contributions in their survey about object detection. In doing so, they cover many general aspects in the field of object detection, which includes, for example, detection frameworks, but also object feature representation and object proposal generation. On top, they address context modelling, training strategies, and, eventually, evaluation metrics for object detection.

A second review paper in the area of computer vision and object detection is from *Zhao et al. (2019)*, which first provides a short introduction on convolutional neural networks, deep learning, and their history. They focus on typical generic object detection architectures and briefly survey various particular tasks. These cover salient object detection, but also pedestrian and face detection. In addition, an experimental analysis is provided, which allows the comparison of different approaches and hence, draw constructive conclusions amongst them.

Finally, *Jiao et al. (2019)* review existing approaches of general detection models and additionally introduce a common benchmark dataset for them. The authors also outline a comprehensive and systematic overview of numerous approaches for object detection, which include one-stage and multi-stage object detectors. Furthermore, they list and analyse established, as well as new, applications of object detection and its most representative branches.

### *Image segmentation*

Image segmentation is usually the first step in different computer vision applications, like scene understanding, video surveillance, and robotic perception. Further applications can be medical image analysis, augmented reality and image compression, to name a few. *Garcia-Garcia et al. (2018)* provide a survey of deep learning approaches for semantic segmentation that can be translated and applied to numerous areas. In doing so, datasets, but also challenges, are outlined to guide researchers in the decision which method is most suitable for their needs and aims. Subsequently, the existing approaches and methods, like CNNs, are surveyed in the contribution. Additionally, they review common loss functions and error metrics and provide quantitative results for the introduced methods, but also the datasets that have been used for an evaluation.

*Minaee et al. (2020)* present a comprehensive review that covers a wide spectrum of contributions in the area of semantic and instance-level segmentation. This includes fully convolutional pixel-labelling networks and encoder-decoder architectures. Moreover, recurrent networks, multi-scale and pyramid-based methods. Further, visual attention and generative models in an adversarial setting. They studied the strengths and challenges of the proposed deep learning models, but also their similarity. Finally, the authors investigated the most commonly applied datasets and present performance results for them.

### Face recognition

Face recognition is a significant biometric method for identity authentication that has been applied to numerous application areas. This includes public security, daily life, military, but also finance. *Masi et al. (2018)* introduce the main benefits of face recognition with deep learning, also called deep face recognition. They focus on identification and verification by learning representations of the face. The review gives a structured overview of works from the past years, covering the principals and state-of-the-art in face recognition methods.

*Li & Deng (2020)* provide a survey on facial expression recognition with deep learning, including datasets and algorithms. They present datasets that are available and have been commonly applied in previous works. Further, they outline commonly recognised data selection and evaluation concepts that have been used for these datasets. Next, they outline the general pipeline and workflow for a deep facial expression recognition approach, covering the corresponding background knowledge, and finally propose, for each stage, a feasible implementation.

*Wang & Deng (2018)* also provide a survey of the latest trends on deep facial expression detection, including the design of algorithms, but also possible applications, protocols and databases. They outline various network architectures and loss functions that have been introduced in the general field of the deep facial expression. They categorized the face processing approaches into two different classes: "one-to-many augmentation" and "many-to-one normalization". They also give a summarized overview of common databases and compare them concerning model training and model evaluation. Finally, they explored further scenarios for deep facial expression, like cross-factor, heterogeneous, industrial and multiple-media.

### Action and motion recognition

Understanding human actions and motions in visual data, like surveillance videos, is closely connected to research fields like object recognition, semantic segmentation, human dynamics and domain adaptation. *Herath, Harandi & Porikli (2017)* review notable steps that have been taken towards recognizing human actions. Therefore, they start with a discussion of first approaches that applied handcrafted representations. Subsequently, they review deep learning-based approaches suggested in this field.

*Wang et al. (2018b)* give an outline of latest trends and improvements of motion recognition in RGB-D images. They categorized the surveyed approaches into four groups. The groups are based on the particular modality used for recognition, and can be RGB-, skeleton, depth-, or RGB+D-based. Finally, they discuss the advantages and limitations, with a focus on approaches that encode spatial-temporal-structural information, which is inherent in video sequences.

### Biometric recognition

Biometric recognition, or biometrics, studies the identification of people utilizing their unique phenotypical characteristics, like fingerprints or the iris, for applications ranging from cell phone authentication to airport security systems. *Sundararajan & Woodard*

*(2018)* review one hundred distinct methods that study recognizing individuals with deep learning applying different biometric modalities. They conclude that the majority of research in biometrics based on deep learning has been conducted around face recognition and speaker recognition so far.

*Minaee et al. (2019)* conduct a survey of more than 120 works on biometric recognition, including face, fingerprint, iris, palm print, ear, voice, signature, and gait recognition. For each biometric recognition task, they present the available datasets used in the literature and their characteristics and outline the performance on popular public benchmarks.

### Image super-resolution

Image super-resolution is a basic task in image processing that has seen a rise in popularity with the advent of deep learning. Image super-resolution methods and algorithms are used to improve the resolution of (low-resolution) images and videos. *Wang, Chen & Hoi (2020)* give a comprehensive overview on latest trends and advances in the field of image super-resolution focusing on deep learning methods. They divide the existing papers on image super-resolution methods into three main categories, namely supervised image super-resolution, unsupervised image super-resolution and finally, domain-specific image super-resolution. Moreover, they cover other topics in the field of image super-resolution, like public accessible benchmark data collections and metrics for a performance evaluation.

### Image captioning

Image captioning refers to the generation of a description for an image. Therefore, their primary objects, but also their attributes and their relationships to each other within the image need to be recognized. Furthermore, image captioning must produce sentences that are syntactically and semantically correct. *Hossain et al. (2019)* introduce a broad survey of works for image captioning based on deep learning. They analyse their main strengths, performances, but also their limitations. In addition, they explore the datasets and the evaluation metrics that have been used for automatic image captioning with deep learning.

### Data augmentation

Data augmentation can be used for the expansion of (limited) datasets to obtain larger training and evaluation sets. *Shorten & Khoshgoftaar (2019)* review image augmentation algorithms that cover geometric transformations, but also colour space augmentations and feature space augmentation. Further, techniques like kernel filters, mixing images, random erasing, neural style transfer and meta-learning. They also cover generative adversarial network-based augmentation methods. In addition, they explore and study further characteristics in the area of data augmentation, like test-time augmentation, final size of the dataset, the impact of the resolution, but also curriculum learning. Finally, they give an overview of available approaches for meta-level decisions for implementing data augmentation.

### Generative adversarial networks

Generative adversarial networks or GANs belong to the field of generative models in machine learning. GANs have experienced an in-depth exploration during the last few years with the most significant impact in the field of computer vision. Wang, She & Ward (2019) survey three real-world problems that have been approached with GANs: the generation of high-quality images, diversity of image generation, and stable training. They give a detailed overview of the current state-of-the-art in generative adversarial networks. Furthermore, they structure their review using a specific taxonomy, which they have adopted based on variations in generative adversarial network-based architectures and loss functions.

### Going deeper: common architectures, methods, evaluations, pros, cons, challenges and future directions in computer vision

Table 3 gives more details about the presented methods, pros, cons, evaluations and challenges and future directions of the surveys in the category computer vision. The two surveys about general computer vision from Voulodimos et al. (2018) and Guo et al. (2016) introduce mainly CNNs, Boltzmann and autoencoder architectures. The authors list several pros and cons, regarding which networks can learn features automatically, generalize well, are computationally demanding during training, can be trained in real-time and do not perform well on small training sets. A challenge is, that the underlying theory of the models is not well understood, which leads to the problem of selecting an optimal or effective architecture or algorithm for a given task and that there is no clear understanding of what kind of architectures should perform better than other ones. Further challenges and future directions are the training with limited data, reducing the time complexity, the development of more powerful models and a better understanding in evolving and optimizing CNN architectures.

The three surveys about object detection with deep learning (Liu et al., 2020; Zhao et al., 2019; Jiao et al., 2019), all outline the main object detectors in deep learning, like R-CNN, Fast R-CNN, Faster R-CNN, Mask R-CNN, *etc*. in a temporal sequence. In doing so, showing the improvement from one detector to the other over time, which are for example a more arcuate object detection, improved detection speed and quality, variable image sizes, memory consumption, small-size object detection and localization, and end-to-end training possibilities. Future common challenges are even better and more efficient detection frameworks (*e.g*. the research field of generic object detection is still far from complete), unsupervised and weakly supervised learning, network optimization and combination, video object detection, 3D object detection and multi-domain object detection.

The two surveys about image segmentation (Garcia-Garcia et al., 2018; Minaee et al., 2020) review numerous deep learning architectures and variants, like CNN, AlexNet, VGG-16, GoogLeNet, ResNet, ReNet and further custom architectures. Garcia-Garcia et al. (2018) list the pros and cons for accuracy, efficiency, training, instance sequences, multi-modal and 3D. Minaee et al. (2020) consider the quantitative accuracy, speed (inference time), storage requirements (memory footprint) and further present a metrics

**Table 3 Architectures, pros, cons, evaluations, challenges and future directions in deep learning in the category *computer vision*.**

| **General computer vision** | |
|---|---|
| *Voulodimos et al. (2018)* | |
| Architectures/Methods | CNN, Boltzmann (DBN and DBM), SdA |
| Pros/Evaluations | Automatic feature learning (CNN), invariant to transformations (CNN); can work in an unsupervised fashion (DBN, DBM, SdA); can be trained in real time (SdA) |
| Cons/Evaluations | Needs labelled data (CNN); computationally demanding training (CNN, DBN, DBM) |
| Challenges and future directions | Optimal selection of model type and structure for a given task; why specific architecture or algorithm is effective in a given task or not |
| *Guo et al. (2016)* | |
| Architectures/Methods | CNN, RBM, AutoEncoder, Sparse coding |
| Pros/Evaluations | Generalization (CNN, RBM, AutoEncoder, Sparse coding); Unsupervised learning (RBM, AutoEncoder, Sparse coding); Feature learning (CNN, RBM, AutoEncoder); Real-time training (CNN, RBM); Real-time prediction (CNN, RBM, AutoEncoder, Sparse coding); Biological understanding (Sparse coding); Theoretical justification (CNN, RBM, AutoEncoder, Sparse coding); Invariance (CNN, Sparse coding) |
| Cons/Evaluations | Unsupervised learning (CNN); Feature learning (Sparse coding); Real-time training (AutoEncoder, Sparse coding); Biological understanding (CNN, RBM, AutoEncoder); Invariance (RBM, AutoEncoder); Small training set (CNN, RBM, AutoEncoder, Sparse coding) |
| Challenges and future directions | Underlying theory is not well understood; no clear understanding of which architectures should perform better than others; training with limited data; time complexity; more powerful models; better understanding in evolving and optimizing the CNN architectures |
| **Object detection** | |
| *Liu et al. (2020)* | |
| Architectures/Methods | Region Based (Two Stage) Frameworks (RCNN, SPPNet, Fast RCNN, Faster RCNN, RPN, RFCN, Mask RCNN, Chained Cascade Network and Cascade RCNN, Light Head RCNN), Unified (One Stage) Frameworks |
| Pros/Evaluations | Improved detection speed and quality (Fast RCNN); end-to-end detector training (Fast RCNN); efficient and accurate generating region proposals (Faster RCN); efficient region proposal computation (RPN); fully convolutional over the entire image (RFCN); pixelwise object instance segmentation (Mask RCNN); simple to training (Mask RCNN); end-to-end learning of more than two cascaded classifiers (Chained Cascade Network and Cascade RCNN); reduce the RoI computation (Light Head RCNN); single-stage object detectors based on fully convolutional deep networks (OverFeat); uses features from an entire image globally (YOLO); real time detection (YOLOv2); faster than YOLO (SSD); retaining high detection quality (SSD); outperforming all previous one stage detectors (CornerNet) |
| Cons/Evaluations | Slow and hard to optimize (RCNN); expensive in disk space and time (RCNN); Testing is slow (RCNN); slow (DetectorNet); less accurate than RCNN (OverFeat); makes localization errors (YOLO); fail to localize small objects (YOLO); slower than SSD (CornerNet) |
| Challenges and future directions | Open world Learning; better and more efficient detection frameworks; compact and efficient CNN features; automatic neural architecture search; object instance segmentation; weakly supervised detection; few/zero shot object detection; object detection in other modalities; universal object detection; the research field of generic object detection is still far from complete |
| *Zhao et al. (2019)* | |
| Architectures/Methods | Region Proposal-Based Framework (R-CNN, SPP-Net, Fast R-CNN, Faster R-CNN, R-FCN, FPN, Mask R-CNN, Multitask Learning, Multiscale Representation, and Contextual Modeling), Regression/Classification-Based Framework (Pioneer Works, YOLO, SSD) |
| Pros/Evaluations | Hierarchical feature representation, exponentially increased expressive capability, jointly optimize several related tasks together, large learning capacity (CNN); mid-level representations (SPP-Net); improve the quality of candidate BBs, extract high-level features (R-CNN); grouping and saliency cues to provide more accurate candidate boxes of arbitrary sizes quickly and to reduce the searching space in object detection (Region Proposal Generation); high-level, semantic, and robust feature representation for each region proposal can be obtained (CNN-Based Deep Feature Extraction); single stage training (Fast R-CNN); saves storage space (Fast R-CNN); end-to-end training (Faster R-CNN); object detection in a fully convolutional architecture (R-FCN); extract rich semantics from all levels and be trained end to end with all scales (FPN); predict segmentation masks in a pixel-to-pixel manner (Mask R-CNN); representation requires fewer parameters (Mask R-CNN); flexible and efficient framework for instance-level recognition (Mask R-CNN) |

(Continued)

**Table 3 (continued)**

| | |
|---|---|
| **General computer vision** | |
| Cons/Evaluations | Fixed size input image (CNN); Training is a multistage pipeline (R-CNN); training is expensive in space and time (CNN); redundant obtained region proposals (CNN); fixed-size input (SPP-Net); additional expense on storage space (SPP-Net); alternate training algorithm is very time-consuming (Faster R-CNN); training time and memory consumption increase rapidly (FPN); struggles in small-size object detection and localization (Mask R-CNN) |
| Challenges and future directions | Multitask joint optimization and multimodal information fusion; scale adaption; spatial correlations and contextual modeling; cascade network; unsupervised and weakly supervised learning; network optimization; 3D object detection; video object detection |
| *Jiao et al. (2019)* | |
| Architectures/Methods | Two-Stage Detectors (R-CNN, Fast R-CNN, Faster R-CNN, Mask R-CNN), One-Stage Detectors (YOLO, YOLOv2, YOLOv3, SSD, SSD, RetinaNet, M2Det), Latest Detectors (Relation Networks for Object Detection, DCNv2, NAS-FPN) |
| Pros/Evaluations | One-stage end-to-end training (Fast R-CNN); efficiently predict region proposals (Faster R-CNN); accurate object detector (Mask R-CNN); real-time detection (YOLO); improved speed and precision (YOLOv2); multi-label classification (YOLOv3); detects small objects (YOLOv3); single-shot detector for multiple categories within one-stage (SSD); adds prediction module and deconvolution module (DSSD); train unbalanced positive and negative examples (RetinaNet); effective feature pyramids (M2Det); an end-to-end training (RefineDet); better predict hard detected objects (RefineDet); considers the interaction between different targets in an image (Relation Networks for Object Detection); utilizes more deformable convolutional layers (DCNv2); deformable layers are modulated by a learnable scalar (DCNv2); top-down and bottom-up connections to fuse features (NAS-FPN) |
| Cons/Evaluations | Input vectors of fixed length (CNN); no shared computation (R-CNN); worse performance on medium and larger sized objects (YOLOv3); slower than RetinaNet800 (M2Det) |
| Challenges and future directions | Combining one-stage and two-stage detectors; video object detection; efficient post-processing methods; weakly supervised object detection methods; multi-domain object detection; 3D object detection; salient object detection; unsupervised object detection; multi-task learning; multi-source information assistance; constructing terminal object detection system; medical imaging and diagnosis; advanced medical biometrics; remote sensing airborne and real-time detection; GAN-based detector |
| **Image segmentation** | |
| *Garcia-Garcia et al. (2018)* | |
| Architectures/Methods | Variants of CNN, AlexNet, VGG-16, GoogLeNet, ResNet, ReNet, and custom architectures |
| Pros/Evaluations | Relatively simple (AlexNet); less parameters and easy to train (VGG); reduced numbers of parameters and operations (GoogLeNet); addresses problem of training deep networks (ResNet); overcoming vanishing gradients (ResNet); Accuracy, Efficiency, Training, Instance Sequences, Multi-modal, 3D |
| Cons/Evaluations | Complex (GoogLeNet); depth (ResNet); Accuracy, Efficiency, Training, Instance Sequences, Multi-modal, 3D |
| Challenges and future directions | Evaluation metrics; execution time; memory footprint; accuracy; reproducibility; 3D datasets; sequence datasets; point cloud segmentation; context knowledge; real-time segmentation; temporal coherency on sequences; multi-view integration |
| *Minaee et al. (2020)* | |
| Architectures/Methods | Fully convolutional networks; Convolutional models with graphical models; Encoder-decoder based models; Multi-scale and pyramid network based models; R-CNN based models (for instance segmentation); Dilated convolutional models and DeepLab family; Recurrent neural network based models; Attention-based models; Generative models and adversarial training; Convolutional models with active contour models; Other models |
| Pros/Evaluations | Quantitative accuracy, speed (inference time), and storage requirements (memory footprint) |
| Cons/Evaluations | Quantitative accuracy, speed (inference time), and storage requirements (memory footprint) |
| Challenges and future directions | More challenging datasets; interpretable deep models; weakly-supervised and unsupervised learning; real-time models for various applications; memory efficient models; 3D point cloud segmentation; application scenarios |
| **Face recognition** | |
| *Masi et al. (2018)* | |
| Architectures/Methods | DCNN, Deep-Face, VGG16, ResNet-50, FacePoseNet, STN, DREAM |
| Pros/Evaluations | Simplify classification (DREAM); reduce intra-class variance (CenterLoss); reduce the within-class variability (L2-constrained SoftMax); huge number of subjects (Hierarchical SoftMax); training very large pool of subjects (deep metric learning); learning deep embeddings (margin-based contrastive loss) |

**Table 3** (continued)

| General computer vision | |
|---|---|
| Cons/Evaluations | Handling pose variations (DCNN); does not explicitly minimize the intra-class variation of each subject (SoftMax layer based on cross-entropy); unclear parameter selection (L2-constrained SoftMax); increased complexity (deep metric learning losses) |
| Challenges and future directions | Automatically generated template for unknown individuals; video-based face recognition; multi-target tracking and recognition; automatic self-organization of a large *corpus* of unlabeled faces; adaptation and model tuning; biases in large datasets |
| *Li & Deng (2020)* | |
| Architectures/Methods | CNN, RBM, DBN, GAN, face alignment detectors (holistic, part-based, cascaded regression, deep learning), deep FER networks for static images, deep FER networks for dynamic image sequence, AlexNet, VGG, VGG-face, GoogleNet, frame aggregation, expression intensity, RNN, C3D, FLT, CN, NE, others |
| Pros/Evaluations | Real-time, speed, performance (holistic, part-based, cascaded regression, deep learning); network size, preprocessing, data selection, additional classifier, performance (CNN, RBM, DBN, GAN); data size, spatial and temporal information, frame length, accuracy, efficiency (frame aggregation, expression intensity, RNN, C3D, FLT, CN, NE) |
| Cons/Evaluations | Real-time, speed, performance (holistic, part-based, cascaded regression, deep learning); network size, preprocessing, data selection, additional classifier, performance (CNN, RBM, DBN, GAN); data size, spatial and temporal information, frame length, accuracy, efficiency (frame aggregation, expression intensity, RNN, C3D, FLT, CN, NE) |
| Challenges and future directions | Facial expression dataset (illumination variation, occlusions, non frontal head poses, identity bias and the recognition of low-intensity expression, age, gender and ethnicity, employ crowd-sourcing models, fully automatic labeling tool); dataset bias and imbalanced distribution (generalization on unseen test data, performance in cross-dataset settings, imbalanced class distribution); Incorporating other affective model (capture the full repertoire of expressive behaviors, different facial muscle action parts, dealing with continuous data, learn expression-discriminative representations); multimodal affect recognition (multimodal sentiment analysis, processing these diverse modalities, multi-sensor data fusion methods, intra-modality and inter-modality dynamics, infrared images, depth information from 3D face models, physiological data) |
| *Wang & Deng (2018)* | |
| Architectures/Methods | Backbone network (AlexNet, VGGNet, GoogleNet, ResNet, SENet, light-weight architectures, adaptive architectures, joint alignment-recognition architectures), assembled networks (multipose, multipatch, multitask), DeepFace, DeepID2, DeepID3, FaceNet, Baidu, VGGface, light-CNN, loss related |
| Pros/Evaluations | Number of networks, training set, accuracy (DeepFace, DeepID2, DeepID3, FaceNet, Baidu, VGGface, light-CNN, loss related) |
| Cons/Evaluations | Number of networks, training set, accuracy (DeepFace, DeepID2, DeepID3, FaceNet, Baidu, VGGface, light-CNN, loss related) |
| Challenges and future directions | Security issues; privacy-preserving face recognition; understanding deep face recognition; remaining challenges defined by non-saturated benchmark datasets; ubiquitous face recognition across applications and scenes; pursuit of extreme accuracy and efficiency; Fusion issues |
| **Action/motion recognition** | |
| *Herath, Harandi & Porikli (2017)* | |
| Architectures/Methods | Spatiotemporal networks, multiple stream networks, deep generative networks, temporal coherency networks, VGG, Decaf, RNN, LSTM, LRCN, Dynencoder, autoencoder, adversarial models, Siamese Networks, others |
| Pros/Evaluations | Accuracy on seven challenging action datasets (CNN, ClarifaiNet, GoogLeNet, VGG, others) |
| Cons/Evaluations | Accuracy on seven challenging action datasets (CNN, ClarifaiNet, GoogLeNet, VGG, others) |
| Challenges and future directions | Training video data; knowledge transfer; heterogeneous domain adaptation; boost the performance; generic form of deep architectures for spatiotemporal learning; carefully engineered approaches; data augmentation techniques; foveated architecture; distinct frame sampling strategies; more realistic activities; real-life scenarios; deeper understanding in action recognition |
| *Wang et al. (2018b)* | |
| Architectures/Methods | CNN, LSTM, RNN, Autoencoder, DDNN, IDMM, others |
| Pros/Evaluations | Accuracy, Jaccard Index, cross-subject setting, cross-view setting (CNN, LSTM, RNN, Autoencoder, DDNN, IDMM, others) |
| Cons/Evaluations | Accuracy, Jaccard Index, cross-subject setting, cross-view setting (CNN, LSTM, RNN, Autoencoder, DDNN, IDMM, others) |

(Continued)

**Table 3** (continued)

| **General computer vision** | |
| --- | --- |
| Challenges and future directions | Better results on large complex dataset; practical intelligent recognition systems; encoding temporal information; Small training data; viewpoint variation and occlusion; execution rate variation and repetition; cross-datasets; online motion recognition; action prediction; hybrid networks; simultaneous exploitation of spatial-temporal-structural information; fusion of multiple modalities; large-scale datasets; zero/one-shot learning; outdoor practical scenarios; unsupervised learning/self-learning; online motion recognition and prediction |
| **Biometric recognition** | |
| *Sundararajan & Woodard (2018)* | |
| Architectures/Methods | Deep Boltzmann Machines, Restricted Boltzmann Machines, Deep Belief Networks, Autoencoders, CNN, RNN |
| Pros/Evaluations | (recognition) accuracy, Equal Error Rate, Mean Absolute Error, False Alarm Rate, False Rejection Rate (CNN, DNN, DBN, RNN, others) |
| Cons/Evaluations | (recognition) accuracy, Equal Error Rate, Mean Absolute Error, False Alarm Rate, False Rejection Rate (CNN, DNN, DBN, RNN, others) |
| Challenges and future directions | Real-world applicability; beyond face and voice recognition; scaling up in terms of identification; large-scale datasets; dataset quality; computing resources; training speed-up; large-scale identification; behavioral biometrics; robust to data noise; modeling biometric aging; biometric segmentation; fusion of multiple modalities |
| *Minaee et al. (2019)* | |
| Architectures/Methods | CNN, AlexNet, VGGNet, GoogleNet, ResNet, SphereFace, FingerNet, SCNN, RSM, variants |
| Pros/Evaluations | Equal Error Rate, accuracy (Rank1 identification, verification accuracy), performance (accuracy, Equal Error Rate, R1-ACC) |
| Cons/Evaluations | Equal Error Rate, accuracy (Rank1 identification, verification accuracy), performance (accuracy, Equal Error Rate, R1-ACC) |
| Challenges and future directions | More challenging datasets; interpretable deep models; few shot learning, and self-supervised learning; biometric fusion; real-time models for various applications; memory efficient models; security and privacy issues |
| **Image super-resolution** | |
| *Wang, Chen & Hoi (2020)* | |
| Architectures/Methods | SRCNN, DRCN, FSRCNN, ESPCN, LapSRN, DRRN, SRResNe, SRGAN, EDSR, EnhanceNet, MemNet, SRDenseNet, DBPN, DSRN, RDN, CARN, MSRN, RCAN, ESRGAN, RNAN, Meta-RDN, SAN, SRFBN |
| Pros/Evaluations | Performance, PSNR (FSRCNN, LapSR, SRCNN, CARN-M, FALSR-B, FALSR-C, BTSRN, CARN, FALSR-A, OISR-RK2-s, OISR-LF-s, VDSR, MemNet, MSRN, OISR-RK2, MDSR, DBPN, RDN, SAN, RCAN, DRRN, DRCN, EDSR, OISR-RK3) |
| Cons/Evaluations | Performance, PSNR (FSRCNN, LapSR, SRCNN, CARN-M, FALSR-B, FALSR-C, BTSRN, CARN, FALSR-A, OISR-RK2-s, OISR-LF-s, VDSR, MemNet, MSRN, OISR-RK2, MDSR, DBPN, RDN, SAN, RCAN, DRRN, DRCN, EDSR, OISR-RK3) |
| Challenges and future directions | Combining local and global information; combining low- and high-level information; context-specific attention; more efficient architectures; upsampling methods; learning strategies; more accurate metrics; blind IQA methods; unsupervised super-resolution; towards real-world scenarios; dealing with various degradation; domain-specific applications |
| **Image captioning** | |
| *Hossain et al. (2019)* | |
| Architectures/Methods | Image encoder (AlexNet, VGGNet, GoogLeNet, ResNet, Inception-V3), Language model (LBL, LSTM, SC-NLM, RNN, DTR, MELM, Language CNN) |
| Pros/Evaluations | BLEU, PPLX, R@K, mrank, METEOR, CIDEr, PPLX, AP, IoU, PPL, Human Evaluation, E-NGAN, E-GAN, SPICE, SPIDEr |
| Cons/Evaluations | BLEU, PPLX, R@K, mrank, METEOR, CIDEr, PPLX, AP, IoU, PPL, Human Evaluation, E-NGAN, E-GAN, SPICE, SPIDEr |
| Challenges and future directions | Detect prominent objects and attributes and their relationships; generating accurate and multiple captions; open-domain dataset; generate high-quality captions; adding external knowledge; generate attractive image captions; supervised learning needs a large amount of labeled training data; focus on unsupervised learning and reinforcement learning |
| **Data augmentation** | |
| *Shorten & Khoshgoftaar (2019)* | |
| Architectures/Methods | Deep Learning-based (adversarial training, neural style transfer, GAN data augmentation), CNN, LeNet-5, AlexNet, GAN, NAS, DCGAN, CycleGAN, Progressively-Growing GAN, WGAN, variants |

**Table 3** (continued)

| **General computer vision** | |
|---|---|
| Pros/Evaluations | C10, C10+, C100, C100+, SVHN (ResNetl8, ResNet18, WideResNet, Shake-shake regularization); accuracy (original testing data, FGSM, PGD); Visual Turing Test (DCGAN, WGAN); validation accuracy (None, Traditional, GAN, Neural, Neural + loss, Control); AutoAugmen, ARS (Wide-ResNet, Shake-Shake, AmoebaNet, PyramidNet); Rank, score, class (Deep Image) |
| Cons/Evaluations | C10, C10+, C100, C100+, SVHN (ResNetl8, ResNet18, WideResNet, Shake-shake regularization); accuracy (original testing data, FGSM, PGD); Visual Turing Test (DCGAN, WGAN); validation accuracy (None, Traditional, GAN, Neural, Neural + loss, Control); AutoAugmen, ARS (Wide-ResNet, Shake-Shake, AmoebaNet, PyramidNet); Rank, score, class (Deep Image) |
| Challenges and future directions | Establishing a taxonomy of augmentation techniques; improving the quality of GAN samples; combine meta-learning and data augmentation; explore relationships between data augmentation and classifier architecture; extending data augmentation principles to other data types; impact on video data; translation to text, bioinformatics, tabular records; establish benchmarks for different levels of limited data; super-resolution networks; test-time augmentation; meta-learning GAN architectures; practical integration of data augmentation into deep learning software tools; common data augmentation APIs |
| **Generative adversarial networks** | |
| *Wang, She & Ward (2019)* | |
| Architectures/Methods | FCGAN, SGAN, BiGAN, CGAN, InfoGAN, AC-GAN, LAPGAN, DCGAN, BEGAN, PROGAN, SAGAN, BigGAN, rGANs, YLG, AutoGAN, MSG-GAN, Loss-Variant GANs, WGAN, WGAN-GP, LSGAN, f-GAN, UGAN, LS-GAN, MRGAN, Geometric GAN, RGAN, SN-GAN, RealnessGAN, Sphere GAN, SS-GAN, variants |
| Pros/Evaluations | Fast sample generation, handles sharp probability distribution (FCGAN); mode diversity, stabilizes training (MRGAN); unified framework (f-GAN); solves vanishing gradient, image quality, solves mode collapse (WGAN); converges fast, stable model training, complex functions (WGAN-GP); vanishing gradient, stabilized training, mode diversity, easy implementation (LSGAN); vanishing gradient, mode collapse (LS-GAN); mode collapse, stable training, converges to Nash equilibrium (Geometric GAN); mode collapse, high order gradient information, training stability (Unrolled GAN); vanishing gradient, unified framework (IPM GANs), mode collapse (RGAN); computationally light, easy implementation, image quality, mode collapse, stable training, vanishing gradient (SN-GAN); stable training, accurate results, no additional constraints (Sphere GAN); self-supervision, competitive results (SS-GAN); discriminator distribution as a measure of realness, image quality (RealnessGAN); time (DCGAN, BEGAN, PROGAN, RFACE); accuracy (DCGAN, BEGAN, PROGAN); score (DCGAN, BEGAN, PROGAN); performance (FCGAN, BEGAN, PROGAN, LSGAN, DCGAN, WGAN-GP, SN-GAN, Geometric GAN, RGAN, AC-GAN, BigGAN, RealnessGAN, MSG-GAN, SS-GAN, YLG, Sphere GAN) |
| Cons/Evaluations | Vanishing gradient for G, mode collapse, low image resolution (FCGAN); low image resolution, vanishing gradient for G, limited testing (MRGAN); stability (f-GAN); convergence time, vanishing gradient, complex learning hard to converge (WGAN); no batch normalization (WGAN-GP); image quality (LSGAN); difficult implementation, image quality (LS-GAN); vanishing gradient, limited testing (Geometric GAN); image quaility (Unrolled GAN); added relativism, performance comparison (RGAN); limited testing (SN-GAN); limited investigation (Sphere GAN); self-supervised architecture (SS-GAN); model diversity (RealnessGAN); time (DCGAN, BEGAN, PROGAN, RFACE); accuracy (DCGAN, BEGAN, PROGAN); score (DCGAN, BEGAN, PROGAN); performance (FCGAN, BEGAN, PROGAN, LSGAN, DCGAN, WGAN-GP, SN-GAN, Geometric GAN, RGAN, AC-GAN, BigGAN, RealnessGAN, MSG-GAN, SS-GAN, YLG, Sphere GAN) |
| Challenges and future directions | Limited GAN research in other non-computer vision areas; GANs are hard to apply to the natural language application field; generating comments to live streaming (NLP); significant impact on neuroscience by tackling privacy issues; limited exploration of time-series data generation; lack of efficient evaluation metrics in some areas; society, safety concerns, *e.g.* generation of tampered videos; detector for AI-generated images; GPU memory problems for large batched images; loss functions important for stable training; video application is still limited |

evaluation for pixel accuracy/mean pixel accuracy (MPA), intersection over union (IoU)/Jaccard Index and Dice coefficient. Both surveys see future challenges in more memory efficient models, real-time segmentations and diverse datasets, like 3D datasets, sequence datasets, and in general more challenging datasets. Both also point out point cloud segmentation. Further challenges are for example execution time, accuracy, reproducibility and weakly-supervised and unsupervised learning.

The three surveys in face recognition (*Masi et al., 2018*; *Li & Deng, 2020*; *Wang & Deng, 2018*) have all a strong focus on datasets and loss functions. A reason for that is that there exist many large public facial databases, which make comparable evaluations among the methods feasible. In this regard, *Li & Deng (2020)* focus on databases and evaluations specifically for Facial Expression Recognition (FER) in their survey. However, the surveys see the common challenges in boarded databases, handling biases in large datasets, like ethnicity, gender, age or other factors, having more non-saturated benchmark datasets and ubiquitous face recognition across applications and scenes. Further open challenges are privacy and security issues, pursuing extreme accuracy and efficiency, understanding deep face recognition and fusion issues. Finally, *Masi et al. (2018)* state reducing intra-class variance and increasing the margin between classes while training as two main challenges, and that video processing and clustering are currently the next frontiers for face recognition.

Table 3 presents two surveys in Action/motion recognition (*Herath, Harandi & Porikli, 2017*; *Wang et al., 2018b*), the latter one specific on RGB-D-based human motion recognition with deep learning. Both surveys present numerous deep learning-based methods, like VGG, Decaf, RNN, LSTM, LRCN, autoencoder, adversarial models and others, and present accuracy evaluations on numerous public available benchmark datasets. Common challenges of the surveys point out the need for practical (carefully engineered) systems that can handle more realistic activities and (outdoor) real-life scenarios. Further challenges are the development of more generic architectures that can handle also multiple modalities, have a better performance and can better train video data. Moreover, the surveys point out a deeper understanding in action recognition, better results on smaller training data, but also better results on large complex datasets, solutions for online motion recognition and prediction, and zero/one-shot learning.

In the sub-category biometric recognition, we present two surveys (*Sundararajan & Woodard, 2018*; *Minaee et al., 2019*). Both survey publications are dominated by CNN architectures and variants. *Sundararajan & Woodard (2018)* divide their survey by biometric modalities, namely physiological biometrics (face, fingerprint, palmprint, and iris) and behavioral biometrics (voice, signature, gait, and keystroke), and present evaluation results on public and private datasets. *Minaee et al. (2019)* arrange their survey by the following biometric recognitions: fingerprint, iris, palmprint, ear, voice, signature and gait. However, *Sundararajan & Woodard (2018)* conclude, that deep learning in biometrics has so far been very little explored beyond face and speaker recognition. Among others, *Sundararajan & Woodard (2018)* point out that future challenges are computing resources and training speed-up, also in regards to large-scale identification. Furthermore, there is a need for models that are more robust, the modeling of biometric aging and the fusion of multiple modalities. *Minaee et al. (2019)* see future directions, for example, in more challenging datasets, interpretable deep learning models, few shot and self-supervised learning, real-time and memory efficient models, and the dealing with security and privacy issues.

*Wang, Chen & Hoi (2020)* present in their survey specific deep learning architecture and methods, like SRCNN, DRCN, FSRCNN, ESPCN, LapSRN, DRRN, SRResNe, that have been developed for the field of super-resolution. For evaluation, they show the performance (PSNR) of various models like FSRCNN, LapSR, SRCNN, CARN-M, FALSR-B, FALSR-C, BTSRN, CARN, on four benchmark datasets. They see one future challenge in more efficient architectures and better learning strategies in regards to loss functions and normalization, but also unsupervised super-resolution. Further, they state that the evaluation metrics need more exploration, specifically they want to see more accurate metrics and the development of blind IQA methods in the field of super-resolution. Moreover, they request for combining low- and high-level information, but also combining local and global information. Final challenges concern real-world scenarios dealing with various degradation and domain-specific applications.

A survey about image captioning with deep learning is presented by *Hossain et al. (2019)*. In doing so, they survey image encoder methods, like AlexNet, VGGNet, GoogLeNet, ResNet, Inception-V3, and language models, like LBL, LSTM, SC-NLM, RNN, DTR, MELM, Language CNN, in the field of image captioning. They also list the used evaluation metrics applied in the survey publications for most commonly and public benchmark datasets in that field, like +BLEU, PPLX, R@K, mrank, METEOR, CIDEr, PPLX, AP, IoU, PPL, Human Evaluation, E-NGAN, E-GAN, SPICE, SPIDEr. The authors see future challenges in detecting prominent objects and attributes, and their relationships, and generating accurate and multiple captions. They also point out that open-domain datasets are needed. Future directions are the generation of high-quality captions, adding external knowledge and the generation of attractive image captions. Finally, they state that supervised learning needs a large amount of labeled training data, so the focus will be on unsupervised learning and reinforcement learning in image captioning.

*Shorten & Khoshgoftaar (2019)* present a survey on data augmentation for deep learning. Thereby, dividing deep Learning-based methods in adversarial training, neural style transfer and GAN data augmentation. They introduce several deep learning architectures in the field of data augmentation, like CNN, LeNet-5, AlexNet, GAN, NAS, DCGAN, CycleGAN, Progressively-Growing GAN, WGAN and variants. Evaluation of the data augmentation methods are outlined for a wide range of tasks and datasets with the task specific metrics and scores, like (validation) accuracy, visual Turing test, ARS, C10, C10+, C100, C100+, SVHN, rank, class, *etc*. Future works are seen in establishing a taxonomy of augmentation techniques. Further, improving the quality of GAN samples, combing meta-learning and data augmentation, exploring the relationships between data augmentation and classifier architectures. Moreover, extending the data augmentation principles to other data types, identify the impact on video data and a translation to text, bioinformatics, and tabular records. Finally, the authors call for a practical integration of data augmentation into deep learning software tools and common data augmentation APIs.

In their survey, *Wang, She & Ward (2019)* give a comprehensive overview of generative adversarial networks, like FCGAN, SGAN, BiGAN, CGAN, InfoGAN, AC-GAN, LAPGAN, DCGAN, BEGAN, PROGAN, SAGAN, BigGAN, rGANs, YLG, AutoGAN, MSG-GAN, Loss-Variant GANs, WGAN, WGAN-GP, LSGAN, f-GAN, UGAN, LS-GAN, MRGAN, Geometric GAN, RGAN, SN-GAN, RealnessGAN, Sphere GAN, SS-GAN, variants. They also provide a compact overview of the pros and cons of different GAN-based methods, outlining which methods can, for example, handle the vanishing gradient problem, generate low quality images, have not been tested or investigated enough, and which are easy or hard to implement. In addition, they present several evaluation tables and images, about time performance (DCGAN, BEGAN, PROGAN, RFACE), accuracy (DCGAN, BEGAN, PROGAN), scoring (DCGAN, BEGAN, PROGAN) and further performances (FCGAN, BEGAN, PROGAN, LSGAN, DCGAN, WGAN-GP, SN-GAN, Geometric GAN, RGAN, AC-GAN, BigGAN, RealnessGAN, MSG-GAN, SS-GAN, YLG, Sphere GAN). The authors conclude, that GAN-based research is still limited in other non-computer vision areas and that GANs are hard to apply to the natural language application field. However, GANs could be used in NLP for generating comments to live streaming. They also state that GANs could have a significant impact on neuroscience by tackling privacy issues. On the other hand, they outline society and safety concerns when tampered videos are generated with GANs. Hence, a future challenge is to develop detectors for AI-generated images. Future directions are solving the lack of efficient evaluation metrics in some areas, GPU memory problems for large batched images and video application are in general still under-researched.

To conclude this section, most of the surveys reviewed in this category have also a similar structure: starting with a general introduction of machine/deep learning techniques followed by a discussion of the reviewed publications categorized using different strategies, among which task- or application-based categorization is the most widely used. The works usually conclude by stating challenges and future directions of deep learning in their fields from a high-level point of view. The future challenges are very broad depending on the field and application. However, despite the successes and advances of deep learning in the field of computer vision, there is still research needed to make them more reliable for real world applications.

## Deep learning reviews about natural language processing

This sub-section deals with the deep learning reviews in the area of natural language processing. It is divided into eight sub-categories and the number of references, and citations (according to Google Scholar and status as of mid-August 2020) for each of these categories is given in Table 4:

- general language processing,
- language generation and conversation,
- named entity recognition,
- sentiment analysis,
- text summarization,
- answer selection,
- word embedding,
- and financial forecasting.

The commonality in the field of natural language processing is to analyse and understand natural language data. Examples include text on documents or actual spoken language. Real-world application can be optical character recognition (OCR), which tries to determine the text printed on an image, or a smartphone that can perform instructions by processing voice commands. Again, the major challenge in natural language processing is generalization. Every voice, handwriting or pronunciation is at least slightly different, which makes it difficult to generalize for algorithms. Equivalent to objective evaluations in computer vision, common scores are used to evaluate natural language processing algorithms, like the percentage of the text that has been correctly determined with an OCR approach.

**Table 4** List of published reviews in deep learning in the category *language processing*.

| Language processing | Publications | Number of references | Citations (until August 2020) | Preprints |
|---|---|---|---|---|
| General language processing | *Young et al. (2018)* | 164 | 922 | No |
| Language generation and conversation | *Gatt & Krahmer (2018)* | 548 | 270 | No |
| | *Santhanam & Shaikh (2019)* | 137 | 8 | Yes |
| | *Gao, Galley & Li (2018)* | 20 | 215 | No |
| | *Chen et al. (2017)* | 111 | 202 | No |
| Named entity recognition | *Li et al. (2020)* | 211 | 43 | No |
| | *Yadav & Bethard (2019)* | 83 | 145 | Yes |
| Sentiment analysis | *Zhang, Wang & Liu (2018)* | 150 | 398 | No |
| | *Do et al. (2019)* | 135 | 86 | No |
| Text summarization | *Shi et al. (2018)* | 131 | 30 | Yes |
| Answer selection | *Lai, Bui & Li, 2018* | 56 | 33 | No |
| Word embedding | *Zhang et al. (2016)* | 187 | 27 | Yes |
| | *Almeida & Xexéo (2019)* | 48 | 18 | Yes |
| Financial forecasting | *Xing, Cambria & Welsch (2018)* | 128 | 93 | No |
| Sum | – | 2,109 | 2,490 | – |

### *Natural language processing*

In general, natural language processing is a theory-inspired variety of computational methods and algorithms for the automatic studying of the human language that can be used, for example, for voice commands in all kind of applications. *Young et al. (2018)* survey several important deep learning-based approaches that have been utilized for various tasks of natural language processing. They provide a walk-through of their evolution during the last years and overview, compare and contrast the numerous models. Finally, they provide an in-depth overview of the past, the present and future role of deep learning in natural language processing.

### *Language generation and conversation*

A task of natural language processing is the generation (NLG) of text or speech from a non-linguistic input. This can be the generation of new texts from (often human-written) existing ones from one language to another language by machine translation, or a summarization and fusion of texts with the goal to make them more concise. *Gatt & Krahmer (2018)* provide an overview of the published research in common NLG tasks and the corresponding neural architectures. They explore common research areas between NLG and general artificial intelligence and outline the particular challenges in evaluation in the field.

*Santhanam & Shaikh (2019)* provide an overview of classical methods, statistical methods, and methods that utilize deep neural networks and review publications on open domain dialogue systems. Thereby, they recognize three further research directions for the development of more effective dialogue systems. These are incorporating conversation context and world knowledge, but also including larger contexts. Further, they highlight the problem of generic or dull responses and mention ways to improve the quality of NLG systems, for example by incorporating personae or personality information. *Gao, Galley & Li (2018)* present a tutorial that surveys neural approaches to conversational artificial intelligence and arrange conversational systems into three different categories. These are agents for question-answering, task-oriented dialogues and social bots. For each of these categories, they introduce the state-of-the-art overview of neural methods, but also make connections between these methods and classical methods. Finally, they outline the current progress in this field and present the remaining challenges by applying certain models and systems in case studies.

*Chen et al. (2017)* propose a survey overview of the current progress in the area of dialogue systems from different angles and explore future research areas and topics. Established dialogue systems are grouped by the authors into two model categories: Task-oriented and non-task-oriented. Then, they outline how deep learning-based approaches can support these with specific algorithms. Finally, they highlight several further research areas, which can support and advance the field of dialogue-systems.

### Named entity recognition

Named entity recognition deals with the identification of named entities (for example real-world objects, like persons or locations) and furthermore, their classification into specific categories. It functions as foundation for natural language approaches, like the summarization of text, answering questions or machine translation. *Li et al. (2020)* provide a large panoramic of established deep learning-based approaches for named entity recognition. They introduce named entity recognition resources, including tagged named entity recognition corpora, and further off-the-shelf named entity recognition applications. Available works are systematically arranged, depending on a taxonomy along three axes, namely the distributed representations for the input, the context encoder, and the tag decoder. Finally, they review the main approaches for deep learning-based techniques that have recently been applied and outline the particular challenges for named entity recognition systems in their contribution.

*Yadav & Bethard (2019)* present a broad overview of deep neural network approaches for the field of named entity recognition. They compare them with existing methods for named entity recognition that use feature engineering and further supervised or semi-supervised learning methods. Finally, they outline the advantages that have been gained by neural networks and depict how including specific works on feature-based named entity recognition systems can yield further improvements.

### *Sentiment analysis*

Sentiment analysis is the (automatic) recognition and analysis of people's opinions, sentiments, emotions and appraisals. This can be used in data mining applications for the exploration of this subjective information source and, therefore, opinions of specific entities, like products, events or services, but also topics, individuals or organizations. In their contribution, *Zhang, Wang & Liu (2018)* start with an introduction of deep learning, then they provide an extensive review of its recent sentiment analysis applications. They divide the area of recent sentiment analysis in different sub-categories, like a sentiment classification on a document-level, sentence-level and aspect-level, and a further opinion expression extraction.

*Do et al. (2019)* wrote a comprehensive and context-based overview of deep learning methods that have been used in aspect-based sentiment analysis. For their review contribution, they categorised and summarised 40 approaches by their main deep learning architecture and their specific classification tasks. The review works consisting of general, but also adaptions of common convolutional neural networks, long-short term memory methods, and gated recurrent units.

### *Text summarization*

Text summarization targets the summarization of (long) documents into shorter ones while, at the same time, keeping the key meaning and information of the original text documents. *Shi et al. (2018)* give a broad technical literature review on various seq2seq (sequence-to-sequence learning) methods for the summarization of text from the angle of training strategies, network structures and summary generation methods. In addition to the review, they implemented an open-source library, called the Neural Abstractive Text Summarizer (NATS) toolkit, which can be used for abstractive text summarization. Further, they conducted a set of experiments on the common CNN/Daily Mail dataset to evaluate the capabilities and results of diverse neural network components. They conclude by benchmarking two implemented NATS methods on the Newsroom and Bytecup datasets.

### *Answer selection*

The aim of (automatic) answer selection is identifying correct (and incorrect) answers. For example, for a given question and a number of possible candidate answers, answer selection identifies which of the candidates answered the question correctly (and, concurrently, which of the candidates do not). In their survey, *Lai, Bui & Li (2018)* outline an extensive, systematic analysis of numerous deep learning-based approaches for answer selection along two main dimensions: (1) neural network architectures, such as attentive architecture, siamese architecture and compare-aggregate architecture, and (2) learning strategies, such as listwise, pairwise, and pointwise. Moreover, they examined the most common datasets for answer selection and their evaluation metrics, and present various possible research directions for the future in this field.

### Word embedding

Word embedding, or short embedding, covers feature learning and modelling strategies for representing words or phrases as numbers or vectors. In their work, *Zhang et al. (2016)* review the state-of-the-art of neural information retrieval research, with a focus on the usage of queries and document representations that have been learned, like neural embeddings. In doing so, they outline the achievements in the field of neural information retrieval, but also point out limitations for a broader usage, and conclude by proposing possible and favourable future research directions.

The survey contribution of *Almeida & Xexéo (2019)* depicts and delineates the main recent strategies in the field of word embedding. The authors introduce two main categories for word embeddings and the corresponding publications: prediction-based models and count-based models.

### Financial forecasting

Financial forecasting tries to (automatically) predict financial market trends, like stock market predictions. Financial forecasting can, for example, be based on financial statements and reports, but also news articles and press releases, with the goal to keep a competitive business advantage. *Xing, Cambria & Welsch (2018)* present in their work the scope of natural language-based financial forecasting (NLFF) research by arranging and organizing the methods and approaches from the reviewed works. Their review publication targets on providing a greater knowledge of the advancements and NLFF hotspots.

### Going deeper: architectures, evaluations, pros, cons, challenges and future directions in language processing

Table 5 provides more details about the presented methods, pros, cons, evaluations and challenges and future directions of the surveys in the category language processing.

General language processing is covered by *Young et al. (2018)*, where at first word and character embeddings are described, highlighting the challenge of out of vocabulary (OOV) words, which are partly addressed by character embeddings. Still, for embeddings it has been found to be limited in terms of perceptual understanding, which could motivate grounded learning. For CNNs the limitation of capturing long distance dependencies has been noted, which also motivated the development of dynamic convolutional neural networks and Recurrent Neural Networks. The survey also mentions the role of the pooling layer, where in the case of max pooling an information loss can be observed, calling for modified techniques, such as dynamic multi-pooling CNNs. The fixed size representation of traditional Sequence-to-Sequence models can be overcome by means of the Attention Mechanism. Memory-augmented networks capture a similar intuition to model the complex dependencies and relationships in text. Finally, the survey provides a good overview of the performance of a range of approaches on various tasks, including POS tagging, Semantic Role Labelling, and Sentiment Classification. In general, recurrent network architectures, like LSTMs, are ranked among the highest across the tasks.

**Table 5** **Architectures, pros, cons, evaluations, challenges and future directions in deep learning in the category *language processing*.**

| | |
|---|---|
| **General language processing** | |
| *Young et al. (2018)* | |
| Architectures/Methods | Word and character embeddings (*e.g.*, word2vec); CNN (including temporal windows, custom pooling methods), LSTM, GRU; Attention Mechanism; Recursive NN; VAE; Deep Reinforcement Learning; Memory-Augmented NN |
| Pros/Evaluations | Generalization *via* distributed representation (*e.g.*, word2vec); ability to summarize sentences (Recurrent Neural Networks); improved performance over traditional RNNs (LSTM, GRU); modelling of the structure of text (Recursive Neural Networks) |
| Cons/Evaluations | Out of vacabulary words (word embeddings); not suitable for long distance relationships (CNN), max-pooling to restrictive (CNN); vanishing gradients (traditional RNN); fixed sized vector (Seq2Seq models) |
| Challenges and future directions | Lack of labelled data limits supervised learning; expect to see more Reinforcement Learning in NLP; expect to see more models with internal memory; combination of symbolic and sub-symbolic AI |
| **Language generation and conversation** | |
| *Gatt & Krahmer (2018)* | |
| Architectures/Methods | LSTM, Encoder-Decoder Architectures, Sequence-to-Sequence, Evaluation measures (BLUE, NIST, ROUGE, METEOR, GTM, CIDEr, WMD, Edit Distance, TER, TERP, TERPA, DICE, JACCARD, MASI, PYRAMID, SPICE). |
| Pros/Evaluations | Suitable for machine translation (Sequence-to-Sequence); modelling long-ranging dependencies (LSTM); efficiency (of data-driven approaches over rule-based) |
| Cons/Evaluations | Limited suitability in commercial application (data-driven methods); limited availability of data (data-driven methods) |
| Challenges and future directions | Stylistic control, social media data, situated language generation, generation from structured knowledge bases and ontologies |
| *Santhanam & Shaikh (2019)* | |
| Architectures/Methods | MLP, LSTM, GRU, Encoder-Decoder Architecture, Memory Networks, Transformer, evaluation measures (BLEU, METERO, Perplexity, Distinct, Word Error Rate, F1-Score) |
| Pros/Evaluations | Well researched model (Sequence-to-Sequence) |
| Cons/Evaluations | Bottleneck of fixes size vector (Encoder-Decoder networks); not mature (transformer architectures for open domain dialogue systems) |
| Challenges and future directions | Encoding contextual information (*e.g.*, from knowledge bases), incorporating personality, dull and generic responses; cognitive architectures, encoding emotional content |
| *Chen et al. (2017)* | |
| Architectures/Methods | LSTM, GRU, CNN, Encoder-Decoder, Sequence-to-Sequence, Attention mechanism, End-to-End systems, key-value memory networks, evaluation measures (BLUE, METEOR, ROUGE) |
| Pros/Evaluations | Allow to sample from distribution to avoid generic responses (VAE), make use of large amounts of data (deep learning) |
| Cons/Evaluations | No information about uncertainty (query-based systems), not differential (query-based systems) |
| Challenges and future directions | Trivial and generic answers, inconsistencies in the training data due to multiple speakers, warm-up for new domains, privacy-preservation |
| **Named entity recognition** | |
| *Li et al. (2020)* | |
| Architectures/Methods | Word-level and character level representations, CNN, RNN, LSTM, CRF-based neural system, ELMo, BERT, Recursive Neural Networks, Transformer, GPT, Pointer Networks |
| Pros/Evaluations | Rule-base system perform well (if lexicon is exhaustive); effective use of past and future information (bidirectional RNNs); promising performance of traditional embeddings and language model embedding; Transformer encoder is better suited than LSTMs |
| Cons/Evaluations | Later words influence the representation more (RNNs); external knowledge is labour intensive; external knowledge hurts generality of DL-bases systems; pointer networks and RNNs (due to greedy decoding); computationally expensive (CRF, for many entity classes) |

(Continued)

**Table 5** (continued)

| General language processing | |
|---|---|
| Challenges and future directions | Multiple languages (including transfer learning between languages), data annotation (quality and consistency of annotations), informal text and unseen entities, scaleability of DL-based NER |
| *Yadav & Bethard (2019)* | |
| Architectures/Methods | NER datasets, NER evaluation measures (*e.g.*, relaxed, strict F1), word level and character level architectures, CNN, LSTM, GRU, BRAT, Sequence-to-Sequence |
| Pros/Evaluations | Feature-inferring systems outperform feature-engineered systems; word+character-based systems perform better than word or character-based systems, |
| Cons/Evaluations | Need of domain experts (for constructing and maintaining knowledge resources for knowledge-based systems) |
| Challenges and future directions | Taking insights from past approaches into developing DL-based approaches (*e.g.*, usage of affix features) |
| **Sentiment analysis** | |
| *Zhang, Wang & Liu (2018)* | |
| Architectures/Methods | Word Embeddings, (Denoising) Autoencoder, CNN, RNN, LSTM, Attention Mechanism, Memory Networks, Recursive Neural Networks |
| Pros/Evaluations | Recurrent attention network (capture sentiment of complicated contexts); RecNN (capture aspects of compositionality of language) |
| Cons/Evaluations | Bag-of-Word representation (no word sequence, lack of encoding of semantics of words) |
| Challenges and future directions | Multimodal data (*e.g.*, acoustic data, visual sentiment detection), research on resource-poor languages |
| *Do et al. (2019)* | |
| Architectures/Methods | Word embeddings (word2vec, GloVe), CNN, RNN (including bidirectional), LSTM, GRU, RecNN, Hybrid Models, Attention, MemNet |
| Pros/Evaluations | Extraction of local patterns (CNN); LSTMs superior to CNNs (same performance with fewer training data) |
| Cons/Evaluations | Key phrases need to be of limited length (CNN); large training data required (CNN); unable to capture broader contextual information or sentence dependencies (CNN, due to fixed size of hidden layer); heavily dependent on parser (RecNN) |
| Challenges and future directions | Domain adaptation (most of work on customer reviews); multilingual application (especially variation in languages); technical challenges (training data sets, computational resources); linguistic complications (ambigious semantics, implicit aspects) |
| **Text summarization** | |
| *Shi et al. (2018)* | |
| Architectures/Methods | RNN, Sequence-to-Sequence, Pointer-Generator Networks, Encoder-Decoder Architecture, LSTM, CNN, Attention |
| Pros/Evaluations | Copy OOV words (Pointer-Generator Networks); computational complexity, *i.e.*, can be parallelized, fixed input sequence hence upper bound for computation, short path between input and output (CNN); handling of long documents (hierarchical attention) |
| Cons/Evaluations | Cannot handle salient information, OOV words, suffer from repetitions (abstractive text summarization); exposure bias, inconsistency of training and testing measurements (Sequence-to-Sequence models), exploding gradients (RNN, incl. LSTM), cannot be parallelized (RNN); lack in diversity (beam-search-based approaches) |
| Challenges and future directions | Large transformers, reinforcement learning, summary generation (*i.e.*, language generation), more diverse datasets, improved evaluation (*e.g.*, measures) |
| **Answer selection** | |
| *Lai, Bui & Li (2018)* | |
| Architectures/Methods | Word Embeddings (*e.g.*, GloVe); Siamese architecture: CNN, (bidirectional) LSTM; Attentive architecture: (bidirectional) LSTM; Compare-Aggregate Architecture: BiMPM (bidirectional LSTM) |
| Pros/Evaluations | Improved performance for Transfer Learning; treats task as prediction of list of answers (listwise approach); improved performance (listwise approach over pointwise and pairwise approaches) |

| **Table 5** (continued) | |
|---|---|
| **General language processing** | |
| Cons/Evaluations | Do not consider answer selection as a prediction task on list of candidate answers (pointwise and pairwise approaches); no interaction between input sentences (Siamese architecture); limited interaction between input sentences (Attension architecture); improved performance (Compare-Aggregate architecture over Siamese architecture, Attentive architecture) |
| Challenges and future directions | Potential for further Transfer Learning approaches; Connection to other tasks, such as open domain question answering and community question answering; Application to real-world applications, *e.g.* in the context of truth discovery methods |
| **Word embedding** | |
| *Zhang et al. (2016)* | |
| Architectures/Methods | Word2vec (CBOW, skip-gram), Latent Semantic Indexing, HAL, Neural Language Translation Model, Dual Embedding Space Model (DESM); query-log embeddings |
| Pros/Evaluations | Query language models outperform embedding-based document models; Sigmoid for transforming similarity values (instead of directly using cosine or Euclidean distance); increased expressiveness and improved performance for larger context window and embeddings dimensions |
| Cons/Evaluations | Poor ranking model for larger sets of document (DESM); computational costs of local embeddings during query time |
| Challenges and future directions | Consider word order for embeddings; selection of suitable datasets for computing embeddings |
| *Almeida & Xexéo (2019)* | |
| Architectures/Methods | Prediction-based methods: Hierarchical Softmax, Log-linear models, Negative sampling; Count-based methods: Pre-DL methods (*e.g.*, SVD, LSA, HAL, LR-MVL), GloVe |
| Pros/Evaluations | Good performance of word2vec (skip-gram with negative sampling); embeddings good for composition (n-gram-based, hierarchical softmax); improved performance over other count-based and word2vec models (GloVe) |
| Cons/Evaluations | High dimensionality in combination with discrete joint distributions (early n-gram language models) |
| Challenges and future directions | Adapting embeddings for task-specific work; combination of prediction-based and count-based; composition of word embeddings for higher-levels entities |
| **Financial forecasting** | |
| *Xing, Cambria & Welsch (2018)* | |
| Architectures/Methods | Representations (semantic, seniment, event representation), word embeddings; traditional ML methods (*e.g.*, SVM, SVR); Recurrent Neural Networks, deep believe networks (DBN), self-organizing fuzzy neural networks (SOFNN), adaptive neuro-fuzzy inference system (ANFIS), neural tensor network (NTN), CNN; evaluation measures (accuracy, closeness, trading simulation) |
| Pros/Evaluations | SOFNN are faster than ANFIS networks |
| Cons/Evaluations | SVM paying only attention to classification accuracy (in practice the loss needs to be taken in consideration); Social media text contains a lot of noise (corporate disclosures and financial reports are better-structured) |
| Challenges and future directions | Domain-specific resource building (including ontologies); online predictive models (*e.g.*, fuzzy rules for interpretability); comprehensive evaluation measures (unification of measures to allow comparisons) |

For machine translation the generic encoder-decoder architecture has been mentioned to be well suited (*Gatt & Krahmer, 2018*). In particular, Sequence-to-Sequence models are considered to provide good performance on this task. Another important point raised is in relation to suitability in practical settings, as the availability of data is often limited. Hence, the application of data-driven methods, especially deep learning, is limited in relation to rule-based approaches. The task of open domain dialogue systems is highlighted by *Santhanam & Shaikh (2019)*. Here Sequence-to-Sequence models are a popular choice. Making use of the attention mechanism enhanced the performance.

As key challenges they identified the encoding of contextual information and the incorporation of personality into the text generation. They mention the risk of generating dull and generic responses. As future directions they foresee taking insights from cognition into account and ways to encode emotional content. In Chen et al. (2017) a difference is made between task-oriented and non-task-oriented dialogue systems. For task-oriented systems, often external knowledge bases are queried. Such procedure has two main downsides. First, there is no information about the uncertainty, and second, such approaches are not differential. As a response, techniques like key-value memory networks might address these issues. It has been noted that trivial responses is one of the key challenges, which can be tackled *via* improved objective functions. Another noteworthy approach being mentioned are hybrid methods, combining the strength of multiple approaches. Challenges for future research include shortening the warm-up phase required to adapt to novel domains. A deeper understanding of learning should also facilitate improved dialogue systems. Finally, protecting the privacy of users should be a relevant aspect of future systems.

For Named Entity Recognition, Li et al. (2020) point out the importance of the input representation. When using language model embeddings (*e.g.*, Transformer), the performance of NER systems improved. Also, leveraging external knowledge improves performance. Yadav & Bethard (2019) highlights that feature-inferring systems outperform systems that rely on feature engineering. Using a combination of word embedding and character embedding performs better than each of them individually. What exact architecture to choose depends on the data and domain. If large datasets are available, learning RNNs from scratch is an viable option. For cases where data is scarce, transfer learning is a promising approach, with fine tuning on the target domain is often an effective way. Crucially, the manual annotations are required to be of high accuracy and consistency. Taking insights from traditional approaches and lifting them into contemporary DL-based approaches may improve the current state of the art. Fine-grained NER will be of more importance in the future, as well as jointly solving NER and Entity Linking as one combined task.

For sentiment analysis, Zhang, Wang & Liu (2018) highlights the downsides of following a Bag-of-Words representation, including their inability to represent sequence information and the lack of semantic encoding. In Do et al. (2019) the usage and importance of word embeddings for sentiment detection is being covered. They found that CNNs require more training data in contrast to recursive neural networks. Furthermore, CNNs lack the ability to capture broader contextual information. Recursive Neural Networks, at the other hand, are well suited to capture the complex relationships found in text, which are required to capture opinions, emotions and sentiment. Challenges include changes in the domain and languages, where under-resourced languages with different structure are less well researched. As for future trends, the surveys identified multimodal data, including acoustic and visual data, and a stronger focus on resource-poor languages.

For the task of text summarization in Shi et al. (2018) most of the presented approaches are also known to be used for other tasks, for example sentiment detection.

A prominent exception is the development of Pointer-Generator Networks, where the input is directly copied to the output, which is particularity interesting for the challenge of Out-of-Vocabulary words. This challenge is relevant for all abstractive text summarization approaches, which also suffer from repetitions and insufficient handling of salient information. In terms of computation, recurrent neural networks cannot easily be parallelized. Here, CNNs have the advantage of a fixed size input and a short path between the input and output. As challenges and future work for the task of text summarization the survey mentions large transformers, reinforcement learning schemes, the generation of summaries *via* language generation, the adoption of more diverse datasets and improvements in evaluation, for example *via* improved evaluation measures.

The task of answer selection is covered by *Lai, Bui & Li (2018)*. First, three learning approaches are compared. It is found that the listwise approach outperformed the pointwise, as well as the pairwise approach. As explanation it is mentioned that the listwise approach considered the task as a list of candidates, instead of considering all candidates in isolation. Next, multiple architectures are presented. Here the Compare-aggregate architecture has been mentioned to provide improved performance over the Siamese and Attentive architectures. This can be explained by the ability of this architecture to capture interactions between the input sentences. As future research, synergies between closely related tasks have been mentioned, for example open domain question answering and community question answering. Furthermore, it is expected that answer selection will find its way into real-world approaches, for example applications related to truth discovery.

Word embedding is studied as a standalone task in *Almeida & Xexéo (2019)* and in the context of information retrieval in *Zhang et al. (2016)*. Both surveys shed light on a number of approaches that precede word embeddings, which share the motivation of exploiting the distributional semantics. A difference is being made between prediction-based and count-based methods, with word2vec being the most prominent example for the first and GloVe the most popular for the second. It has been mentioned that GloVe is able to outperform many other count-based approaches, as well as prediction-based methods. In the context of information retrieval, a noteworthy approach is the computation of local embeddings, based on the query. While there is a performance improvement associated with such approach, the computational complexity is high and the embeddings need to be computed at query time. As future work, word embeddings are expected to get more rich and integrate more contextual information, for example the word order.

In the domain of financial forecasting *Xing, Cambria & Welsch (2018)* the temporal aspect is being stressed. Hence, approaches including recurrent neural networks have been applied in this domain. Interestingly, the usage of self-organizing networks has been researched, also for the task of event embeddings. Self-organizing fuzzy neural network have been found to be faster than similar approaches like adaptive neuro-fuzzy inference system (ANFIS). Similar to other surveys one of the challenges are domain-specific datasets and resources, including knowledge bases and ontologies. The need for improved evaluation measures is another common topic for future research. Furthermore, the ability to interpret the model is an added benefit in the financial forecasting domain and will attract more research in the future.

**Table 6** List of published reviews in deep learning in the category *medical informatics*.

| Medical informatics | Publications | Number of references | Citations (until August 2020) | Preprints |
|---|---|---|---|---|
| Health informatics | *Ravì et al. (2016)* | 145 | 700 | No |
| Medical image analysis | *Litjens et al. (2017)* | 439 | 3696 | No |
| | *Shen, Wu & Suk (2017)* | 117 | 1,200 | No |
| | *Xing et al. (2017)* | 207 | 94 | No |
| | *Haskins, Kruger & Yan (2020)* | 122 | 49 | No |
| Medical imaging | *Lundervold & Lundervold (2019)* | 359 | 199 | No |
| Health-record analysis | *Shickel et al. (2017)* | 63 | 366 | No |
| Cancer detection and diagnosis | *Hu et al. (2018)* | 144 | 108 | No |
| Bioinformatics | *Lan et al. (2018)* | 127 | 85 | No |
| Radiotherapy | *Meyer et al. (2018)* | 234 | 84 | No |
| Pharmacogenomics | *Kalinin et al. (2018)* | 128 | 44 | No |
| Radiology | *Mazurowski et al. (2019)* | 125 | 97 | No |
| Sum | – | 2,210 | 6,722 | – |

To conclude the majority of surveys in this section highlight the importance and contribution of word embeddings to wide variety of tasks. Currently, there does not appear to exist a consensus, if CNN-based or LSTM-based approaches are better suited for the majority of NLP problems. Dedicated architectures, such as attention-based, encoder-decoders, and especially the transformer architecture gained a lot of popularity recently. Another common emphasis of many surveys is the topic of evaluation measures. Currently used measures were considered to only partly reflect the true performance in relation to the target tasks, with novel measures being considered one of the future research topics. Similarly, there is an emphasis on need of more datasets, especially for dedicated domains and resource-poor languages.

## Deep learning reviews in medical informatics

This sub-section deals with the deep learning reviews in the medical field. It is divided into nine sub-categories and the number of references, and citations (according to Google Scholar and status as of mid-August 2020) for each of these categories is given in Table 6:

- health informatics,
- medical image analysis,
- medical imaging,
- health-record analysis,
- cancer detection and diagnosis,
- bioinformatics,

- radiotherapy,
- pharmacogenomics,
- and radiology.

In the field of medical informatics, processing and gaining information from all sorts of medical data, *e.g.* for treatment decisions or predictions, is the main task. A classic example is to process medical image acquisitions from patients to automatically detect a pathology, like a tumour, or the automatic processing and analysis of health-records. These examples show that the medical informatics field and its data link to the fields of computer vision and natural language processing and its algorithms. In fact, most algorithmic concepts are overtaken from the field of computer vision and have been adapted for medical imaging data, although there are also examples in the other direction, like the U-Net, which was developed for medical image segmentation and has since inspired countless architectures for other applications. Health data, however, has its very own characteristics. For example, computed tomography scans always produce grey value images in a specific range. Similar to images in computer vision, every patient and every possible pathology is at least slightly different, which makes it difficult to generalize for algorithms in medical informatics. For objective evaluations in medical informatics, common scores like the Dice Similarity Coefficient or the Hausdorff distance are used. In contrast to computer vision, however, the collection of data is much more difficult, because of privacy concerns. A trend over the years in medical informatics, are challenges for specific medical tasks, where algorithms compete against each other in performing a certain task, *e.g.* on the Grand Challenge platform (https://grand-challenge.org/), like our recent challenge towards the automatization of cranial implant design in cranioplasty (https://autoimplant.grand-challenge.org/ and the successor https://autoimplant2021.grand-challenge.org/) (*Li & Egger, 2020*; *Li et al. (2021)*). However, designing algorithms that work reliably in clinical practice for the targeted medical scenario is challenging, and examples of successful clinical translation are scarce. This is amplified by the dramatic consequences a misclassification in the medical domain can have on the patients'lives.

### *Health informatics*

Health or healthcare informatics is the field that applies informatics to the medical domain with the aim to improve health care. *Ravì et al. (2016)* provide a broad overview about the application of deep learning for health informatics and cover the areas medical informatics, sensing, bioinformatics, imaging and public health, also listing the application, input data and underlying base method. They start their review with a comprehensive background about deep learning architectures like deep neural networks, deep autoencoder and deep belief networks. They conclude their work by discussing the limitations and challenges of deep learning in healthcare.

### *Medical image analysis*

Medical image analysis is the task of automatically or semi-automatically extracting information from (patient-specific) medical images. For example, this could be an automatic determination of the tumour volume from a patient's magnetic resonance imaging scan with the aim to choose a therapy strategy. The publication of *Litjens et al. (2017)* gives an overview of main deep learning techniques in regard to medical image analysis. It presents over 300 works in that area and reviews the application of deep learning in topics like organ or disease detection, image classification, segmentation, registration, and further tasks. Furthermore, compact outlines of studies are presented by the application areas: abdominal, breast, cardiac, digital pathology, musculoskeletal, neuro, pulmonary and retinal. Finally, they summarize recent works at that time and discuss remaining challenges and areas for future research work.

*Shen, Wu & Suk (2017)* present the basics of deep learning-based approaches and analyse the reported results in fields like medical image registration, tissue segmentation, anatomical and cell structure detection, computer-aided disease diagnosis and prognosis. They conclude their work by raising remaining research questions and proposing future research directions for additional improvements in the field of medical image analysis.

*Xing et al. (2017)* give an overview about microscopy image analysis. They start with a brief introduction of common deep neural networks and provide a compact outline of recent deep learning successes in numerous applications, including detection, classification and segmentation in the area of microscopy image analysis. They present the background of (fully) convolutional neural networks, deep belief networks, recurrent neural networks and stacked autoencoders, and connect their basic principles and modelling to certain application in different microscopy images. They conclude by discussing remaining research challenges and outline possible directions for future research in deep learning-based microscopy image analysis.

The review of *Haskins, Kruger & Yan (2020)* present the progress of deep learning-based approaches for the field of medical image registration by outlining research challenges, but also significant advancement in the last years. They divide their article in three main categories, namely unsupervised transformation estimation, supervised transformation estimation and deep iterative registration. Each main category is then divided again in sub-categories. They conclude with surveying highlights of future research directions in this field.

### Medical imaging

Medical imaging deals with the acquisition of human and animal images from cellular to body scale. Common examples include computed tomography (CT) and magnetic resonance imaging (MRI), which allow to acquire images at vascular and organ level for diagnostic and therapeutic reasons (*Pepe et al., 2019*; *Gsaxner et al., 2019a*; *Gsaxner et al., 2018*). In this context, *Lundervold & Lundervold (2019)* provide an analysis of deep learning-based methods in medical imaging with a focus on MRI acquisitions. The goal of their review is threefold: First, they provide a compact background introduction of deep learning and its main contributions. Then, they outline how deep learning-based methods have been used for the whole MRI workflow from image acquisition to diagnosis, and give a starting point for researchers in the area of deep learning-based medical imaging. Finally, they point to open-source code repositories, educational resources, and further related data sources that are of interest for medical imaging.

### Health record analysis

Health record analysis explores the digital information stored in electronic health records. Originally intended to archive patient information and performing administrative tasks in healthcare, such as billing, researchers starred to utilize these records also for numerous other applications in clinical informatics. Shickel et al. (2017) survey the applications of deep learning for the analysis of health-record data. They report several deep learning-based methods and techniques that have been used for different clinical applications, such as information extraction, outcome prediction, representation learning, but also de-identification and phenotyping. In their review, they identify a number of limitations for the current research regarding the data heterogeneity, model interpretability, but also missing universal benchmarks. They conclude their review by outlining the field and proposing directions for upcoming deep health-record analysis research.

### Cancer detection and diagnosis

As previously mentioned, medical image acquisitions, like MRI or CT, can be used for the diagnosis of pathologies, such as tumours, and internal injuries like bone fractures. However, the manual processing of these images can be cumbersome and time-consuming, even for experts (Hahn et al., 2020). This led to the investigation of automatic approaches. Hu et al. (2018) review deep learning-based applications for cancer detection and diagnosis. In their survey, they start with a background on deep learning and common architectures applied to the detection and diagnosis of cancer. They focus on four common architectures in deep learning, such as (fully) convolutional neural networks, deep belief networks, but also auto-encoders. Additionally, they present a review on studies that exploit deep learning for cancer detection and diagnosis, grouped by cancer type. Finally, they provide a summary and personal comments to the reviewed works and suggest future research directions.

### Bioinformatics

Bioinformatics is an interdisciplinary field developing approaches and software tools for the understanding of biological data with a strong focus on large and complex data sets. Lan et al. (2018) concentrate on the review of research contributions using deep learning and data mining approaches for the analysis of domain-specific knowledge in bioinformatics. Their review article provides a summary of several data mining methods that have been utilized for pre-processing, classification and clustering along with different optimized neural network architectures and deep learning methods. Furthermore, they present the advantages and disadvantages of such methods in practical applications, discuss, and compare them in terms of their industrial usage.

### Radiotherapy

Radiotherapy can be used, for example, to treat cancer patients. However, planning and delivering radiotherapy treatment is a complicated procedure, which artificial intelligence tries to automate and therefore facilitate. In their review, *Meyer et al. (2018)* introduce the basics of deep learning and its position in the overall context of machine learning. They introduce popular neural architectures with a particular emphasis on classic convolutional neural networks. Subsequently, they give an overview of contributions on deep learning-based methods that can be utilized for radiotherapy. Thereby, they classify them into seven different categories in regards to the overall patient workflow.

### Pharmacogenomics

Pharmacogenomics (pharmaco- + genomics) is an interdisciplinary field between pharmacology and genomics that studies the role of the genome in drug response. In their review, *Kalinin et al. (2018)* introduce recent works and future applications for deep learning in the area of pharmacogenomics. This includes the exploration of new regulatory variants situated in noncoding genomic regions and their role not only in pharmacoepigenomics, but also in patient stratification from clinical records. They aim at the application of deep learning for the prediction of patient-specific drug responses to optimize the process of drug selection and dosing. This process is automated by applying data-driven deep learning algorithms on large and complex data collections, which can provide different sets of information, ranging from the micro- to the macroscopic level, such as from molecular to epidemiological and from clinical to demographic domains.

### Radiology

Radiology is the medical field that deals with the extraction of useful information from images, like CT or MRI patient acquisitions, for diagnosis and treatment of humans and animals. *Mazurowski et al. (2019)* give an overview of the common fields of radiology and present options and chances for deep learning-based approaches there. They also present fundamental deep learning concepts, such as convolutional neural networks before presenting research contributions focused on deep learning and its application to radiology. The reviewed works are grouped by application task. They conclude their work discussing opportunities and challenges for the inclusion of deep learning-based approaches into the clinical practice.

### Going deeper: architectures, evaluations, pros, cons, challenges and future directions in medical informatics

Table 7 gives more details about the presented methods, pros, cons, evaluations and challenges and future directions of the surveys in the category medical informatics. *Ravì et al. (2016)* presented a review on using deep learning in health informatics, where the authors specifically focused on applications in translational bioinformatics (*e.g.*, genomics, pharmacogenomics, epigenomics), medical imaging (CT, X-Ray, MRI, fMRI, positron emission tomography (PET)), pervasive sensing (signal abnormality detection (*e.g.*, ECG, EEG), assistive device for physically impaired patient and energy and activity monitoring for obesity prevention), medical informatics (aggregated medical data such

**Table 7 Architectures, pros, cons, evaluations, challenges and future directions in deep learning in the category *medical informatics*.**

| | |
|---|---|
| **Health informatics** | |
| *Ravì et al. (2016)* | |
| Architectures/Methods | DNN, DAE, DBN, DBM, RNN, CNN, LSTM, RNN |
| Pros/Evaluations | Widely adopted in medical domain; unsupervised learning (DAE); can deal with sequential data (RNN); many variants available for different tasks (DAE, CNN). |
| Cons/Evaluations | Require large quantity of labelled data (CNN); computationally expensive. |
| Challenges and future directions | A deep learning model remains to be a black box and uninterpretable to humans, which is undesirable especially in medical domains; for rare diseases, it remains difficult and sometimes impossible to collect enough data for training deep learning models. It is costly to collect and label medical data; the parameter tuning (*e.g*., hyper parameters) of deep learning models remains to be experience and task dependent; small deviations added to the input could alter the output of deep learning models completely, making deep learning models unreliable and susceptible to manipulations; future directions: interactive deep learning (human-machine collaboration); multi-modality/multi-source health informatics data fusion. |
| **Medical image analysis** | |
| *Litjens et al. (2017)* | |
| Architectures/Methods | MLP: Multi-layer perceptron CNN, RNN, Auto-encoders (AEs) and Stacked AE (SAEs), RBM: Restricted Boltzmann Machine, DBN, VAE: Variational Auto-encoder, GAN: Generative Adversarial Network |
| Pros/Evaluations | End to end training (CNN); freely available pre-trained deep learning models |
| Cons/Evaluations | Hyper-parameter tuning remains to be empirical; medical image annotation is subjective and susceptible to inter-annotator variability and uncertainty |
| Challenges and future directions | Challenges: medical image annotation remains to be time-consuming and expensive; future directions: task-specific pre-processing and data augmentation techniques; incorporating prior knowledge of the specific domain into the training of deep learning models, radiological reports (by radiologists) could be utilized for the efficient annotation of medical imagesl learning from limited data and sparse annotation; leverage non-expert annotation through crowd-sourcing, unsupervised learning (*e.g*., VAE) using unlabeled data in medical image analysis; interpretable deep learning in medical image analysis |
| *Shen, Wu & Suk (2017)* | |
| Architectures/Methods | SAE: Stacked Auto-encoder, DBN: Deep Belief Network, DBM: Deep Boltzmann Machine, CNN: Convolutional neural network |
| Pros/Evaluations | Compared to traditional machine learning algorithms, where image features are handcrafted and domain knowledge is required, a deep learning network can learn features through labelled data by itself, allowing experts outside of medial domain to use the algorithms to tackle medical problems easily. |
| Cons/Evaluations | Overfitting due to limited training samples. |
| Challenges and future directions | Image features learnt by deep learning are difficult to understand and interpret; future directions: building a medical equivalent of ImageNet to facilitate the training of deep learning networks; incorporating domain-specific knowledge in the design and/or training of deep learning networks; develop a universal algorithmic technique that is compatible with various image scanning protocols and modalities. |
| *Xing et al. (2017)* | |
| Architectures/Methods | CNN, FCN: fully convolutional network, RNN: recurrent neural network, SAE: stacked autoencoder |
| Pros/Evaluations | Can be trained in an unsupervised manner (SAE); the input size of FCN is not fixed and can be arbitrary; the training of a CNN can be easily parallelized. |
| Cons/Evaluations | Obtaining large number of annotated microscopy images is expensive; NN requires a fixed input size |
| Challenges and future directions | It is difficult to interpret the results and behavior of deep neural networks, which could be an issue in medical domains; processing high volumes of medical data requires computational (algorithmic, hardware) acceleration; future directions: develop deep learning-based methods for WSI (whole slide imaging) image analysis; use a patch-based strategy for high dimensional microscopy image analysis to reduce computational expenses; fusing different types of patients' data (images, diagnostic reports) as the input of deep neural networks is promising for the analysis of microscopy images; design task-specific deep learning architecture based on domain knowledge by, *e.g*., combining learnt and handcrafted features together; develop unsupervised or semi-supervised learning algorithms in order to make best use of the large quantities of unlabeled data available. |

(Continued)

**Table 7** (continued)

| | |
|---|---|
| **Health informatics** | |
| *Haskins, Kruger & Yan (2020)* | |
| Architectures/Methods | Deep iterative registration, supervised (including partially/weakly supervised) transformation estimation, unsupervised transformation estimation, Reinforcement learning |
| Pros/Evaluations | Deep reinforcement learning based registration methods are intuitive to understand; supervised transformation-based methods are fast and can work in real-time; weakly/partially supervised transformation-based methods rely less on the ground truth, while maintaining high registration accuracy; unsupervised image registration are independent from the need of large number of labeled data. |
| Cons/Evaluations | Iterative registration techniques are difficult to be used in real-time registration; the labeling of the ground truth for supervised transformation estimation requires expertise and thus the ground truth is difficult to obtain; unsupervised registration is in generally limited to the unimodal case. |
| Challenges and future directions | If synthetic ground truth was used for supervised transformation estimation, the synthetic data has to be closely similar to real data; future directions: deep reinforcement leaning and generative adversarial network (GAN) based registration will be gaining more attention in the near future; extend the unsupervised transformation methods to multimodal registration. |
| **Medical imaging** | |
| *Lundervold & Lundervold (2019)* | |
| Architectures/Methods | ANNs: Artificial neural networks, CNNs: Convolutional neural networks |
| Pros/Evaluations | Deep learning has been successfully used in the whole MRI processing chain, including acquisition, reconstruction, registration, segmentation, *etc*. |
| Cons/Evaluations | processing high-dimensional medical images using CNN requires ample computational powers; edical data used for research are usually of high quality while real-world clinical data tend to be messy, the performance of deep learning models is expected to drop in a production environment. |
| Challenges and future directions | the privacy and protection of medical data makes it difficult for deep learning models to be trained and openly deployed; the problem of incorporating deep learning models into already well-established clinical practice has not been fully addressed; future directions: more focus should be put on the reproducibility of machine learning for medical imaging researches, *e.g*., through data and code sharing; the current established peer-review system cannot keep up with the fast development of machine learning in medical imaging researches. To tackle the issue, posting papers to arXiv for fast publication and open-sourcing codes and dataset is a direction worthy of exploring; biomedical challenges/competitions held publicly should be encouraged; ederated learning and differential privacy |
| **Health-record analysis** | |
| *Shickel et al. (2017)* | |
| Architectures/Methods | MLP: Multilayer Perceptron, CNN: Convolutional Neural Networks, RNN: Recurrent Neural Networks, AE: Autoencoders, RBM: Restricted Boltzmann Machine |
| Pros/Evaluations | LSMT, RNNs and their variants are able to process sequential data, which is desirable for tasks involving EHR dataset processing. |
| Cons/Evaluations | Deep learning models lack transparency and are difficult to be interpreted, which is undesirable in clinical domain. |
| Challenges and future directions | the data from electronic health record (EHR) are heterogeneous; extracting information from Clinical notes are challenging, due to the various writing styles of the authors/clinicians; current deep EHR researches lack reproducibility and no universal benchmark datasets are available for inter-institution; future directions: including robust mechanism to handle the EHR irregularity is worthy of exploration for future deep learning researches; clinical notes contain essential information about the patients, which are however not made good use of. Natural language processing (NLP) techniques (*e.g*., LSTM, RNN) should be focused more on the clinical notes for future researches; various types of patients' data should be considered as a whole and converted to a unified representation; patient deidentification using deep learning is another area of research in the future; future deep learning research should be focused on increasing the model interpretability, which is of utmost importance for clinical applications including EHR analysis. |
| **Cancer detection and diagnosis** | |
| *Hu et al. (2018)* | |
| Architectures/Methods | FCNs: fully convolutional networks, AEs: Auto-Encoders, DBNs: Deep belief networks |

**Table 7 (continued)**

| **Health informatics** | |
|---|---|
| Pros/Evaluations | Deep learning-based approaches automate feature engineering, compared to traditional machine learning techniques where image features are usually hand-crafted based on domain knowledge. |
| Cons/Evaluations | large quantities of data are needed to train deep learning models. For rare cancers, it is difficult to collect enough images for training; the training of deep learning models are time and computation-consuming; deep learning models are *black boxes* and hard to be interpreted. |
| Challenges and future directions | Even though large quantities of image data are accumulated in the PACS in hospitals, more efforts are needed to make these datasets publicly available for research purposes; there lacks a common large cancer dataset, based on which various deep learning related cancer image analysis studies can be compared and evaluated directly; the datasets are usually imbalanced, with the negative (non-cancer) cases far exceeding positive cases; the trained deep learning models on cancer image analysis cannot generalize well across institutions; future directions: increase the detection and diagnosis accuracy when it comes to blurry and low sign-to-noise ratio images; multi-modality information could be utilized for cancer detection. |
| **Bioinformatics** | |
| *Lan et al. (2018)* | |
| Architectures/Methods | KNN: K nearest neighbor, Naïve Bayes, decision tree, SVM: support vector machine, (Deep) neural networks, clustering, CNN: convolutional neural networks, SAEs: Stacked auto-encoders, DBN: Deep belief network, RNN: Recurrent neural network |
| Pros/Evaluations | Deep learning is able to learn knowledge from massive amount of data automatically. |
| Cons/Evaluations | Deep learning requires large datasets for training and is dependent on high-end hardware; compared to traditional machine learning algorithms, deep learning models lack interpretability. |
| Challenges and future directions | Data imbalance is prevalent in medical domain; future directions: aggregation of different machine learning algorithms and fusion of data from different modalities; development of semi-supervised and reinforcement learning algorithms. |
| **Radiotherapy** | |
| *Meyer et al. (2018)* | |
| Architectures/Methods | DNN: Deep neural network, RNN: Recurrent neural network, AE: Auto-encoder, CNN: Convolutional neural network |
| Pros/Evaluations | the availability of large amount of training data and the increasing power of graphics processing unit (GPU) contributed to the success of deep learning in many field, including the medical domain. |
| Cons/Evaluations | Deep learning theories are mostly empirically and experimentally obtained, which induces criticism; small noise, imperceptible to humans, added to the input could alter the output completely, which makes the reliability of deep learning models questioned. |
| Challenges and future directions | building coherent, large and balanced medical datasets that represent real-world scenarios for deep learning algorithms remains to be challenging; the interpretation of deep learning models is difficult |
| **Pharmacogenomics** | |
| *Kalinin et al. (2018)* | |
| Architectures/Methods | Deep learning in general |
| Pros/Evaluations | Deep learning can take the advantage of ever-increasing amount of medical data available. |
| Cons/Evaluations | Current deep learning algorithms are dependent on large datasets; when the dataset is small, deep learning models are susceptible to overfitting; deep learning models lack interpretability. |
| Challenges and future directions | Future directions: In pharmacogenomics, deep learning will revolutionize from 'prediction' to 'prescription'. |
| **Radiology** | |
| *Mazurowski et al. (2019)* | |
| Architectures/Methods | ANNs: Artificial neural networks, CNN: convolutional neural network |
| Pros/Evaluations | Deep learning-based methods have been verified effective in many radiological tasks such as medical image classification, segmentation, detection, reconstruction and registration and become the state of the art. |
| Cons/Evaluations | Within the large realm of radiology, current deep learning models have reported to outperform human experts in only a minority of radiological tasks; introducing deep learning models in clinical practice will induce legal and ethical issues. |

(Continued)

| Table 7 (continued) | |
|---|---|
| **Health informatics** | |
| Challenges and future directions | Compared to natural images, medical datasets are smaller and often imbalanced (in case of rare diseases), which can lead to suboptimal training of deep learning models; proper clinical validation of a deep learning model is essential for its clinical usability, which is however often overlooked in medical deep learning researches; future directions: more studies are needed in the future to optimally incorporate deep learning models in existent radiology workflow, so that the current radiological practice can be improved. |

as Electronic health records (EHR)) and public health (*e.g.*, pandemic disease control and surveillance, social public behavior modelling). In each of the categories covered, the authors reviewed the representative publications for each specific task such as using deep learning for RNA binding protein prediction, myocardium identification, calorie measurement using CNN and EEG anomaly detection based on a deep belief network (DBF). Where applicable, the limitations of the reviewed publications and methods for the respective tasks are briefly analyzed. The survey is concluded by presenting the limitations and challenges of using deep learning in healthcare and future perspectives. For example, iterative deep learning that enabled the collaboration between human experts and the machine would benefit the health informatics field, which would at the same time overcome the shortcomings of deep learning models such as lack of reliability and interpretability.

Four surveys were included in the 'Medical image analysis' category, where two of them (*Litjens et al., 2017*; *Shen, Wu & Suk, 2017*) specifically used 'deep learning in medical image analysis' as the titles and the survey from *Xing et al. (2017)* focused on the analysis of microscopy images, and *Haskins, Kruger & Yan (2020)* focused on medical image registration. In *Litjens et al. (2017)*, the reviewed publications were grouped/categorized by two means: First, by different medical tasks such as classification (*e.g.*, image, lesion), detection (*e.g.*, organ, landmark, lesion), segmentation (*e.g.*, organ, lesion), registration and other medical tasks (image retrieval, generation, enhancement, medical reports). Second, by anatomical areas such as brain, eye, chest, breast, cardiac, *etc*. Different medical tasks and anatomical areas constitute the subcategories of the survey. For each subcategory, the authors selected and reviewed the corresponding publications, providing information, if available, about the deep learning architecture used, image modalities and the limitations of the respective studies. The authors concluded the survey by summarizing the challenges and outlook of deep learning in medical image analysis and the desirable properties deep learning models should have in medical tasks. *Shen, Wu & Suk (2017)* used a similar task-based grouping strategy to that of *Litjens et al. (2017)*, where the authors reviewed deep learning methods that were applied to structure detection (*e.g.*, organ, body part, cell), image segmentation, computer-aided detection (CADe) and computer-aided diagnosis (CADx). *Xing et al. (2017)* also grouped the reviewed publications by applications in microscopy images, such as detection (*e.g.*, mitosis detection in histology images, cell and nuclei detection), segmentation (*e.g.*, nucleus, cell, neural membrane segmentation) and classification (*e.g.*, human epithelial-2 cell images classification). Due to that microscopy images are distinct from other medical image modalities such as CT and MRI, tailored pre/post-processing methods and sometime network architectures were involved. *Haskins, Kruger & Yan (2020)* divided the reviewed publications based on the registration techniques (iterative registration, supervised and unsupervised transformation estimation) used. High expense of obtaining the training data and the labels have been identified as the most prevalent challenges of using deep learning for medical image analysis.

The survey from *Lundervold & Lundervold (2019)* specifically focused on the application of deep learning in magnetic resonance imaging (MRI), from the acquisition and reconstruction of raw MRI data to the registration, restoration, synthesis, retrieval and segmentation of MRI images. Limited data availability, difficulty of interpreting deep learning models and integrating deep learning models into existent clinical workflows has been identified by the authors as the main challenges of deep learning in MR image analysis.

*Shickel et al. (2017)* reviewed the recent development of electronic heath record (EHR) analysis using deep learning, which is a relatively new and less touched area compared to image-based deep learning applications. Due to the distinct characteristics of EHR, LSTM, RNNs and their variants (mostly NLP techniques) are among the most widely used deep learning architectures in this field, compared to medical image analysis where CNNs were prevalently adopted. The survey starts by reviewing techniques regarding the representation learning of EHR, which is fundamental for all deep learning-based EHR applications. Then, the survey goes into application-specific publications such as outcome prediction and the deidentification of clinical data. The authors concluded the survey by pointing out aspects specific to EHR data, such as data heterogeneity and irregular measures worthy of future investigation. Challenges in this field are, for example, the EHR data even present greater heterogeneity compared to other well-structured and standardized medical data such as CT and MRI, making the learning of their representation using deep learning more challenging.

The survey by *Hu et al. (2018)* is focused on the detection and diagnosis of cancers based on medical images. The reviewed publications in this survey are grouped according to the areas where the cancer occurs, such as breast cancer, lung cancer, skin cancer, prostate cancer, brain cancer, colonial cancer and other types of cancers. For each type of cancer, the deep learning-based methods used, including the architectures like FCN and AE were discussed. The authors pointed out that, even though large amount of imaging data of cancers has been accumulated over the years, it is still a challenge to make these valuable datasets publicly available for the development of deep learning models.

*Lan et al. (2018)* reviewed data mining and deep learning techniques that can be used in bioinformatics. Unlike previous surveys, which are built upon publications about specific medical applications, the survey from *Lan et al. (2018)* focused more on the introduction of the data mining and deep learning methodologies, such as traditional machine learning algorithms (*e.g.*, nearest neighbor, decision tree, clustering) and various deep learning structures and compositions (CNN, SAE, DBN, RNN). Another large part of the overall content in the survey is about a comparison between data mining and deep learning. Class imbalance is common in many medical datasets, which is pointed out to be one of the main challenges.

*Meyer et al. (2018)* focused on a specific application—radiotherapy used for cancer treatment. They have identified the common steps in the radiology workflow including patient consultation, image acquisition, target structure segmentation, treatment planning and delivery and follow-up. The deep learning methods that have been applied in these steps were reviewed and discussed. These include, for example, that some researchers have used CNNs to reduce the artifacts on CT images acquired from limited angles and some researchers used a CNN to reconstruct MRI images at 7T fidelity level given an image taken under 3T. Building coherent, large and balanced medical datasets is identified by the authors as the main challenges, which should be paid more attention to in the future.

*Kalinin et al. (2018)* restrict their survey to pharmacogenomics. The publications in the survey were grouped by area of applications such as drug discovery, patient stratification, toxicology. The authors stated that when the dataset used for training deep learning models is small, deep learning is susceptible to overfitting. Lack of interpretability of the deep learning models is identified as another limitation of deep learning based methods in pharmacogenomics. The authors envisaged that future deep learning in pharmacogenomics will revolutionize from 'prediction' to 'prescription'.

*Mazurowski et al. (2019)* gave an overview of deep learning in radiology, where the imaging modality is restricted to MRI. Like the survey from *Lundervold & Lundervold (2019)*, *Mazurowski et al. (2019)* grouped the reviewed papers in terms of their applications including classification, segmentation, detection, registration, reconstruction, *etc*. For classification of radiological data, the most adopted classifier is composed of several convolutional layers, followed by a fully connected layer. For segmentation fully convolutional neural network (FCNN) is a popular choice. The survey concludes by stating challenges and outlooks of deep learning in radiology. For example, the authors pointed out that more studies are needed in the future to seamlessly integrate deep learning models in existent radiology workflows, so that the current radiology practice can be improved. Another challenge is that medical image datasets are usually much smaller than natural image datasets and training on small datasets could lead to overtraining. Class imbalance (healthy cases far outnumber pathological cases) in medical image datasets is another factor that could lead to suboptimal training of deep learning models.

To conclude, most of the surveys reviewed in this category followed a similar structure: a general introduction of machine/deep learning techniques followed by the discussion of the publications categorized using different strategies, among which task- or application-based categorization is the most widely used. The surveys usually conclude by stating challenges and future perspective of deep learning in the medical field from a high level. It should be noted that, the topics in subcategories 'health informatics', 'medical image analysis', 'medical imaging' and 'radiology', are highly overlapping and too broad, such that the respective surveys are unable to provide ample low-level details regarding the deep learning models. Some surveys such as the work of *Shickel et al. (2017)*, focused on a very specific and narrow topic (Electronic Health-record analysis), so that they can provide some low-level insights of deep learning usage for the task.

**Table 8** List of published reviews in deep learning in the category additional works.

| Additional works | Publications | Number of references | Citations (until August 2020) | Preprints |
|---|---|---|---|---|
| Big data | *Zhang et al. (2018)* | 102 | 383 | No |
| | *Mohammadi et al. (2018)* | 229 | 340 | No |
| | *Emmert-Streib et al. (2020)* | 154 | 3 | No |
| Reinforcement learning | *Mousavi, Schukat & Howley (2016)* | 45 | 57 | No |
| | *Li (2017)* | 604 | 463 | Yes |
| | *Arulkumaran et al. (2017)* | 100 | 438 | No |
| Mobile and wireless networking | *Zhang, Patras & Haddadi (2019)* | 574 | 372 | No |
| Mobile multimedia | *Ota et al. (2017)* | 111 | 70 | No |
| Multimodal learning | *Ramachandram & Taylor (2017)* | 103 | 114 | No |
| Remote sensing | *Ball, Anderson & Chan (2017)* | 419 | 184 | No |
| Graphs | *Zhang, Cui & Zhu (2020)* | 170 | 142 | No |
| Anomaly detection | *Kwon et al. (2019)* | 45 | 196 | No |
| Recommender systems | *Zhang et al. (2019)* | 210 | 583 | No |
| Agriculture | *Kamilaris & Prenafeta-Boldú (2018)* | 72 | 612 | No |
| Multiple areas | *Pouyanfar et al. (2018)* | 181 | 188 | No |
| | *Dargan et al. (2019)* | 87 | 22 | No |
| | *Raghu & Schmidt (2020)* | 275 | 4 | Yes |
| Sum | – | 3,481 | 4,171 | – |

## Additional deep learning reviews

This sub-section deals with additional deep learning reviews, which were not covered within the other categories. It is divided into eleven sub-categories and the number of references, and citations (according to Google Scholar and status as of mid-August 2020) for each of these categories is given in Table 8:

- big data,
- reinforcement learning,
- mobile and wireless networking,
- mobile multimedia,
- multimodal learning,
- remote sensing,
- graphs,
- anomaly detection,
- recommender systems,
- agriculture,
- and multiple areas

These additional works did not fit precisely into one of the three previous main sections, because they are from a completely different field, like agriculture, or span over several fields and data sources. Multimedia data, for example, can consist of images and text, which means it connects to computer vision and natural language processing. However, we still wanted to present these works, because of their influences and to show how widely deep learning has already been applied. At this position, the interested reader is referred to the comprehensive and interdisciplinary monograph by *Deng & Yu (2014)*, which provides an in-depth overview of deep learning in computer vision and language processing.

Note that reinforcement learning, equivalent to data augmentation and generative adversarial networks, is actually a universal technique that can also be used in several areas, like computer vision and language processing. However, the presented surveys review deep reinforcement learning in the context of common deep learning architectures, with a wide range of applications like images, text or robotics. Hence, we present them in this additional section and not in a specific one, like computer vision.

### Big data

Big data is the field that analyses data that is too comprehensive, or too complex, to handle using classic data-processing tools. The aim of big data is to systematically extract information from large and complex data sets. Application areas include e-commerce, industrial control, and precision medicine. *Zhang et al. (2018)* review in their contribution the works on emerging deep learning models for feature learning with big data. They review deep learning-based methods and models that have been used with large data collections, heterogeneous data, but also real-time and low-quality data.

*Mohammadi et al. (2018)* give an overview on deep learning-based approaches that have been used to support the learning and analytics in the domain of internet of things (IoT). They begin with the characteristics of IoT data, but also the treatments of IoT data, namely the analytics and streaming of big data in the domain of IoT. Next, they review the promising aspects of deep learning-based methods for getting certain results in analytics regarding these data types' applications. In addition, they outline the potential of upcoming deep learning-based methods for data analytics in the IoT domain. Besides a comprehensive background on different deep learning methods, they review further research efforts, which affected the IoT domain by applying deep learning. Furthermore, they state how smart IoT devices have incorporated deep learning and review methods for fog computing and cloud computing in aiding IoT approaches.

*Emmert-Streib et al. (2020)* start their contribution with a background analysis in deep learning-based methods, like convolutional neural networks, deep feed-forward neural networks, but also deep belief networks, long short-term memory networks and autoencoders, because, according to the authors, these are currently the most commonly used architectures. Additionally, they introduce related concepts such as restricted Boltzmann machines and resilient backpropagation and discuss the differences when dealing with big data *vs*. small data and specific data types. They state that the adaptiveness of these (network) architectures enables a "Lego-like" generation of countless new neural networks.

### *Reinforcement learning*

Reinforcement learning is an algorithmic learning strategy where the algorithm tries to maximize an agent's performance *via* rewards, based on observations within the task environment. Reinforcement learning-based approaches have proven to be successful in numerous fields, like the robotics domain. The article of *Mousavi, Schukat & Howley (2016)* surveys the recent advances in supervised and unsupervised deep reinforcement learning, putting an emphasis on most commonly applied deep architectures, like convolutional neural networks, recurrent neural networks, but also autoencoders that have effectively been incorporated within reinforcement learning-based frameworks. They structure their review in three main categories, these are: supervised reinforcement learning, unsupervised reinforcement learning, and deep reinforcement learning in environments that allow to observe parts of the process as Markov decisions.

*Li (2017)* gives also an overview of deep reinforcement learning strategies, discussing six main elements, six significant mechanisms, but also twelve related applications. After presenting the fundaments of machine and deep learning, he also introduces the main elements of reinforcement learning, such as value function, policy, reward, and exploration strategies. Afterwards, the mechanisms, including attention and memory, transfer learning, unsupervised learning, multiagent reinforcement learning, but also hierarchical reinforcement learning and learning to learn, are presented. Finally, numerous possible applications are outlined, such as games (like AlphaGo), natural language processing, covering dialogue systems, text generation, machine translation, but also computer vision, robotics, finance, business management, education, healthcare, Industry 4.0, intelligent transportation systems, smart grids and further computer systems.

*Arulkumaran et al. (2017)* start their review by giving a general overview of the reinforcement learning field and continuing afterwards to the central areas of value-based methods, but also policy-based methods. The review covers the main approaches and methods in the field of deep reinforcement learning, such as deep q-networks, asynchronous advantage actor critic and trust region policy optimization. Additionally, they outline the main benefits of deep neural networks, in particular on visual understanding, with reinforcement learning.

### *Mobile and wireless networking*

Mobile and wireless networking, or short networking, has rapidly evolved during the last years, thanks to the spread of mobile devices and mobile applications (*Bevilacqua et al., 2015*; *Labini et al., 2019*). In addition, the recently released 5G technology is expected to massively increase the mobile traffic volumes. *Zhang, Patras & Haddadi (2019)* present a comprehensive survey in deep learning-based research in mobile and wireless networking. They start with an introduction about fundamentals on deep learning-based methods that could lead to networking applications and review certain approaches and platforms with the potential to promote the progression of mobile systems using deep learning. Next, they give an overview on deep learning-based research in mobile and wireless networking, by categorizing it in several domains.

### Mobile multimedia

Mobile multimedia refers to various applications that can be accessed or created by portable devices, like smartphones (*Karner et al., 2020*). This includes mobile applications such as audio and video players, games and e-healthcare. *Ota et al. (2017)* introduce the basics of deep learning for multimedia, thereby focusing on the core parts of deep learning in regard to mobile environments, namely low-complexity deep learning methods, software tools and frameworks for mobile and other resource-constrained environments, but also specific hardware available in mobile devices, which can be used to facilitate the computationally intensive training and inference of deep networks. In addition, they present numerous deep learning-based, mobile applications to show possible real-life scenarios for such a technology.

### Multimodal learning

Multimodal learning uses data of different modalities in a learning strategy. An example for data from different modalities are the acquisitions from positron emission tomography-computed tomography (PET-CT) scanners, where the tissue data from the CT and metabolically active regions from the PET are acquired from a patient (*Gsaxner et al., 2019b*). In their survey, *Ramachandram & Taylor (2017)* first classify the architectures for deep multimodal learning. Afterwards, they introduce certain methods to combine multimodal representations that have been learned with these deep learning-based architectures. In particular, they outline two main research fields for potential upcoming works, namely regularization methods and strategies that learn and optimize structures in the domain of multimodal fusion.

### Remote sensing

Remote sensing covers technologies for the remote analysis of objects or scenes. Examples are satellite-based imaging, aerial imaging, crowdsourcing (such as tweets or phone imagery), but also advanced driver-assistance systems and unmanned aerial vehicles. *Ball, Anderson & Chan (2017)* provide a compact analysis of recent deep learning-based research in the domain of remote sensing. They introduce the theories and tools, but also challenges within the remote sensing field. Thereafter, they present remaining research questions and opportunities, like modelling physical phenomena with human-understandable solutions, inadequate data sets and big data. Furthermore, they focus on specific non-traditional data sources that are heterogeneous, deep learning-based architectures and algorithms to learn spatial, spectral and temporal data, but also transfer learning. Further, they provide a fundamental insight into deep learning-based systems, and outline obstacles for training, but also optimizing deep learning-based methods.

### Graphs

Graphs are representations of objects and of their relationships with other objects. Common examples include social networks, traffic networks, e-commerce networks, but also biological networks. *Zhang, Cui & Zhu (2020)* provide a survey on various types of deep learning-based approaches on graphs. They split the existing approaches into five different categories in regards to the underlying model architectures, but also training strategies, namely graph convolutional networks, graph recurrent neural networks, graph reinforcement learning, graph autoencoders and graph adversarial methods. They propose a systematic outline of these techniques, mostly by following the historical appearance and review the structures and differences. They conclude their review by outlining the applications in this area.

### Anomaly detection

Anomaly detection is a strategy used to detect unexpected events or items in data sets. It can be used in areas like signal processing, statistics, finance, manufacturing, econometrics, networking, but also data mining. *Kwon et al. (2019)* propose an outline of deep learning-based methods, covering deep neural networks, restricted Bolzmann machine-based deep belief networks and recurrent neural networks. They also cover machine learning techniques that are related to network anomaly detection. Furthermore, they present the latest work from the literature that used deep learning-based techniques with an emphasis on network anomaly detection. Finally, they outline their own results with deep learning-based methods for the analysis of network traffic.

### Recommender systems

Recommender systems are utilized to predict the preferences of a user, for example, to provide web users with personalized information about products and services, like movies, insurances, or restaurants. *Zhang et al. (2019)* propose a survey of latest deep learning-based research for recommender systems. They formulate and introduce a deep learning-based taxonomy for recommendation models, together with an outlining of recent contributions from the literature. They conclude their contribution by outlining current trends and propose new perspectives in regard to the development in the field.

### Agriculture

Agriculture is the scientific process that deals with the cultivation of plants and livestock to produce products like food, feed and fibres. *Kamilaris & Prenafeta-Boldú (2018)* review 40 research works in deep learning, which have been used for numerous agricultural, but also food production problems. For each problem, they explored the specific agricultural challenge, the frameworks and models that have been used, but also the data source, data nature and data pre-processing methods. They also report and present the overall performance results for the selected metrics. In addition, they survey the comparison of deep learning-based approaches with other existing common methods, with a focus on the differences in the performances for classification and regression.

### Multiple areas

At last, this sub-section presents deep learning review contribution that span over multiple disciplines and applications with no main focus on a specific area. *Pouyanfar et al. (2018)* present a more general review in the field of deep learning-based algorithms and techniques, but also their applications. Their survey proposes an in-depth analysis of historical, but also novel methods in visual processing, audio processing, and text processing, but also the analysis of social networks, and further the processing of natural language. Next, they provide a comprehensive review on improvements in deep learning-based approaches and survey deep learning challenges, like unsupervised learning and online learning, but also black-box models, and show how these remaining demands may be addressed in the upcoming works.

*Dargan et al. (2019)* focus in their contribution on common deep learning concepts, including fundamental and more sophisticated architectures, characteristics, techniques, limitations and motivational aspects. They introduce several main differences amongst classical machine learning, deep learning and approaches in conventional learning, but also main challenges that still need to be solved. They chronologically analyse and conduct an extensive review of significant deep learning-based applications, thereby including numerous fields, techniques, methods and architectures that have been applied, and discuss the works of their application for usage in a real world scenario.

In their review of deep learning-based scientific discovery, *Raghu & Schmidt (2020)* provide an analysis of several deep learning-based models that have been applied to topics like sequential, visual, but also graph structured data. Further, they present related tasks and varying training approaches, together with methods that enable the usage of deep learning-based methods for sparse data, and how to get a better understanding of complex models. They also give different outlines of overall design processes, hints for implementations, and point to tutorials, open-sourced pipelines in deep learning, research summaries, but also pretrained models that have been implemented within the community, with the aim to speed up the application of deep learning-based approaches across multiple scientific fields and domains.

### Going deeper: architectures, evaluations, pros, cons, challenges and future directions in additional works

Table 9 provides more details about the presented methods, pros, cons, evaluations and challenges and future directions of the surveys in the category additional works. The three surveys about big data from *Zhang et al. (2018)*, *Mohammadi et al. (2018)* and *Emmert-Streib et al. (2020)* introduce common architectures, like DBN, CNN, RNN, AutoEncoders, whereby *Mohammadi et al. (2018)* and *Emmert-Streib et al. (2020)* introduced further models and variants, like LSTM. *Zhang et al. (2018)* and *Mohammadi et al. (2018)* have also a focus on real-time data. *Zhang et al. (2018)* and *Mohammadi et al. (2018)* make a complexity analysis, which includes for both works the forward passes. *Emmert-Streib et al. (2020)* do not perform such an analysis, but extract the pros and cons when going over the different deep learning models, for example that CNN can easily process high-dimensional input and is good at extracting local information. On the downside CNN can overfit, be hard to train and has the vanishing gradient problem. RNNs can be used for problems with sequential data, but have problems with long-term dependencies. ResNets can be used to tackle the degradation problem. A property of AutoEncoder is that the input and output layer have the same size and they can be used

**Table 9** **Architectures, pros, cons, evaluations, challenges and future directions in deep learning in the category *additional works*.**

| | |
|---|---|
| **Big data** | |
| *Zhang et al. (2018)* | |
| Architectures/Methods | DBN, CNN, RNN, AutoEncoder |
| Pros/Evaluations | Computational complexity, storage complexity (FC forward pass, TT forward pass, FC backward pass, TT backward pass) |
| Cons/Evaluations | Computational complexity, storage complexity (FC forward pass, TT forward pass, FC backward pass, TT backward pass) |
| Challenges and future directions | Large-scale deep learning models for huge amount of data; multi-modal deep learning models; new learning frameworks and computing infrastructures; compress the large-scale deep learning models; current multi-modal deep learning models only bi-modal data; exploring effective fusion methods; reduce the computational complexity of deep computation models; optimize the network structures; reliable deep learning models for low-quality data |
| *Mohammadi et al. (2018)* | |
| Architectures/Methods | AutoEncoder, RNN, RBM, DBN, LSTM, CNN, VAE, GAN, Ladder Net, variants |
| Pros/Evaluations | Sizes and complexities in different applications (LSTM, RNN, RBM) |
| Cons/Evaluations | Sizes and complexities in different applications (LSTM, RNN, RBM) |
| Challenges and future directions | Service discovery; model and task distribution; design factors for fog environments; energy management of mobile edge devices; mobile edge computing environments |
| *Emmert-Streib et al. (2020)* | |
| Architectures/Methods | FNNN, RNN, Hopfield Network, DBM, RBM, CNN, AutoEncoder, LSTM, variants |
| Pros/Evaluations | Can easily process high-dimensional input (CNN); good at extracting local information (CNN); outperformed other methods for predicting the toxicity of drugs (D-FFNN); used for problems with sequential data (RNN); degradation problem (ResNet); same size of the input and output layer (AutoEncoder); dimensionality reduction (AutoEncoder); long-term dependencies (LSTM) |
| Cons/Evaluations | Overfitting (CNN); hard to train (CNN); vanishing gradient (CNN); adequate amount of data (AutoEncoder); long-term dependencies (RNN) |
| Challenges and future directions | Finding the right application for a deep learning model; understandable decisions (non black boxes); small data sizes; transfer of knowledge between such models; exploring further advanced models (reinforcement learning, graph CNN, VAE) |
| **Reinforcement learning** | |
| *Mousavi, Schukat & Howley (2016)* | |
| Architectures/Methods | AutoEncoder, CNN, RBM, RNN, FNN, LSTM, DQN, MDRNN, DFQ, DRQN, combinations and variants |
| Pros/Evaluations | Target output of the network is the same of the input (AutoEncoder); could outperform state of the art algorithms in pattern recognition (CNN); useful for applications, which have temporal and sequential data (RNN); learn very long-term dependencies (LSTM); very fast supervised leaning method (NFQ); learning optimal policy (DQN); better stability (DQN); can play a variety of games (MDRNN+LSTM); vanishing gradient problem (MDRNN+LSTM); predict action-conditional frames (CNN+RNN+RL); continuous grid-world tasks (DFQ); noisy and incomplete data (LSTM+DQN); can have better performance (DRQN) |
| Cons/Evaluations | Large amount of data (CNN); high performance computing power (CNN); correlation between temporally long-term events (RNN); some stability issues (model-free reinforcement learning algorithms, like Q-learning); needs sufficient data (DQN); vanishing gradient problem (RNN); real world applications (Markov assumption); not always significant superiority (DRQN) |
| Challenges and future directions | Learn optimal control policies in problems with raw visual input; still challenges in real application such as robotics; investigating deep architectures for end to end leaning and deep state representation; end to end learning in real world applications; transfer learning |
| *Li (2017)* | |
| Architectures/Methods | MLP, CNN, ResNets, RNN, AlexNet, Seq2Seq, DQN, D-DQN, combinations and variants |

(Continued)

| Table 9 (continued) | |
|---|---|
| **Big data** | |
| Pros/Evaluations | Can outperform DQN (PGQ); stability of policy gradients (Q-Prop); stabilized the learning (DQN); achieved outstanding results (DQN); end-to-end RL approach (DQN); perform well on many different tasks (DQN); tackle over-estimate (D-DQN); better policies on Atari games (D-DQN); faster converge (dueling network architecture); avoids the optimization of action at every time step (DDPG); fully differentiable CNN (VIN) |
| Cons/Evaluations | Reinforcement learning vs deep reinforcement learning; unstable training of action value function approximation (CNN); over-estimate (DQN); policies on Atari games (DQN) |
| Challenges and future directions | Better understanding of how deep learning works; investigate comments/criticisms, *e.g.*, from cognitive science; considering perspectives of government, academia and industry; few products so far; need products and market validation |
| *Arulkumaran et al. (2017)* | |
| Architectures/Methods | DQN, NFQ, CNN, model-free DRL, model-based DRL, combinations and variants |
| Pros/Evaluations | High-dimensional problems (DRL); powerful function approximation properties (DRL); learn directly from raw, high-dimensional visual inputs (CNN); comparable to professional video games tester (DQN); compactly represent both high-dimensional observations and the Q-function (DQN); addresses instability problem (DQN); easier-to-learn relative values (duelling DQN) |
| Cons/Evaluations | Black box optimization methods (DRL); difficult policy search (DRL); suffer from severe local minima (DRL); require many samples (DNN); cannot adapt to new situations (ALVINN); can be unable to recover from situations (ALVINN) |
| Challenges and future directions | Improving the data efficiency of neural networks; not yet ready for complex real-world problems; deeper integration with other traditional AI approaches; better sample complexity, generalization, and interpretability; better understanding |
| **Mobile and wireless networking** | |
| *Zhang, Patras & Haddadi (2019)* | |
| Architectures/Methods | MLP, RBM, AutoEncoder, CNN, RNN, LSTM, GAN, DRL, tailored Deep Learning methods for Mobile Networks |
| Pros/Evaluations | Naïve structure and straightforward to build (MLP); can generate virtual samples (RBM); powerful and effective unsupervised learning (AutoEncoder); weight sharing, affine invariance (CNN); expertise in capturing temporal dependencies (RNN); can produce lifelike artifacts from a target distribution (GAN); ideal for high-dimensional environment modeling (DRL) |
| Cons/Evaluations | High complexity, modest performance and slow convergence (MLP); difficult to train well (RBM); expensive to pretrain with big data (AutoEncoder); high computational cost, challenging to find optimal hyper-parameters, requires deep structure for complex tasks (CNN); high model complexity, gradient vanishing and exploding problems (RNN); training process is unstable (convergence difficult) (GAN); slow in terms of convergence (DRL) |
| Challenges and future directions | Serving deep learning with massive high-quality data; deep learning for spatio-temporal mobile data mining; deep learning for geometric mobile data mining; deep unsupervised learning in mobile networks; deep reinforcement learning for mobile network control |
| **Mobile multimedia** | |
| *Ota et al. (2017)* | |
| Architectures/Methods | CNN, DBN, RCN, NMT, GNMT, variants |
| Pros/Evaluations | Performance, energy consumption and cost (GPU, FPGA, ASIC, VPU, Distributed Computing, High Performance Computing) |
| Cons/Evaluations | Performance, energy consumption and cost (GPU, FPGA, ASIC, VPU, Distributed Computing, High Performance Computing) |
| Challenges and future directions | Resource-constrained mobile platforms; optimizing techniques for mobile deep learning-based architectures by "compressing"; developing new hardware for mobile deep learning architectures; energy constraints in the mobile world; coupling deep network architectures and the mobile computing capabilities |
| **Multimodal learning** | |

**Table 9** (continued)

| **Big data** | |
|---|---|
| *Ramachandram & Taylor (2017)* | |
| Architectures/Methods | Sparse RBM, DBN, DBM, FCNN, CNN, RNN, AutoEncoders, MLP, LSTM, variants |
| Pros/Evaluations | Multiple modalities into the learning problem almost always results in much better performance; deep learning methods facilitate a flexible intermediate-fusion approach; makes it simpler to fuse modality-wise representations and learn a joint representation; allows multimodal fusion at various depths in the architecture |
| Cons/Evaluations | Deep learning-based architectures still involve a great deal of manual design for multimodal learning; architecture learning can be extremely compute-intensive in multimodal learning |
| Challenges and future directions | Full exploration of fusion architectures; truly generic learning methods, with minimal or no human intervention; take advantage of advances in hardware acceleration and distributed deep learning; few multimodal medical data sets; small medical data sets, involving only between ten and 50 subjects; high class imbalances (normal vs abnormal cases); make data sets publicly available |
| **Remote sensing** | |
| *Ball, Anderson & Chan (2017)* | |
| Architectures/Methods | AutoEncoder, DBN, RNN, CNN, variants |
| Pros/Evaluations | Overall accuracy results in percent for hyperspectral images (Indian Pines, Kennedy Space Center, Pavia City Center, Pavia University, Salinas, Washington DC Mall) |
| Cons/Evaluations | Overall accuracy results in percent for hyperspectral images (Indian Pines, Kennedy Space Center, Pavia City Center, Pavia University, Salinas, Washington DC Mall) |
| Challenges and future directions | Inadequate data sets; human-understandable solutions for modeling physical phenomena; big data; nontraditional heterogeneous data sources; deep learning-based architectures and learning algorithms for spectral, spatial, and temporal data; transfer learning; an improved theoretical understanding of deep learning-based systems; high barriers to entry; training and optimizing the deep learning approaches |
| **Graphs** | |
| *Zhang, Cui & Zhu (2020)* | |
| Architectures/Methods | Graph recurrent neural networks, graph convolutional networks, graph autoencoders, graph reinforcement learning, graph adversarial methods |
| Pros/Evaluations | Time complexity (Graph RNN, GCN, GAE, Graph Reinforcement Learning, Graph Adversarial Methods) |
| Cons/Evaluations | Time complexity (Graph RNN, GCN, GAE, Graph Reinforcement Learning, Graph Adversarial Methods) |
| Challenges and future directions | New models for unstudied graph structures; compositionality of existing models; dynamic graphs; interpretability and robustness |
| **Anomaly detection** | |
| *Kwon et al. (2019)* | |
| Architectures/Methods | RBM, DBM, DBN, DNN, AutoEncoder, RNN, variants |
| Pros/Evaluations | Accuracy, precision, recall, F1-score (FCN); FCN has improved detection accuracy when compared to SVM, random forest, and Adaboosting |
| Cons/Evaluations | Accuracy, precision, recall, F1-score (FCN) |
| Challenges and future directions | New deep learning techniques, like GANs, need to be studied for anomaly detection; train missing data through prediction with the reinforced learning algorithms in the GAN model; GAN can work with machine learning to generate multi-modal outputs; improved performance for network anomaly detection with greater accuracy |
| **Recommender systems** | |
| *Zhang et al. (2019)* | |
| Architectures/Methods | Recommendation with Neural Building Blocks (MLP, AutoEncoder, CNN, RNN, RBM, Neural Autoregressive Distribution Estimation, Attentional Models, Adversarial Network, Deep Reinforcement Learning); Recommendation with Deep Hybrid Models (RNN+CNN, AutoEncoder+CNN, RNN+AutoEncoder, *etc.*) |

(Continued)

**Table 9** (continued)

| | |
|---|---|
| **Big data** | |
| Pros/Evaluations | Approximate any measurable function to any desired degree of accuracy, widely used in many areas (MLP); almost all the AutoEncoder variants can be applied to the recommendation tasks; powerful in processing unstructured multimedia data (CNN); suitable for sequential data processing (RNN); can be easy to design (RBM); can process long and noisy inputs (Attentional Models); can outperform CNN and RNN (Attentional Models); tractable (NADE); personalized recommendation (DRL); performance in real-world applications (DRL); personalized citation recommendation task (GAN); significantly improve performance (GAN) |
| Cons/Evaluations | Might not be as expressive as AutoEncoder, CNN, and RNN (MLP); fails to deal with non-integer ratings and can lead to worse prediction accuracy (AutoEncoder); only slightly performance improvement (CNN); trivial methods can achieve the same or even better accuracy results (RNN); can be limited to binary values and not be tractable (RBM); sparsity (NADE) |
| Challenges and future directions | Joint representation learning from user and item content information; explainable recommendation with deep learning; going deeper for recommendation; machine reasoning for recommendation; cross domain recommendation with deep neural networks; deep multi-task learning for recommendation; scalability of deep neural networks for recommendation; better, more unified and harder evaluation |
| **Agriculture** | |
| *Kamilaris & Prenafeta-Boldú (2018)* | |
| Architectures/Methods | CNN, AlexNet, VGG16, ResNet, DRNN, LSTM, DBN, AutoEncoder, variants |
| Pros/Evaluations | Classification Accuracy, Precision, Recall, F1 score, LifeCLEF metric, Quality Measure, Mean Square Error, Root Mean Square Error, Mean Relative Error, Ratio of total fruits counted, L2 error, Intersection over Union, combined scores |
| Cons/Evaluations | Classification Accuracy, Precision, Recall, F1 score, LifeCLEF metric, Quality Measure, Mean Square Error, Root Mean Square Error, Mean Relative Error, Ratio of total fruits counted, L2 error, Intersection over Union, combined scores |
| Challenges and future directions | Need of large datasets; data augmentation techniques with label-preserving transformations; low variation among the different classes; need for experts in order to annotate input images; experts or volunteers are susceptible to errors during data labeling; cannot generalize beyond the "boundaries of the dataset's expressiveness"; difficulty in detecting heavily occluded and distant objects; not many publicly available datasets; some agricultural-related problems are under-researched; usage of aerial imagery; higher performance classification or prediction; combining hand-crafted features with automatic features; demonstrating the ability of the models to generalize to various real-world situations; researchers should make their datasets publicly available; commercial usage in the future |
| **Multiple areas** | |
| *Pouyanfar et al. (2018)* | |
| Architectures/Methods | DNN, RvNN, RNN, CNN, DBN, DBM, GAN, VAE, variants |
| Pros/Evaluations | Is able to generate very high-level data representations from massive volumes of raw data (DNN); solution to many real-world applications (DNN); major breakthroughs in different applications (DNN); scalable architecture (DLAU) |
| Cons/Evaluations | Black box machines that inhibit development at a fundamental level (DNN); no decision understanding (DNN); weak statistical interpretability (DNN); needs extensive datasets (DNN); majority of the existing implementations are supervised algorithms (DNN); require considerable amounts of computational resources (DNN); no FPGA test beds (DNN) |
| Challenges and future directions | Interpretability needs to be investigated; modeling multiple complex data modalities at the same time; one-shot learning and zero-shot learning needs to be further explored; more unsupervised and semi-supervised learning to handle real-world data without the need of manual human labels; many applications are still under-researched, *e.g.* disaster information management and finance; reduction of dimensionality without losing critical information; more research in deep reinforcement learning; handling noisy and messy data |
| *Dargan et al. (2019)* | |
| Architectures/Methods | DNN, AutoEncoder, CNN, RBM, LSTM, RNN, GAN, variants |

| Table 9 (continued) | |
|---|---|
| **Big data** | |
| Pros/Evaluations | Most effective, supervised, time and cost efficient machine learning approach (DNN); not a restricted learning approach (DNN); learns the illustrative and differential features in a very stratified way (DNN); appreciable performance in a wide variety of applications (DNN); powerful tool in many fields (DNN); strong learning ability (DNN); learn feature extraction methods from the data (DNN); little engineering by hand (DNN); optimized results (DNN); solve highly computational tasks (DNN); do not rely on prior data and knowledge (DNN); high-level abstraction (DNN) |
| Cons/Evaluations | Large amount of data (DNN); requires hardware with very high performance (DNN); creates new features by its own processes and techniques (DNN); difficult to understand (DNN); high training requirement (DNN); need high-end graphical processing units (DNN) |
| Challenges and future directions | Manage input data continuously; transparency of the conclusion; demanding resources; improved methods for big data analytics; needs very high computation power; suffer from local minima; needs large amount of data; no strong theoretical foundation; difficult to find the topology and training parameters; handle noisy input; raising the performance; unsupervised learning; maintenance of wide repository of data; fully autonomous driving |
| *Raghu & Schmidt (2020)* | |
| Architectures/Methods | CNN, Graph Neural Networks, Neural Networks for Sequence Data, RNN, LSTM, GAN, VAE, variants |
| Pros/Evaluations | Fundamental breakthroughs in core problems in machine learning (DNN); supervised learning highly successful (DNN); end-to-end system (DNN); encapsulates complex functions (DNN); efficient analysis and automated processing of complex data (DNN); enormous number of resources developed and shared by the community (DNN); open sourced code (DNN) |
| Cons/Evaluations | Determine the scientific problems (DNN); where to start (DNN); breadth and diversity of different techniques (DNN); which combination of methods is most promising (DNN); supervised learning dependence on large amounts of labelled data (DNN); interpretation (DNN); not always the best technique for a problem (DNN) |
| Challenges and future directions | Black box; determining how the neural network model makes a specific prediction |

for dimensionality reduction. On the downside AutoEncoder often need an adequate amount of data. LSTM, for example, can handle long-term dependencies. *Zhang et al. (2018)* see future challenges in the development of large-scale deep learning models for huge amount of data and multi-modal deep learning models. They state that the gain in computational performance is lagging far behind the growth rate of big data. Further challenges are the development of new learning frameworks and computing infrastructures, compressing the large-scale deep learning models and the focus on more than two modalities, exploring effective fusion models, reduce the computational complexity, optimize network structures and reliable deep learning models for low-quality. *Mohammadi et al. (2018)* see the main challenges in: (1) Service discovery; (2) Model and task distribution; (3) Design factors for fog environments; (4) Energy management of mobile edge devices; and (5) mobile edge computing environments. *Emmert-Streib et al. (2020)* see future works in finding the right application for a deep learning model, but also understand the decisions of deep learning models (black boxes), especially for the healthcare domain. Further challenges are small data sizes, the transfer of knowledge between models and exploring more advanced models, like reinforcement learning, graph CNN and VAE.

*Mousavi, Schukat & Howley (2016)*, *Li (2017)* and *Arulkumaran et al. (2017)* present all surveys about deep reinforcement learning. They also have in common that all three surveys cover various types of input data like, images, videos, text and language, for a range of applications, like natural language processing or robotics. The contributions go over classic concepts in reinforcement learning, like Q-learning, policy search and actor critic, and how these have been handled in deep reinforcement learning, thereby introducing deep learning methods, like CNN, AutoEncoder, RBM, RNN, LSTM, and their combinations with classic reinforcement learning concepts, resulting in deep reinforcement learning methods and concepts, like DQN, DFQ, DRQN, or DRQN. All surveys also introduce and discuss the performance of deep reinforcement learning methods in playing Atari games, which seems to be one of the main benchmarks for evaluating (deep) reinforcement learning methods. Beside pros and cons from classic deep learning methods, *Mousavi, Schukat & Howley (2016)* identified several pros and cons in deep reinforcement learning, like DQN can learn an optimal policy and can have a good stability, but needs sufficient data. They also state that model-free reinforcement learning algorithms, like Q-learning, can have stability issues and DRQN can have better performance for some tasks. They also extract several pros for combined methods, for example MDRNN+LSTM can play a variety of games, MDRNN+LSTM tackles, the vanishing gradient problem, CNN+RNN+RL can predict action-conditional frames, and LSTM+DQN can handle noisy and incomplete data. *Li (2017)* identified several pros and cons in deep reinforcement learning, *e.g.* that DQN stabilized the learning, achieved outstanding results, is an end-to-end RL approach and performs well on many different tasks. However, *Li (2017)* also states that PGQ can outperform DQN. D-DQN on the other hand can tackle the problem of over-estimation in DQN and have better policies on Atari games than DQN. The author further states, that some specific extensions, like dueling network architectures can converge faster, but also that DDPG can avoid the optimization of action at every time step, or that VIN are fully differentiable CNN. A mature con is that deep reinforcement learning is a black box compared to many classic reinforcement learning techniques. *Arulkumaran et al. (2017)* also state that DRL are black box optimization methods. However, they can handle high-dimensional problems and have powerful function approximation properties. On the downside, the policy search can be difficult, they can suffer from severe local minima and DNN require in general many samples. They find that DQNs are comparable to professional video games tester, can compactly represent both high-dimensional observations and the Q-function, and address the instability problem. Further duelling DQNs are able to easier-to-learn relative values, Finally, ALVINN cannot adapt to new situations and can be unable to recover from specific situations they ran into. *Mousavi, Schukat & Howley (2016)* see future challenges in learning optimal control policies in problems with raw visual input, investigating deep architectures for end to end leaning and deep state representation and transfer learning. They also see challenges in real application such as robotics and end to end learning in real world applications. Future challenges of *Li (2017)* are a better understanding of how deep learning works and the investigation of comments/criticisms, *e.g.*, from cognitive science, also considering perspectives of government, academia and industry. *Li (2017)* states that there are only a few products so far and there is a need for products and market validation. *Arulkumaran et al. (2017)* see future challenges in improving the data efficiency of neural networks and that they are not yet ready for complex real-world problems. Further open issues are a deeper integration with other traditional AI approaches for a better sample complexity, generalization, and interpretability. Finally, they hope for a better theoretical understanding of neural networks within deep reinforcement learning. Note, that the survey of *Li (2017)* seems to be unfinished.

*Zhang, Patras & Haddadi (2019)* give a comprehensive survey in the domain of deep learning for mobile and wireless networking. In doing so, they introduce common deep learning architectures used in this field, like MLP, RBM, AutoEncoder, CNN, RNN, LSTM, GAN, DRL, but also tailored deep learning methods for mobile networks. They provide a compact overview of the pros and cons of these architectures, *e.g.* if they are straightforward to build, can generate virtual samples, if they are ideal for high-dimensional environment modeling or which architectures have modest performance, slow convergence, are difficult to train or expensive to pretrain with big data, to name a few. In this regards, they also state the learning scenarios: supervised, unsupervised, reinforcement, and suitable problems, like modeling data with simple correlation or extracting robust representations. They conclude their survey by listing five main and important remaining research issues that should be addressed in the future in mobile and wireless networking: (1) Serving deep learning with massive high-quality data, for example high-quality and large-scale labeled datasets still lack for mobile network applications; (2) Deep learning for spatio-temporal mobile data mining, by means of spatio-temporal distribution of mobile traffic and application popularity are difficult to understand; (3) Deep learning for geometric mobile data mining, because of inherent complexities of representations in mobile and wireless networking, traditional machine learning tools can struggle to interpret geometric data and make reliable inferences; (4) Deep unsupervised learning in mobile networks, because data labeling is costly and requires domain-specific knowledge, hence unsupervised learning becomes essential in extracting insights from unlabeled data; (5) Deep reinforcement learning for mobile network control by making no strong assumptions (*e.g.* about the objective functions, like function convexity, or data distribution, like Gaussian or Poisson distributed) about the target system and employ function approximation.

*Ota et al. (2017)* went through the basics of deep learning for multimedia, focusing on the main components of deep learning for mobile environments, namely the low-complexity deep learning algorithms, the software frameworks that are optimized for mobile and resource constrained environments, and the specific hardware for mobile devices supporting the computationally expensive processes of training deep networks and inference. They also highlight several mobile deep learning applications and give an overview of real-life usage in this field. In doing so, the authors introduce several deep learning-based architectures, like CNN, DBN, RCN, NMT, GNMT and variants. Because of the hardware constraints of mobile devices, the authors focus on different hardware acceleration solutions for DNNs (GPU, FPGA, ASIC, VPU, Distributed Computing, High Performance Computing) in the mobile domain and provide the pros and cons for performance, energy consumption and cost. Future challenges are the resource-constrained mobile platforms and the development of techniques for optimizing such architectures by "compressing" them and further developing new hardware that can optimally cater to the needs of deep learning-based architectures. Further challenges are the energy constraints that exist in the mobile world and coupling deep network-based architectures and the mobile computing capabilities in an optimal way. They foresee that more and more mobile applications can run deep learning-based engines, as mobile devices become more powerful, resulting in a new world of possibilities where we have seen only the tip of the iceberg so far.

*Ramachandram & Taylor (2017)* survey deep learning in the area of multimodal learning. Therefore, they introduce various deep learning-based architectures, like Sparse RBM, DBN, DBM, FCNN, CNN, RNN, AutoEncoder, MLP, and LSTM. In their contribution, they have a strong focus on multimodal fusion, namely, intermediate and late fusion, being implemented with deep architectures. They state that incorporating multiple modalities into the learning problem almost always results in much better performance. Further, they state that deep learning methods facilitate a flexible intermediate-fusion approach, but also that they make it simpler to fuse modality-wise representations and learn a joint representation. Finally, they allow multimodal fusion at various depths in the architecture. As downside, the authors see the involvement of a great deal of manual design for multimodal learning with deep learning-based architectures. In addition, that architecture learning can be extremely compute-intensive in multimodal learning. Future challenges are a full exploration of fusion architectures and truly generic learning methods with minimal or no human intervention. Moreover, taking advantage of advances in hardware acceleration and distributed deep learning. Specifically for the medical domain, they see future challenges in (1) the relatively few multimodal medical data set collections that are available and in addition (2) the relatively small medical data set sizes, involving only between ten and 50 subjects, and on top (3) a high class imbalances (normal vs abnormal cases). Hence, they encourage researches in the medical domain to make data sets publicly available in the future.

*Ball, Anderson & Chan (2017)* provide a comprehensive survey of state-of-the-art remote sensing deep learning research. Thereby, they introduced the most commonly used deep learning architectures in remote sensing, like AutoEncoder, DBN, RNN, CNN and variants of them. They also list some popular deep learning tools, like Caffe and Keras, and the subject areas of deep learning works in remote sensing, like human detection, animal detection or vehicle detection/recognition. For evaluation, the authors focus on the overall accuracy results in percent for hyperspectral images of common open-source data sets (Indian Pines, Kennedy Space Center, Pavia City Center, Pavia University, Salinas, Washington DC Mall). Supplemental material from the authors, in form of more detailed tables, can be found online. According to the authors, the main unsolved challenges and opportunities in remote sensing are: (1) inadequate data sets; (2) human-understandable solutions for modeling physical phenomena; (3) big data; (4) nontraditional heterogeneous data sources; (5) deep learning-based architectures and learning algorithms for spectral, spatial, and temporal data; (6) transfer learning; (7) an improved theoretical understanding of deep learning-based systems, (8) high barriers to entry; and (9) training and optimizing the deep learning approaches.

*Zhang, Cui & Zhu (2020)* provide a deep learning survey on graphs, more specific on deep learning for graph data with its unique characteristics. In doing so, they divide the existing methods into five categories based on their deep learning-based architectures and training strategies, namely graph recurrent neural networks, graph convolutional networks, graph autoencoders, graph reinforcement learning and graph adversarial methods. For evaluation, the authors focus on the time complexity, divided into tables for Graph RNN, GCN, GAE, Graph Reinforcement Learning, and Graph Adversarial Methods. The author identified the following main future challenges for deep learning on graphs: (1) new models for unstudied graph structures, because the existing methods are not suitable for all graph structures due to the extremely diverse structures of graph data; (2) compositionality of existing models, by means of incorporating interdisciplinary knowledge in a more general way, in contrast to a case-by-case basis remains an open problem; (3) most methods focus on static graphs rather than dynamic graphs, however, many real graphs are dynamic in nature, because their nodes, edges, and features can change over time; (4) interpretability and robustness is also a future challenge, because interpreting deep learning-based results on graphs is critical in decision-making problems, for example in the medical domain or disease-related tasks.

*Kwon et al. (2019)* present a survey about anomaly detection with deep learning-based methods. In doing so, they give a general overview about deep learning architectures, like RBM, DBM, DBN, DNN, AutoEncoder, and RNN, but also briefly discuss the classification techniques supervised learning, unsupervised learning and semi-supervised learning. For evaluation, they focus on their own FCN model applying it to the public NSL-KDD dataset. They present the accuracy, precision, recall and F1-score for their experiments. They conclude that the FCN model has an improved detection accuracy when compared to conventional machine learning techniques, like SVM, random forest, and Adaboosting. The authors see future challenges in studying other deep learning techniques, like GANs, for anomaly detection. They foresee that missing data can be trained through prediction with the reinforced learning algorithms in the GAN model. Further, they foresee that GANs can work with machine learning to generate multi-modal outputs, which may lead to an improved performance for network anomaly detection with greater accuracy.

*Zhang et al. (2019)* give a survey on deep learning for recommender systems. Thereby, dividing deep learning-based recommendation models into the following two main categories: (1) Recommendation with Neural Building Blocks (MLP, AutoEncoder, CNN, RNN, RBM, Neural Autoregressive Distribution Estimation, Attentional Models, Adversarial Network, Deep Reinforcement Learning); (2) Recommendation with Deep Hybrid Models (RNN+CNN, AutoEncoder+CNN, RNN+AutoEncoder, *etc.*). By going over these models, they discuss the pros and cons addressing specific recommender task-related problems. For example, MLP can approximate any measurable function to any desired degree of accuracy and has been widely used in many areas, but might not be as expressive as AutoEncoders, CNNs, or RNNs. The authors identified that almost all AutoEncoder variants can be applied to the recommendation tasks and CNNs are powerful in processing unstructured multimedia data. On the downside, AutoEncoder can fail to deal with non-integer ratings and can lead to worse prediction accuracy and CNN can sometimes have only slightly performance improvements over other methods. RNN are suitable for sequential data processing, but trivial methods can sometimes achieve the same or even better accuracy results. RBMs can be easy to design but can also be limited to binary values and are not tractable in contrast to NADE, which however, have problems with sparse data. Attentional models can process long and noisy inputs and outperform CNNs and RNNs. DRL can have advantages in personalized recommendation and in the performance of real-world applications. Finally, GANs can have advantages in personalized citation recommendation tasks and sometimes significantly improve performance. They conclude that existing works have already established a solid foundation for deep recommender systems, but identified the following open issues: (1) Joint representation learning from user and item content information; (2) Explainable recommendation with deep learning; (3) Going deeper (more than three to four layers) for recommendation; (4) Machine reasoning for recommendation; (5) Cross domain recommendation with deep neural networks; (6) Deep multi-task learning for recommendation; (7) Scalability of deep neural networks for recommendation; (8) and finally, they state that the recommender field needs a better, more unified and harder evaluation of the methods.

*Kamilaris & Prenafeta-Boldú (2018)* survey deep learning methods in the domain of agriculture. Thereby, giving a summary of the most commonly used methods and architectures, like CNN, AlexNet, VGG16, ResNet, DRNN, LSTM, DBN, AutoEncoder and further variants of these. For a comparison and evaluation of the surveyed works, they focus on metrics like Classification Accuracy, Precision, Recall, F1 score, LifeCLEF metric, Quality Measure, Mean Square Error, Root Mean Square Error, Mean Relative Error, Ratio of total fruits counted, L2 error, Intersection over Union and combined scores. Future challenges they see in the need of large datasets and data augmentation techniques with label-preserving transformations. Other challenges are the low variation among the different classes, the need for experts in order to annotate input images and that experts or volunteers are susceptible to errors during data labeling. They state also that current methods cannot generalize beyond the "boundaries of the dataset's expressiveness" have difficulties in detecting heavily occluded and distant objects and that there are not many publicly available datasets for researchers in the domain of agriculture. Further, they state that some agricultural-related problems are under-researched, such as seeds identification. Future directions they see in the usage of aerial imagery (*i.e.* by means of drones), higher performance classification or prediction, like a growth estimation of plants, trees, *etc*., combining hand-crafted features with automatic features, and demonstrating the ability of the models to generalize to various real-world situations. Finally, they encourage researchers to make their datasets publicly available and that some works can already be used commercially in the near future, *e.g.* for automatic robots that collect crops, and overall conclude and encourage that deep learning can be used towards smarter, more sustainable farming and more secure food production.

The surveys from *Pouyanfar et al. (2018)*, *Dargan et al. (2019)* and *Raghu & Schmidt (2020)* span over multiple areas and various applications in the context of deep learning

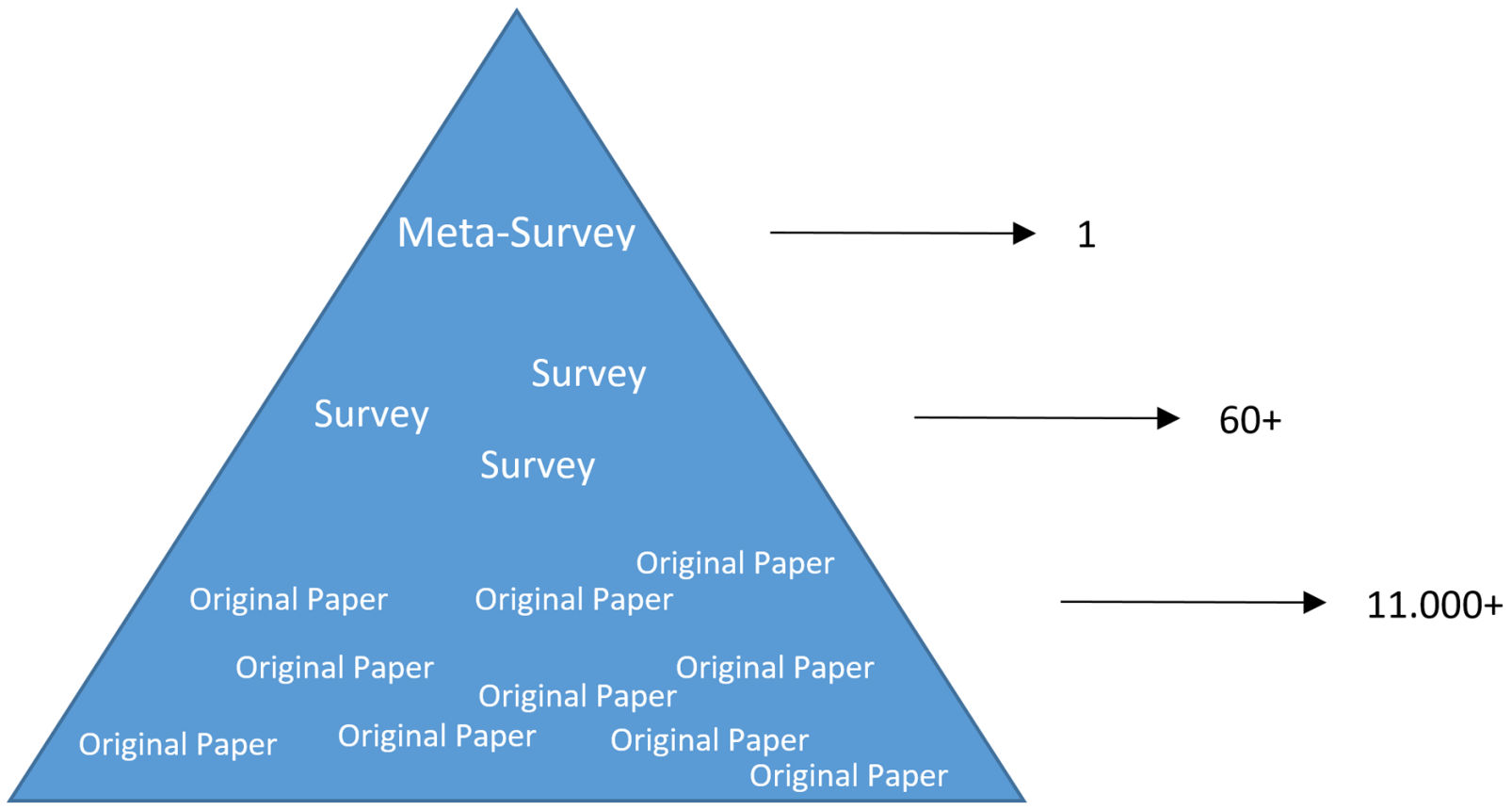

**Figure 18** **Publication pyramid: our meta-survey contribution based on over 60 surveys, which are based on over 11,000 original works (note that surveys can also reference other surveys or further publications that are not reviewed within the surveys).** Full-size DOI: 10.7717/peerj-cs.773/fig-18

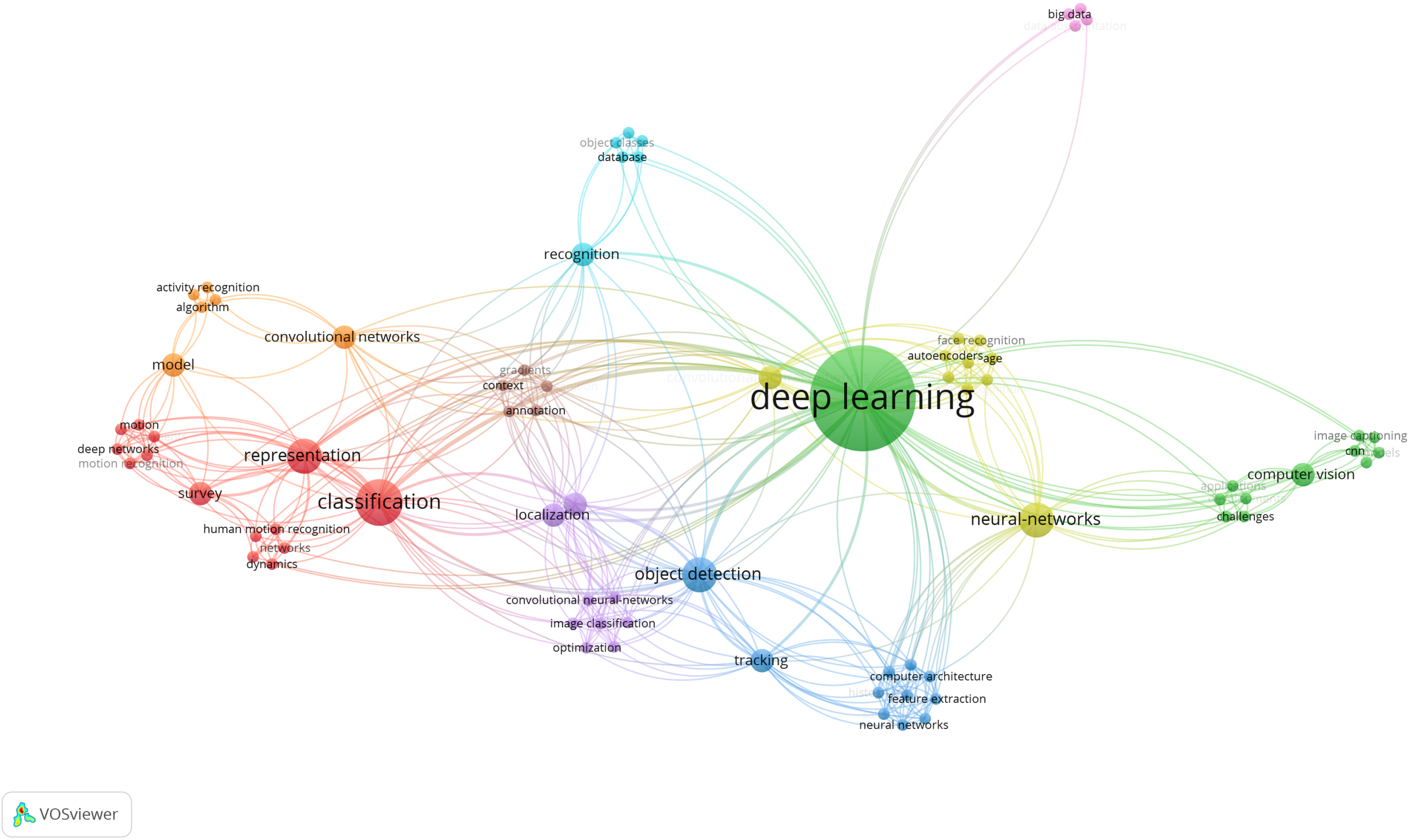

**Figure 19** **Network visualization (VOSviewer) for the survey articles supplied keywords from the category computer vision.** Full-size DOI: 10.7717/peerj-cs.773/fig-19

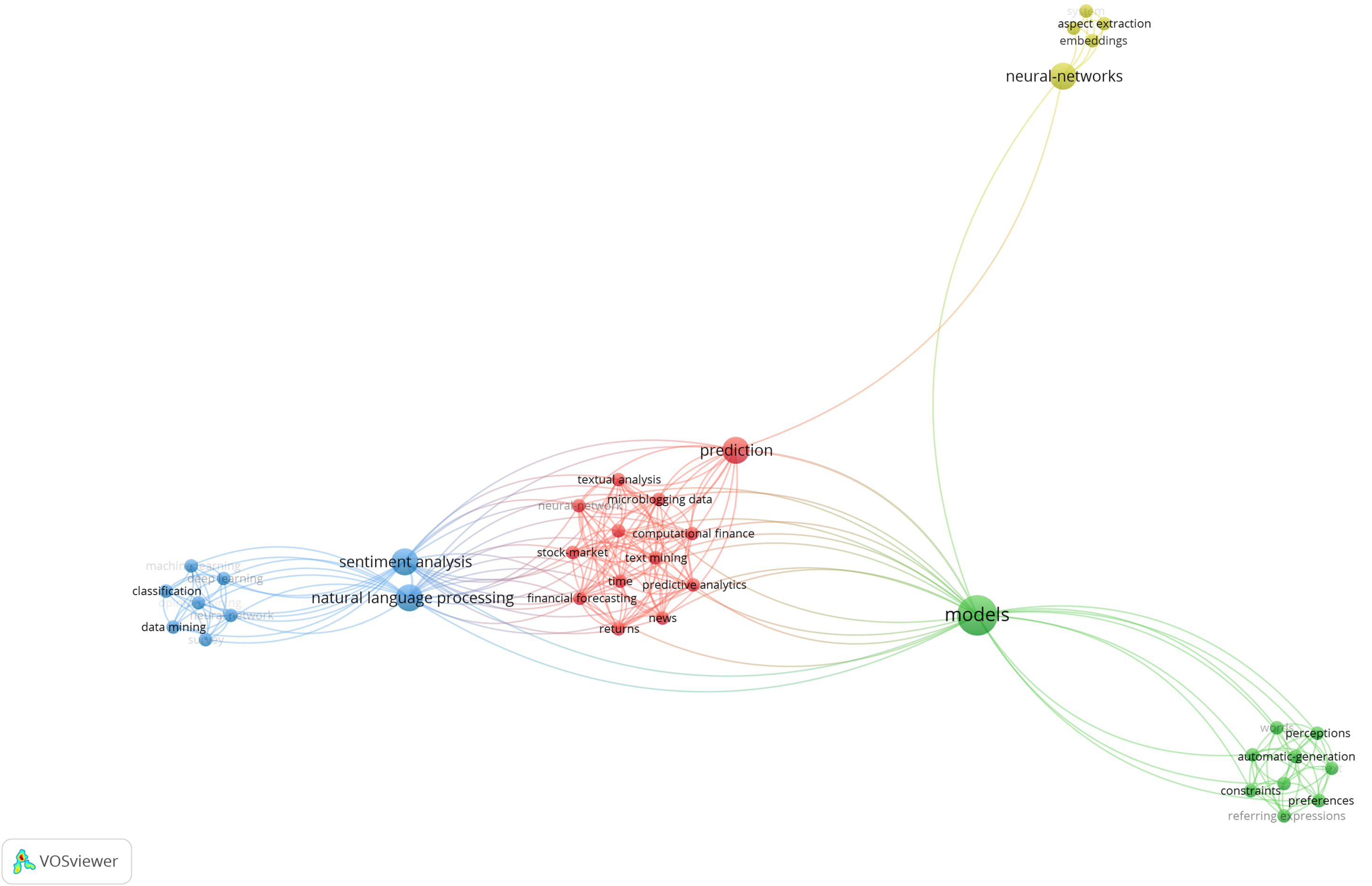

**Figure 20** **Network visualization (VOSviewer) for the survey articles supplied keywords from the category language processing.**

and provide a broad overview. In doing so, they all survey over numerous domains and fields, like visual, audio, and text processing, natural language processing, and applications, like sentiment analysis, social network analysis, parking systems, smart city, stock market analysis, pose estimation, answer selection, *etc*., and how they been tackled with deep learning. All works cover the main deep learning architecture and techniques and some variants and combinations, including DNN, RvNN, RNN, CNN, DBN, DBM, GAN, VAE. By going over the different deep learning techniques and methods, *Dargan et al. (2019)* make a direct comparison to methods from classic machine learning. *Raghu & Schmidt (2020)* introduce an overall general workflow to design and work with deep learning, starting with the data, over the learning to a validation and analysis of the results, and provide an implementation guideline pointing out to even further resources, like tutorials. They also provide specific sections for transfer learning, domain adaptation, multitask learning and weak supervision (distant supervision). Furthermore, they go over several learning and training strategies, like self-supervised learning, semi-supervised learning, self-training (bootstrapping), co-training, also in

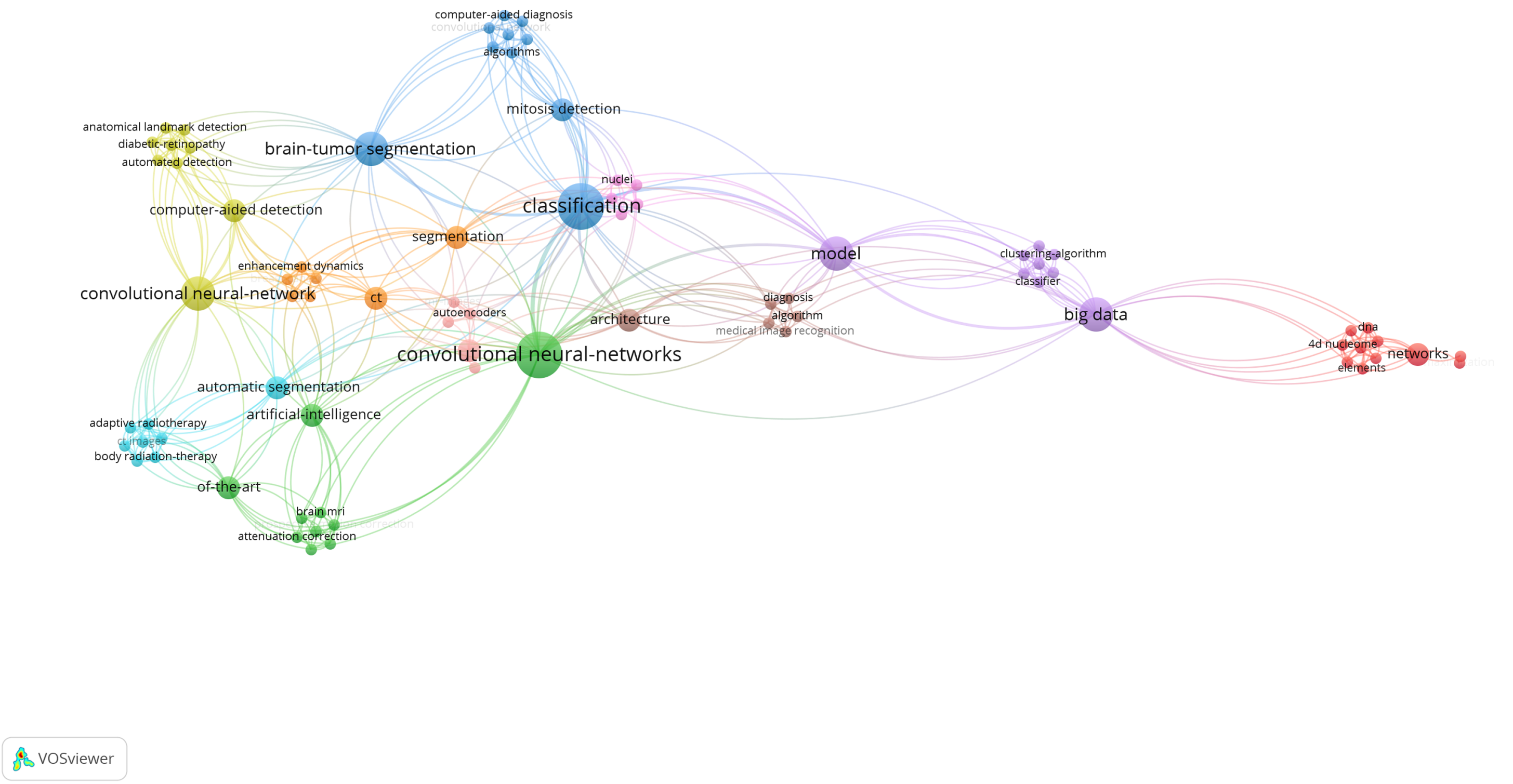

**Figure 21** **Network visualization (VOSviewer) for the survey articles supplied keywords from the category medical informatics.**
Full-size DOI: 10.7717/peerj-cs.773/fig-21

combination, like self-supervision with semi-supervised learning, for different data types and domains, like images and NLP, and discuss data augmentation for image data and sequence data. Because of the broadness of the topics, the authors also provide more high level pros and cons in the area of deep learning. Examples are, that they all agree that deep learning had major breakthroughs in different applications and provides solution to many real-world application, and state that it is the most effective, supervised, time and cost efficient machine learning approach. Further pros in deep learning are that it can solve highly computational tasks, does not rely on prior data and knowledge, requires little engineering by hand, and that there is a lot of open sourced code from the community. All criticize that deep learning is a black box approach, which is difficult to understand (decision understanding and weak statistical interpretability), and that deep learning-based approaches need massive data and computational hardware resources to work well. Moreover, it is hard to know where to start with deep learning, by means of how to determine the optimal method for a scientific problem, because of the breadth and diversity of different techniques, and furthermore which combination of methods may be the most promising. It was also stated that deep learning is not always the best technique for a problem. All surveys discuss deep reinforcement learning and *Dargan et al. (2019)* state that deep reinforcement kind of learning will be the future direction. Other challenges are the large amount of data, on one hand handling these efficiently, on

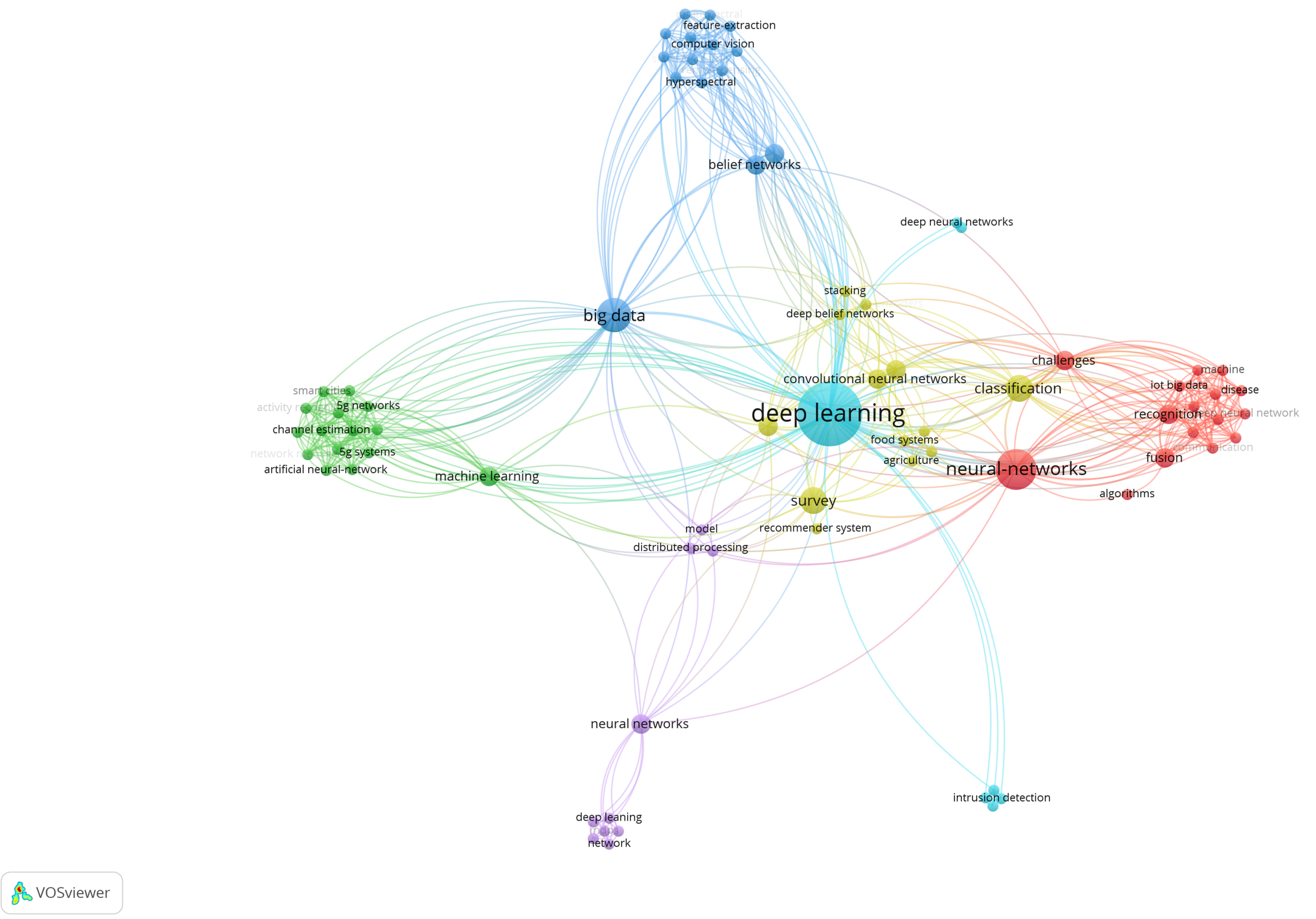

**Figure 22 Network visualization (VOSviewer) for the survey articles supplied keywords from the category additional works.**

Full-size DOI: 10.7717/peerj-cs.773/fig-22

the other hand developing unsupervised methods to avoid the need of manual human labels. *Raghu & Schmidt (2020)* do not provide a section about future challenges, but state the difficulty of determining how the neural network model makes a specific prediction. Hence, all surveys agree that a future challenge lies in enlighten the black box behavior of deep learning.

To conclude this section, most of the surveys reviewed in this category have also a similar structure: a general introduction of machine/deep learning techniques followed by a discussion of the reviewed publications categorized using different strategies, among which task- or application-based categorization is the most widely used. The surveys usually conclude by stating challenges and future perspectives of deep learning in their fields from a high level. The main challenge amongst the majority of the surveys is seen in deep learning being a black box approach, which makes it hard to understand how results and decisions have been achieved. A further common challenge is seen in the lack of (annotated) data, to train deep learning approaches more reliable.

## CONCLUSION AND DISCUSSION

In this contribution, selected reviews and surveys on deep learning have been presented in a *compact* categorized meta-survey. A search has been performed in common libraries and search engines, like in IEEE Xplore Digital Library, Scopus, DBLP, PubMed, Web of Science and Google Scholar, which resulted in over 60 review publications for this meta-survey contribution about deep learning during the last three to 4 years (status as of August 2020). In addition to the identified review publications, which have been arranged in different categories and sub-categories according to their data sources, the references and citations of these reviews have been retrieved and are presented. The over 60 surveys we present in our contribution reference themselves over 11,000 works, and can therefore be seen as a good starting point, especially for unfamiliar researchers, to obtain a first high-level overview of the related topic (Fig. 18).

To get a high-level overview about the commonalities and trends for the four subsections, we extracted and mapped the survey keywords. For the network extraction, we used the VOSviewer (https://www.vosviewer.com/), which is a software tool for constructing and visualizing bibliometric networks, including publication keywords. In summary, we used the text mining functionalities of the VOSviewer to construct and visualize the co-occurrence networks of the publication-provided keywords that have been extracted from the surveys.

Figure 19 shows a network visualization for the survey articles supplied keywords in the category *computer vision*. The visualization reveals the keyword "deep learning" with its connections as the main cluster. Further, main keyword clusters are "classification", "representation" and "object detection", which also reveal the main commonalities and trends for the surveys in computer vision. The left side shows several smaller categories that all belong to "activity" and "motion recognition" and connect mainly to the main clusters "classification" and "representation". The main cluster "object detection" connects mainly to "localization", "image classification", "optimization", "tracking" and "feature extraction". On the right side, we see a cluster around "computer vision" connecting to "image-captioning", "cnn", "applications" and "challenges". Interestingly, there is an own cluster for "face recognition" and "autoencoders" on the right side of the main "deep learning" cluster. Finally, there is a smaller cluster around "big data" and a "database" cluster connecting to "recognition".

As the network visualization in Fig. 20 shows, the keyword "models" is the main cluster for the category *natural language processing*. Smaller clusters of "perceptions", "automatic generation", "preferences" and "constrains" are directly derived from it. The keyword "neural networks" forms a larger sub-cluster, encapsulating the smaller categories "aspect extraction" and "embeddings". The cluster with most sub-categories is "prediction". It contains, amongst others, "textual analysis", "financial forecasting" and "computational finance", "news", "text mining" and "microblogging data", which outlines important applications of deep learning in natural language processing. Deriving from both the "models" and "prediction" clusters are "sentiment analysis" and "natural language processing", which contain the larger sub-categories "classification" and "data mining".

Figure 21 shows a network visualization for the survey articles supplied keywords in the category *medical informatics*. The visualization reveals the keywords "classification" and "convolutional-neural-network(s)" with their connections as the main clusters, and hence the overall commonalities and trends for the surveys in this area. Further, larger isolated clusters are "brain-tumour segmentation", "model" and "big data", which shows the main sub-commonalities and trends for the surveys. Additional sub-clusters center on "radio- and radiation therapy", "diabetic-retinopathy", "computer-aided diagnosis" and "networks" for "dna" and "4d nucleome". The network visualization also reveals that the sub-clusters "ct", "segmentation and automatic segmentation", "artificial intelligence", "architecture", "computer-aided detection" and "mitosis detection" are isolated and connect to several main and sub-cluster. Another smaller cluster centers on "clustering-algorithm" and "classifier" connecting mainly to the topic "model" and "big data".

"Deep learning" is also the biggest keyword cluster for the category *additional works*, as visualized in Fig. 22. Around it, several loosely connected sub-clusters around the keywords "survey", "convolutional neural networks", and "classification" were extracted, as well as some smaller clusters, amongst others, "deep belief networks", "food systems", "agriculture", "deep neural networks", "intrusion detection" and "recommender systems". The two largest sub-clusters in this category are "neural-networks", encapsulating the keywords "fusion", "recognition", "challenges" and "algorithms", just to name a few, as well as "big data", which is mainly connected to the sub-cluster "belief networks" and some smaller groups, for example "hyperspectral", "computer vision" and "feature-extraction". "Machine learning", with sub-clusters containing terms like "artificial neural-network", "channel estimation" and "5G systems and networks", forms a smaller sub-cluster to the main "deep learning" topic, as does "neural networks", which is connected to, *e.g.*, "model" and "network" keywords. This network visualization underlines the broad variety of topics covered in the *additional works* category.

Even if deep learning is still a relatively young scientific field and technology, there have been already a few hundred review and survey contributions within the last years, which can be seen as indicator for the number of breakthroughs that have been achieved with these methods. This categorized meta-survey shows that, based on the selected works of this contribution, on average more than one deep learning review per month was published during this time period, and it can be expected that these numbers will increase in the near future. Moreover, the number of references (>11,000) and citations (>17,000) of the selected works can be seen as an indicator for the current importance of deep learning. Apparently, the medical field has currently the overall highest amount of citations (>6,000), but interestingly, also the second lowest number of overall references (~2,000) included in the reviews (note also that for the medical category the least number of reviews was selected).

A reason for the massive research activities in deep learning is probably also given by the relatively easy usage and extension of these approaches: Comprehensive and user-friendly libraries and toolkits, like TensorFlow (*Abadi et al., 2016*) or PyTorch (*Paszke* *et al., 2017*), do not necessarily require an in-depth education in computer science anymore. This was not the case in the years preceding the diffusion of deep learning, when very good technical skills and programming experience, like C or C++, were necessary to implement efficient algorithms with a reasonable runtime. Additionally, nowadays, many researchers share their source code using online repositories, like GitHub, making it available to the research community. Following this trend, some publication venues started to require that the source code should be made available alongside with the publication of the paper. On top of that, most deep learning libraries and toolkits are built for Python, a high-level programming language with a faster learning curve compared to languages like C++ or Java.

The widespread of graphic processing units (GPUs) also contributed to the raise and impact of deep learning: Most deep learning libraries and toolkits support the training and execution of deep learning on GPUs, which strongly speed up the computation time thanks to their parallel architecture. Additionally, GPUs became less expensive during the last years and GPU clusters are more and more common in universities, research centers, and companies. Moreover, private corporate companies, like Google, Microsoft, and Amazon, offer online cloud computing services for little to no cost for private users or students.

Deep learning certainly has already had a massive impact in the daily life of most people *via* the countless applications that are based on this technique. It will be interesting to see what the future brings for us in the area of deep learning. However, deep learning is not perfect, as seen in tragic real life incidents, like car accidents, racist miss-classification of images, or the machine learning bot Tay from Microsoft (*Hong, 2020*; *Kohl et al., 2018*; *Hong, Choi & Williams, 2020*). In addition, tasks that claim to have outperformed humans with deep learning are often performed under so-called laboratory conditions, which means that a specific sample set for testing is used, but real-life tests are lacking, or have other shortcomings (*Liu et al. (2019)*). There exist of course specific scenarios, where machine learning has undoubtable outperformed humans, like in games with Deep Blue (*Campbell, Hoane & Hsu, 2002*) and AlphaGo (*Chen, 2016*). Here, algorithms were able to surpass the best known living human players. However, these games have fixed rules and strong restrictions, within players/algorithms have to operate. This stands in strong contrast to scenarios and tasks with almost unlimited possibilities. A human face, a spoken sentence, a driving scenario, or a pathology, are always distinct and at least slightly different, which makes it harder to predict. In chess, an algorithm can rely on the fact that the king can only move one block and will not jump over the "chessboard cliff".

In principle, deep learning is trying to imitate the human brain and how it functions and learns, although on a very basic level (*Fan et al., 2020*). This course of action can be seen as a blessing and a curse at the same time, because equivalent to the fact that we cannot take a look into someone's brain, the behavior of a trained deep neural network, with millions of neurons, connections and weights, is not fully understandable in every detail. This, on the other hand, makes it hard to predict exceptions and failures. As a consequence, deep learning is seen as a *black box* approach and this is often difficult to accept (and beforehand get it certified in applications that make critical decisions, like diagnosis decision in a diagnostic system or moral decisions of self-driving cars in unexpected traffic situations (*Kallioinen et al., 2019*)), especially when not all behaviors are foreseeable and there remains uncertainty. Thus, not everyone may feel comfortable in a self-driving car.

We can agree that deep learning is an exciting and relatively new machine learning technique, which has already brought a lot of influence and has infiltrated the life of most humans, like through virtual personal assistants (Amazon's Alexa, Apple's Siri, Google Now, *etc.*) or automatic number-plate recognition for toll roads, parking garages or law enforcement, just to name a few. On the other hand, like most new technologies with such a fast and massive impact, deep learning is not free of failures and controversies. We hope, however, that this very first meta-survey of deep learning provides a quick and comprehensive reference for interested readers. Thereby, readers should gain a high-level overview and a stimulus of this overwhelming field. Summarized, the contribution of this meta-survey covered:

- an overview of current deep learning reviews from various scientific domains;
- a categorized arrangement of the works according to their data sources, for a domain-specific and historical picture;
- an extraction of referenced works and citations to show the research influence of deep learning within these domains;
- the common architectures, methods, pros, cons, evaluations, challenges and future directions for every sub-category;
- a conclusion and critical discussion of past and future directions for deep learning.

## ACKNOWLEDGEMENTS

We want to thank S. M. for the inspiration for this contribution.

## ADDITIONAL INFORMATION AND DECLARATIONS

### Funding

The authors received funding from the Austrian Science Fund (FWF) KLI 678-B31: 'enFaced: Virtual and Augmented Reality Training and Navigation Module for 3D-Printed Facial Defect Reconstructions' and the TU Graz Lead Project (Mechanics, Modeling and Simulation of Aortic Dissection). Moreover, this work was supported by CAMed (COMET K-Project 871132), which is funded by the Austrian Federal Ministry of Transport, Innovation and Technology (BMVIT), and the Austrian Federal Ministry for Digital and Economic Affairs (BMDW), and the Styrian Business Promotion Agency (SFG). The funders had no role in study design, data collection and analysis, decision to publish, or preparation of the manuscript.

## Grant Disclosures

The following grant information was disclosed by the authors:
Austrian Science Fund (FWF): KLI 678-B31.
TU Graz Lead Project.
CAMed (COMET K-Project): 871132.
Austrian Federal Ministry of Transport, Innovation and Technology (BMVIT).
Austrian Federal Ministry for Digital and Economic Affairs (BMDW).
Styrian Business Promotion Agency (SFG).

## Competing Interests

The authors declare that they have no competing interests.

## Author Contributions

- Jan Egger conceived and designed the experiments, performed the experiments, analyzed the data, prepared figures and/or tables, authored or reviewed drafts of the paper, and approved the final draft.
- Antonio Pepe analyzed the data, authored or reviewed drafts of the paper, and approved the final draft.
- Christina Gsaxner analyzed the data, authored or reviewed drafts of the paper, and approved the final draft.
- Yuan Jin analyzed the data, authored or reviewed drafts of the paper, and approved the final draft.
- Jianning Li analyzed the data, authored or reviewed drafts of the paper, and approved the final draft.
- Roman Kern analyzed the data, authored or reviewed drafts of the paper, and approved the final draft.

## Data Availability

The following information was supplied regarding data availability:

There is no additional data or source code from the authors regarding this survey contribution.

# REFERENCES

**Abadi M, Barham P, Chen J, Chen Z, Davis A, Dean J, Devin M, Ghemawat S, Irving G, Isard M, Kudlur M, Levenberg J, Monga R, Moore S, Murray DG, Steiner B, Tucker P, Vasudevan V, Warden P, Wicke M, Yu Y, Zheng X. 2016.** Tensorflow: a system for large-scale machine learning. In: *12th USENIX symposium on operating systems design and implementation (OSDI 16)*. 265–283.

**Almeida F, Xexéo G. 2019.** Word embeddings: a survey. *Available at http://arxiv.org/abs/1901.09069*.

**Arulkumaran K, Deisenroth MP, Brundage M, Bharath AA. 2017.** Deep reinforcement learning: a brief survey. *IEEE Signal Processing Magazine* **34(6)**:26–38 DOI 10.1109/MSP.2017.2743240.

**Ball JE, Anderson DT, Chan CS Sr. 2017.** Comprehensive survey of deep learning in remote sensing: theories, tools, and challenges for the community. *Journal of Applied Remote Sensing* **11(4)**:042609 DOI 10.1117/1.JRS.11.042609.

**Bevilacqua V, Biasi L, Pepe A, Mastronardi G, Caporusso N. 2015.** A computer vision method for the Italian finger spelling recognition. In: *International Conference on Intelligent Computing*. Springer, 264–274.

**Campbell M, Hoane AJ Jr, Hsu FH. 2002.** Deep blue. *Artificial Intelligence* **134(1–2)**:57–83 DOI 10.1016/S0004-3702(01)00129-1.

**Chen H, Liu X, Yin D, Tang J. 2017.** A survey on dialogue systems: recent advances and new frontiers. *ACM SIGKDD Explorations Newsletter* **19(2)**:25–35 DOI 10.1145/3166054.3166058.

**Chen JX. 2016.** The evolution of computing: Alphago. *Computing in Science & Engineering* **18(4)**:4–7 DOI 10.1109/MCSE.2016.74.

**Çiçek Ö, Abdulkadir A, Lienkamp SS, Brox T, Ronneberger O. 2016.** 3D U-net: learning dense volumetric segmentation from sparse annotation. In: *International Conference on Medical Image Computing and Computer-assisted Intervention*. Springer, 424–432.

**Dargan S, Kumar M, Ayyagari MR, Kumar G. 2019.** A survey of deep learning and its applications: a new paradigm to machine learning. *Archives of Computational Methods in Engineering* **27**:1–22 DOI 10.1007/s11831-019-09344-w.

**Deng L, Yu D. 2014.** Deep learning: methods and applications. *Foundations and Trends in Signal Processing* **7(3–4)**:197–387 DOI 10.1561/2000000039.

**Do HH, Prasad P, Maag A, Alsadoon A. 2019.** Deep learning for aspect-based sentiment analysis: a comparative review. *Expert Systems with Applications* **118(6)**:272–299 DOI 10.1016/j.eswa.2018.10.003.

**Egger J, Gsaxner C, Pepe A, Li J. 2020.** Medical deep learning: a systematic meta-review. *Available at http://arxiv.org/abs/2010.14881*.

**Emmert-Streib F, Yang Z, Feng H, Tripathi S, Dehmer M. 2020.** An introductory review of deep learning for prediction models with big data. *Frontiers in Artificial Intelligence* **3**:4 DOI 10.3389/frai.2020.00004.

**Fan J, Fang L, Wu J, Guo Y, Dai Q. 2020.** From brain science to artificial intelligence. *Engineering* **6(3)**:248–252 DOI 10.1016/j.eng.2019.11.012.

**Gao J, Galley M, Li L. 2018.** Neural approaches to conversational AI. In: *The 41st International ACM SIGIR Conference on Research & Development in Information Retrieval*. 1371–1374.

**Garcia-Garcia A, Orts-Escolano S, Oprea S, Villena-Martinez V, Martinez-Gonzalez P, Garcia-Rodriguez J. 2018.** A survey on deep learning techniques for image and video semantic segmentation. *Applied Soft Computing* **70(9)**:41–65 DOI 10.1016/j.asoc.2018.05.018.

**Gatt A, Krahmer E. 2018.** Survey of the state of the art in natural language generation: core tasks, applications and evaluation. *Journal of Artificial Intelligence Research* **61**:65–170 DOI 10.5555/3241691.3241693.

**Gibson E, Li W, Sudre C, Fidon L, Shakir DI, Wang G, Eaton-Rosen Z, Gray R, Doel T, Hu Y, Whyntie T, Nachev P, Modat M, Barratt DC, Ourselin S, Jorge Cardoso M, Vercauteren T. 2018.** Niftynet: a deep-learning platform for medical imaging. *Computer Methods and Programs in Biomedicine* **158**:113–122 DOI 10.1016/j.cmpb.2018.01.025.

**Goodfellow I, Bengio Y, Courville A, Bengio Y. 2016.** *Deep learning*. Vol. 1. Cambridge: MIT Press.

**Gsaxner C, Pepe A, Wallner J, Schmalstieg D, Egger J. 2019a.** Markerless image-to-face registration for untethered augmented reality in head and neck surgery. In: *International Conference on Medical Image Computing and Computer-Assisted Intervention*. Springer, 236–244.

**Gsaxner C, Pfarrkirchner B, Lindner L, Pepe A, Roth PM, Egger J, Wallner J. 2018.** Pet-train: automatic ground truth generation from pet acquisitions for urinary bladder segmentation in ct images using deep learning. In: *2018 11th Biomedical Engineering International Conference (BMEiCON)*. Piscataway: IEEE, 1–5.

**Gsaxner C, Roth PM, Wallner J, Egger J. 2019b.** Exploit fully automatic low-level segmented pet data for training high-level deep learning algorithms for the corresponding ct data. *PLOS ONE* **14(3)**:e0212550 DOI 10.1371/journal.pone.0212550.

**Guo Y, Liu Y, Oerlemans A, Lao S, Wu S, Lew MS. 2016.** Deep learning for visual understanding: a review. *Neurocomputing* **187**:27–48 DOI 10.1016/j.neucom.2015.09.116.

**Hahn LD, Mistelbauer G, Higashigaito K, Koci M, Willemink MJ, Sailer AM, Fischbein M, Fleischmann D. 2020.** Ct-based true-and false-lumen segmentation in type b aortic dissection using machine learning. *Radiology: Cardiothoracic Imaging* **2(3)**:e190179 DOI 10.1148/ryct.2020190179.

**Haskins G, Kruger U, Yan P. 2020.** Deep learning in medical image registration: a survey. *Machine Vision and Applications* **31(1)**:1–18 DOI 10.1007/s00138-020-01060-x.

**Herath S, Harandi M, Porikli F. 2017.** Going deeper into action recognition: a survey. *Image and Vision Computing* **60(2)**:4–21 DOI 10.1016/j.imavis.2017.01.010.

**Hesamian MH, Jia W, He X, Kennedy P. 2019.** Deep learning techniques for medical image segmentation: achievements and challenges. *Journal of Digital Imaging* **32(4)**:582–596 DOI 10.1007/s10278-019-00227-x.

**Hong J. 2020.** Why is artificial intelligence blamed more? Analysis of faulting artificial intelligence for self-driving car accidents in experimental settings. *International Journal of Human-Computer Interaction* **36(18)**:1768–1774 DOI 10.1080/10447318.2020.1785693.

**Hong J-W, Choi S, Williams D. 2020.** Sexist AI: an experiment integrating CASA and ELM. *International Journal of Human-Computer Interaction* **36(20)**:1928–1941 DOI 10.1080/10447318.2020.1801226.

**Hossain MZ, Sohel F, Shiratuddin MF, Laga H. 2019.** A comprehensive survey of deep learning for image captioning. *ACM Computing Surveys (CsUR)* **51(6)**:1–36 DOI 10.1145/3295748.

**Hu Z, Tang J, Wang Z, Zhang K, Zhang L, Sun Q. 2018.** Deep learning for image-based cancer detection and diagnosis: a survey. *Pattern Recognition* **83(11)**:134–149 DOI 10.1016/j.patcog.2018.05.014.

**Huttenlocher DP, Klanderman GA, Rucklidge WJ. 1993.** Comparing images using the hausdorff distance. *IEEE Transactions on Pattern Analysis and Machine Intelligence* **15(9)**:850–863 DOI 10.1109/34.232073.

**Jiao L, Zhang F, Liu F, Yang S, Li L, Feng Z, Qu R. 2019.** A survey of deep learning-based object detection. *IEEE Access* **7**:128837–128868 DOI 10.1109/ACCESS.2019.2939201.

**Kalinin AA, Higgins GA, Reamaroon N, Soroushmehr S, Allyn-Feuer A, Dinov ID, Najarian K, Athey BD. 2018.** Deep learning in pharmacogenomics: from gene regulation to patient stratification. *Pharmacogenomics* **19(7)**:629–650 DOI 10.2217/pgs-2018-0008.

**Kallioinen N, Pershina M, Zeiser J, Nosrat Nezami F, Pipa G, Stephan A, König P. 2019.** Moral judgements on the actions of self-driving cars and human drivers in dilemma situations from different perspectives. *Frontiers in Psychology* **10**:2415 DOI 10.3389/fpsyg.2019.02415.

**Kamilaris A, Prenafeta-Boldú FX. 2018.** Deep learning in agriculture: a survey. *Computers and Electronics in Agriculture* **147(2)**:70–90 DOI 10.1016/j.compag.2018.02.016.

**Karner F, Gsaxner C, Pepe A, Li J, Fleck P, Arth C, Wallner J, Egger J. 2020.** Single-shot deep volumetric regression for mobile medical augmented reality. In: *Multimodal Learning for Clinical Decision Support and Clinical Image-Based Procedures*. Springer, 64–74.

**Kohl C, Knigge M, Baader G, Böhm M, Krcmar H. 2018.** Anticipating acceptance of emerging technologies using twitter: the case of self-driving cars. *Journal of Business Economics* **88(5)**:617–642 DOI 10.1007/s11573-018-0897-5.

**Krizhevsky A, Sutskever I, Hinton GE. 2012.** Imagenet classification with deep convolutional neural networks. *Advances in Neural Information Processing Systems* **25**:1097–1105 DOI 10.1145/3065386.

**Kwon D, Kim H, Kim J, Suh SC, Kim I, Kim KJ. 2019.** A survey of deep learning-based network anomaly detection. *Cluster Computing* **22(1)**:949–961 DOI 10.1007/s10586-017-1117-8.

**Labini MS, Gsaxner C, Pepe A, Wallner J, Egger J, Bevilacqua V. 2019.** Depth-awareness in a system for mixed-reality aided surgical procedures. In: *International Conference on Intelligent Computing*. Springer, 716–726.

**Lai T, Bui T, Li S. 2018.** A review on deep learning techniques applied to answer selection. In: *Proceedings of the 27th International Conference on Computational Linguistics*. 2132–2144.

**Lan K, Wang DT, Fong S, Liu LS, Wong KK, Dey N. 2018.** A survey of data mining and deep learning in bioinformatics. *Journal of Medical Systems* **42(8)**:1–20 DOI 10.1007/s10916-018-1003-9.

**LeCun Y, Bengio Y, Hinton G. 2015.** Deep learning. *Nature* **521(7553)**:436–444 DOI 10.1038/nature14539.

**Li J, Egger J. 2020.** *Towards the automatization of cranial implant design in cranioplasty*. Cham: Springer.

**Li J, Pimentel P, Szengel A, Ehlke M, Lamecker H, Zachow S, Estacio L, Doenitz C, Ramm H, Shi H, Chen X, Matzkin F, Newcombe V, Ferrante E, Jin Y, Ellis DG, Aizenberg MR, Kodym O, Španěl M, Herout A, Mainprize JG, Fishman Z, Hardisty MR, Bayat A, Shit S, Wang B, Liu Z, Eder M, Pepe A, Gsaxner C, Alves V, Zefferer U, Von Campe G, Pistracher K, Schäfer U, Schmalstieg D, Menze BH, Glocker B, Egger J. 2021.** Autoimplant 2020 -first miccai challenge on automatic cranial implant design. *IEEE Transactions on Medical Imaging* **40(9)**:2329–2342 DOI 10.1109/TMI.2021.3077047.

**Li J, Sun A, Han J, Li C. 2020.** A survey on deep learning for named entity recognition. *IEEE Transactions on Knowledge and Data Engineering* 1–1 DOI 10.1109/TKDE.2020.3038670.

**Li S, Deng W. 2020.** Deep facial expression recognition: a survey. *IEEE Transactions on Affective Computing* 1–1 DOI 10.1109/TAFFC.2020.2981446.

**Li Y. 2017.** Deep reinforcement learning: an overview. *Available at http://arxiv.org/abs/1701.07274*.

**Litjens G, Kooi T, Bejnordi BE, Setio AAA, Ciompi F, Ghafoorian M, Van Der Laak JA, Van Ginneken B, Sánchez CI. 2017.** A survey on deep learning in medical image analysis. *Medical Image Analysis* **42(13)**:60–88 DOI 10.1016/j.media.2017.07.005.

**Liu L, Ouyang W, Wang X, Fieguth P, Chen J, Liu X, Pietikäinen M. 2020.** Deep learning for generic object detection: a survey. *International Journal of Computer Vision* **128(2)**:261–318 DOI 10.1007/s11263-019-01247-4.

**Liu X, Faes L, Kale AU, Wagner SK, Fu DJ, Bruynseels A, Mahendiran T, Moraes G, Shamdas M, Kern C, Ledsam JR, Schmid MK, Balaskas K, Topol EJ, Bachmann LM, Keane PA, Denniston AK. 2019.** A comparison of deep learning performance against health-care professionals in detecting diseases from medical imaging: a systematic review and meta-analysis. *The Lancet Digital Health* **1(6)**:e271–e297.

**Long J, Shelhamer E, Darrell T. 2015.** Fully convolutional networks for semantic segmentation. In: *Proceedings of the IEEE Conference on Computer Vision and Pattern Recognition*. 3431–3440.

**Lundervold AS, Lundervold A. 2019.** An overview of deep learning in medical imaging focusing on MRI. *Zeitschrift für Medizinische Physik* **29(2)**:102–127 DOI 10.1016/j.zemedi.2018.11.002.

**Masi I, Wu Y, Hassner T, Natarajan P. 2018.** Deep face recognition: a survey. In: *2018 31st SIBGRAPI Conference on Graphics, Patterns and Images (SIBGRAPI)*. Piscataway: IEEE, 471–478.

**Mazurowski MA, Buda M, Saha A, Bashir MR. 2019.** Deep learning in radiology: an overview of the concepts and a survey of the state of the art with focus on MRI. *Journal of Magnetic Resonance Imaging* **49(4)**:939–954 DOI 10.1002/jmri.26534.

**McCulloch WS, Pitts W. 1990.** A logical calculus of the ideas immanent in nervous activity. *Bulletin of Mathematical Biology* **52(1–2)**:99–115 DOI 10.1016/S0092-8240(05)80006-0.

**Meyer P, Noblet V, Mazzara C, Lallement A. 2018.** Survey on deep learning for radiotherapy. *Computers in Biology and Medicine* **98**:126–146 DOI 10.1016/j.compbiomed.2018.05.018.

**Minaee S, Abdolrashidi A, Su H, Bennamoun M, Zhang D. 2019.** Biometric recognition using deep learning: a survey. *Available at http://arxiv.org/abs/1912.00271*.

**Minaee S, Boykov YY, Porikli F, Plaza AJ, Kehtarnavaz N, Terzopoulos D. 2020.** Image segmentation using deep learning: a survey. *Available at http://arxiv.org/abs/2001.05566*.

**Mohammadi M, Al-Fuqaha A, Sorour S, Guizani M. 2018.** Deep learning for IoT big data and streaming analytics: a survey. *IEEE Communications Surveys & Tutorials* **20(4)**:2923–2960 DOI 10.1109/COMST.2018.2844341.

**Mousavi SS, Schukat M, Howley E. 2016.** Deep reinforcement learning: an overview. In: *Proceedings of SAI Intelligent Systems Conference*. Springer, 426–440.

**Ota K, Dao MS, Mezaris V, Natale FGD. 2017.** Deep learning for mobile multimedia: a survey. *ACM Transactions on Multimedia Computing, Communications, and Applications (TOMM)* **13(3s)**:1–22 DOI 10.1145/3092831.

**Paszke A, Gross S, Chintala S, Chanan G, Yang E, DeVito Z, Lin Z, Desmaison A, Antiga L, Lerer A. 2017.** Automatic differentiation in pytorch. *Available at https://openreview.net/pdf?id=BJJsrmfCZ*.

**Pepe A, Li J, Rolf-Pissarczyk M, Gsaxner C, Chen X, Holzapfel GA, Egger J. 2020.** Detection, segmentation, simulation and visualization of aortic dissections: a review. *Medical Image Analysis* **65(1)**:101773 DOI 10.1016/j.media.2020.101773.

**Pepe A, Trotta GF, Mohr-Ziak P, Gsaxner C, Wallner J, Bevilacqua V, Egger J. 2019.** A markerless registration approach for mixed reality-aided maxillofacial surgery: a pilot evaluation. *Journal of Digital Imaging* **32(6)**:1008–1018 DOI 10.1007/s10278-019-00272-6.

**Pouyanfar S, Sadiq S, Yan Y, Tian H, Tao Y, Reyes MP, Shyu M-L, Chen S-C, Iyengar S. 2018.** A survey on deep learning: algorithms, techniques, and applications. *ACM Computing Surveys (CSUR)* **51(5)**:1–36 DOI 10.1145/3234150.

**Raghu M, Schmidt E. 2020.** A survey of deep learning for scientific discovery. *Available at http://arxiv.org/abs/2003.11755*.

**Ramachandram D, Taylor GW. 2017.** Deep multimodal learning: a survey on recent advances and trends. *IEEE Signal Processing Magazine* **34(6)**:96–108 DOI 10.1109/MSP.2017.2738401.

**Ravì D, Wong C, Deligianni F, Berthelot M, Andreu-Perez J, Lo B, Yang G-Z. 2016.** Deep learning for health informatics. *IEEE Journal of Biomedical and Health Informatics* **21(1)**:4–21 DOI 10.1109/JBHI.2016.2636665.

**Ronneberger O, Fischer P, Brox T. 2015.** U-net: convolutional networks for biomedical image segmentation. In: *International Conference on Medical Image Computing and Computer-assisted Intervention*. Springer, 234–241.

**Sampat MP, Wang Z, Markey MK, Whitman GJ, Stephens TW, Bovik AC. 2006.** Measuring intra-and inter-observer agreement in identifying and localizing structures in medical images. In: *2006 International Conference on Image Processing*. Piscataway: IEEE, 81–84.

**Santhanam S, Shaikh S. 2019.** A survey of natural language generation techniques with a focus on dialogue systems-past, present and future directions. *Available at http://arxiv.org/abs/1906.00500*.

**Shen D, Wu G, Suk H-I. 2017.** Deep learning in medical image analysis. *Annual Review of Biomedical Engineering* **19(1)**:221–248 DOI 10.1146/annurev-bioeng-071516-044442.

**Shi T, Keneshloo Y, Ramakrishnan N, Reddy CK. 2018.** Neural abstractive text summarization with sequence-to-sequence models. *Available at http://arxiv.org/abs/1812.02303*.

**Shickel B, Tighe PJ, Bihorac A, Rashidi P. 2017.** Deep EHR: a survey of recent advances in deep learning techniques for electronic health record (ehr) analysis. *IEEE Journal of Biomedical and Health Informatics* **22(5)**:1589–1604 DOI 10.1109/JBHI.2017.2767063.

**Shorten C, Khoshgoftaar TM. 2019.** A survey on image data augmentation for deep learning. *Journal of Big Data* **6(1)**:1–48 DOI 10.1186/s40537-019-0197-0.

**Sundararajan K, Woodard DL. 2018.** Deep learning for biometrics: a survey. *ACM Computing Surveys (CSUR)* **51(3)**:1–34 DOI 10.1145/3190618.

**Sze V, Chen Y-H, Yang T-J, Emer JS. 2017.** Efficient processing of deep neural networks: a tutorial and survey. *Proceedings of the IEEE* **105(12)**:2295–2329 DOI 10.1109/JPROC.2017.2761740.

**Voulodimos A, Doulamis N, Doulamis A, Protopapadakis E. 2018.** Deep learning for computer vision: a brief review. *Computational Intelligence and Neuroscience* **2018**:1–13 DOI 10.1155/2018/7068349.

**Wallner J, Mischak I, Egger J. 2019.** Computed tomography data collection of the complete human mandible and valid clinical ground truth models. *Scientific Data* **6(1)**:1–14 DOI 10.1038/sdata.2019.3.

**Wang J, Ma Y, Zhang L, Gao RX, Wu D. 2018a.** Deep learning for smart manufacturing: methods and applications. *Journal of Manufacturing Systems* **48(2)**:144–156 DOI 10.1016/j.jmsy.2018.01.003.

**Wang M, Deng W. 2018.** Deep face recognition: a survey. *Available at http://arxiv.org/abs/1804.06655*.

**Wang P, Li W, Ogunbona P, Wan J, Escalera S. 2018b.** RGB-D-based human motion recognition with deep learning: a survey. *Computer Vision and Image Understanding* **171(3)**:118–139 DOI 10.1016/j.cviu.2018.04.007.

**Wang Z, Chen J, Hoi SC. 2020.** Deep learning for image super-resolution: a survey. *IEEE Transactions on Pattern Analysis and Machine Intelligence* **43(10)**:3365–3387 DOI 10.1109/TPAMI.2020.2982166.

**Wang Z, She Q, Ward TE. 2019.** Generative adversarial networks in computer vision: a survey and taxonomy. *Available at http://arxiv.org/abs/1906.01529*.

**Wild D, Weber M, Egger J. 2019.** Client/server based online environment for manual segmentation of medical images. *Available at http://arxiv.org/abs/1904.08610*.

**Xing F, Xie Y, Su H, Liu F, Yang L. 2017.** Deep learning in microscopy image analysis: a survey. *IEEE Transactions on Neural Networks and Learning Systems* **29(10)**:4550–4568 DOI 10.1109/TNNLS.2017.2766168.

**Xing FZ, Cambria E, Welsch RE. 2018.** Natural language based financial forecasting: a survey. *Artificial Intelligence Review* **50(1)**:49–73 DOI 10.1007/s10462-017-9588-9.

**Yadav V, Bethard S. 2019.** A survey on recent advances in named entity recognition from deep learning models. *Available at http://arxiv.org/abs/1910.11470*.

**Young T, Hazarika D, Poria S, Cambria E. 2018.** Recent trends in deep learning based natural language processing. *IEEE Computational Intelligence Magazine* **13(3)**:55–75 DOI 10.1109/MCI.2018.2840738.

**Zeiler MD, Fergus R. 2014.** Visualizing and understanding convolutional networks. In: *European Conference on Computer Vision*. Springer, 818–833.

**Zhang C, Patras P, Haddadi H. 2019.** Deep learning in mobile and wireless networking: a survey. *IEEE Communications Surveys & Tutorials* **21(3)**:2224–2287 DOI 10.1109/COMST.2019.2904897.

**Zhang L, Wang S, Liu B. 2018.** Deep learning for sentiment analysis: a survey. *Wiley Interdisciplinary Reviews: Data Mining and Knowledge Discovery* **8(4)**:e1253 DOI 10.1002/widm.1253.

**Zhang Q, Yang LT, Chen Z, Li P. 2018.** A survey on deep learning for big data. *Information Fusion* **42(9)**:146–157 DOI 10.1016/j.inffus.2017.10.006.

**Zhang S, Yao L, Sun A, Tay Y. 2019.** Deep learning based recommender system: a survey and new perspectives. *ACM Computing Surveys (CSUR)* **52(1)**:1–38 DOI 10.1145/3285029.

**Zhang Y, Rahman MM, Braylan A, Dang B, Chang H-L, Kim H, McNamara Q, Angert A, Banner E, Khetan V, McDonnell T, Nguyen AT, Xu D, Wallace BC, Lease M. 2016.** Neural information retrieval: a literature review. *Available at http://arxiv.org/abs/1611.06792*.

**Zhang Z, Cui P, Zhu W. 2020.** Deep learning on graphs: a survey. *IEEE Transactions on Knowledge and Data Engineering* 1-1 DOI 10.1109/TKDE.2020.2981333.

**Zhao ZQ, Zheng P, Xu ST, Wu X. 2019.** Object detection with deep learning: a review. *IEEE Transactions on Neural Networks and Learning Systems* **30(11)**:3212–3232 DOI 10.1109/TNNLS.2018.2876865.